\documentclass{aa}  

\usepackage{graphicx}
\usepackage{txfonts}
\usepackage{mathtools}  % additional packages
\usepackage[breaklinks]{hyperref}
\newcommand{\KC}{$\kappa_{\mathrm{const}}$}
\newcommand{\KBL}{$\kappa_{\mathrm{BL}}$}

\begin{document} 
     \title{Oscillatory migration of accreting protoplanets\\
       driven by a 3D distortion of the gas flow}

%   \titlerunning{}
%   \authorrunning{}

%   \subtitle{}

    \author{O. Chrenko\inst{1}
          \and
	  M. Lambrechts\inst{2}}

   \institute{Institute of Astronomy, Charles University in Prague,
              V Hole\v sovi\v ck\'ach 2, CZ--18000 Prague~8, Czech Republic\\
	      \email{chrenko@sirrah.troja.mff.cuni.cz}
              \and
	      Lund Observatory, Department of Astronomy and Theoretical Physics,
	      Lund University, Box 43, 22100 Lund, Sweden\\
             }

   \date{Received ??; accepted ??} %Received April 25, 2017;}

% \abstract{}{}{}{}{} 
% 5 {} token are mandatory
 
  \abstract
  % context heading (optional)
  % {} leave it empty if necessary
  {
   The dynamics of a low-mass protoplanet accreting solids
   is influenced by the heating torque,
   which was found to suppress inward
   migration in protoplanetary disks with constant opacities.
  }
  {
   We investigate the differences of the heating torque
   between disks with constant and temperature-dependent opacities.
  }
  {
   Interactions of a super-Earth-sized protoplanet with the
   gas disk are explored using 3D radiation hydrodynamic simulations.
  }
  {
     Accretion heating of the protoplanet
     creates a hot underdense region in the surrounding gas,
     leading to misalignment of the local density and pressure
     gradients.
   As a result, the 3D gas flow is perturbed
   and some of the streamlines form a retrograde spiral rising
   above the protoplanet.
   In the constant-opacity disk, the perturbed
   flow reaches a steady state and the underdense gas responsible
   for the heating torque remains distributed
   in accordance with previous studies.
   If the opacity is non-uniform, however, the differences in the
   disk structure can lead to more vigorous streamline distortion
   and eventually to a flow instability.
   The underdense gas develops a one-sided asymmetry 
   which circulates around the protoplanet in a retrograde fashion.
   The heating torque thus strongly oscillates in time and
   does not on average counteract inward migration.
  }
  { 
    The torque variations make the radial drift of the protoplanet oscillatory,
    consisting of short intervals of alternating rapid inward and outward migration.
    We speculate that transitions between the positive and oscillatory
    heating torque may occur in specific disk regions
    susceptible to vertical convection,
    resulting in the convergent migration of multiple planetary embryos.
  }


   \maketitle
%
%________________________________________________________________

\section{Introduction}
\label{sec:intro}

Migration of protoplanets embedded in their natal gas disks
is a key evolutionary step in formation of each planetary system.
Low-mass protoplanets, incapable of gap opening \citep{Crida_etal_2006Icar..181..587C},
undergo Type I migration under the influence of the
gravitational torques exerted by the Lindblad spiral wakes
\citep{Golreich_Tremaine_1979ApJ...233..857G,Ward_1986Icar...67..164W}
and by the gas in their corotation region
\citep{Ward_1991LPI....22.1463W,Masset_2002A&A...387..605M,Tanaka_etal_2002ApJ...565.1257T,Paardekooper_Mellema_2006A&A...459L..17P,Baruteau_Masset_2008ApJ...672.1054B,Paardekooper_Mellema_2008A&A...478..245P,Masset_Casoli_2009ApJ...703..857M,Paardekooper_Papaloizou_2009MNRAS.394.2283P,Baruteau_etal_2011A&A...533A..84B}.
Type I migration
depends in a complicated way on the disk structure and thermophysics
\citep[e.g.][]{Kley_Crida_2008,Paardekooper_etal_2010MNRAS.401.1950P,Paardekooper_etal_2011MNRAS.410..293P,Lega_etal_2015MNRAS.452.1717L}.
Its detailed understanding is therefore
essential for the creation of realistic population synthesis models
\citep[e.g.][]{Coleman_Nelson_2016MNRAS.457.2480C}.

It has been recently discovered that low-mass protoplanets
evolving in radiative disks are subject to
thermal torques
\citep{Lega_etal_2014MNRAS.440..683L,Benitez-Llambay_etal_2015Natur.520...63B,Masset_VelascoRomero_2017MNRAS.465.3175M,Masset_2017MNRAS.472.4204M}
related to the thermal perturbations induced by a protoplanet in its vicinity.
If the protoplanet itself is cold and non-luminous \citep{Lega_etal_2014MNRAS.440..683L}, 
the gas which arrives into its
potential well becomes heated mostly as a result
of compression (by means of the thermodynamic `$P\mathrm{d}V$' term).
The arising temperature excess becomes smoothed out by the
radiative transfer, so when the gas leaves the high pressure region
it lacks some of its internal energy --
compared to the state before the compression --
and therefore becomes colder and overdense.
Two overdense lobes appear along the streamlines
outflowing from the Hill sphere and their asymmetry
makes the total torque felt by the protoplanet more negative,
enhancing the inward migration.
The process is known as the cold-finger effect \citep{Lega_etal_2014MNRAS.440..683L}.

In the opposite limit, the protoplanet is hot,
as a result of the solid material deposition
during its formation \citep[e.g. by pebble accretion;][]{Ormel_Klahr_2010A&A...520A..43O,Lambrechts_Johansen_2012A&A...544A..32L}.
In such a case, the luminous protoplanet acts as a local heat source
for the surrounding gas. Once the gas is heated, it 
becomes underdense compared to the situation without accretion heating.
\cite{Benitez-Llambay_etal_2015Natur.520...63B}
performed 3D radiation hydrodynamic simulations
with the assumption of the constant disk opacity and found
that the hot protoplanet on a fixed circular orbit
creates two underdense lobes of gas, again associated
with the outflow from the Hill sphere.
The rear lobe (positioned behind the protoplanet outwards from its
orbit) is dominant, therefore there is an overabundance of
gas ahead of the protoplanet and the resulting torque
becomes more positive, supporting outward migration.
The positive enhancement was named the heating torque.
It was proposed to be an additional mechanism
\citep[along with other posibilites, see e.g.][]{Rafikov_2002ApJ...572..566R,Paardekooper_Mellema_2006A&A...459L..17P,Morbidelli_etal_2008A&A...478..929M,Li_etal_2009ApJ...690L..52L,Yu_etal_2010ApJ...712..198Y,Kretke_Lin_2012ApJ...755...74K,Bitsch_etal_2013A&A...549A.124B,Fung_Chiang_2017ApJ...839..100F,Brasser_etal_2018ApJ...864L...8B,Miranda_Lai_2018MNRAS.473.5267M,McNally_etal_2019MNRAS.484..728M}
capable of preventing
the destruction of terrestrial-sized planetary embryos by
an overly efficient inward migration
\citep{Korycansky_Pollack_1993Icar..102..150K,Ward_1997ApJ...482L.211W,Tanaka_etal_2002ApJ...565.1257T}.

Moreover, the heating torque has important dynamical consequences
for migrating protoplanets
\citep{Broz_etal_2018A&A...620A.157B,Chrenko_etal_2018ApJ...868..145C}
because it can excite orbital eccentricities
and inclinations by means of the hot-trail effect
\citep{Eklund_Masset_2017MNRAS.469..206E,Chrenko_etal_2017A&A...606A.114C},
which counteracts the otherwise efficient eccentricity and inclination damping
by waves \citep{Tanaka_Ward_2004ApJ...602..388T,Cresswell_etal_2007A&A...473..329C}.

Nevertheless, the heating torque has not
been extensively studied in 3D radiative disks with non-uniform
opacities. However, realistic opacity functions
\citep{Bell_Lin_1994ApJ...427..987B,Semenov_etal_2003A&A...410..611S,Zhu_etal_2012ApJ...746..110Z}
are of a great significance for the disk
structure and planet migration
\citep[e.g.][]{Kretke_Lin_2012ApJ...755...74K,Bitsch_etal_2013A&A...549A.124B}
and this work will show that the heating
torque is affected as well.

In this paper, we reinvestigate the thermal
torques acting on a low-mass protoplanet, with a special emphasis
on the heating torque in a disk with non-uniform opacity.
We examine the streamlines near the protoplanet
and point out the importance of their 3D distortion
for redistribution of the hot underdense gas responsible
for the heating torque.

%__________________________________________________________________

\section{Model}
\label{sec:model}

We consider a protoplanetary system consisting
of a central protostar surrounded by a disk
of coupled gas and dust in which a single protoplanet is embedded.
The fluid part of the disk model accounts only for the gas, assuming
the dust is a passive tracer that acts as the main contributor
to the material opacity.

\subsection{Governing equations}
\label{sec:equations}

The disk is described using Eulerian hydrodynamics
on a staggered spherical mesh centred on the protostar.
The spherical coordinates consist of the radial distance $r$,
azimuthal angle $\theta$ and colatitude $\phi$.
Our model is built on top of the hydrodynamic module of \textsc{fargo3d}\footnote{\raggedright Public version
  of the code is available at \url{http://fargo.in2p3.fr/}.}
  \citep{Benitez-Llambay_Masset_2016ApJS..223...11B}, which solves the equations
of continuity and momentum of a fluid
\begin{equation}
  \frac{\partial\rho}{\partial t} + \left(\vec{v}\cdot\nabla\right)\rho = - \rho\nabla\cdot\vec{v}  \, ,
  \label{eq:continuity}
\end{equation}
\begin{equation}
  \frac{\partial\vec{v}}{\partial t} + \left(\vec{v}\cdot\nabla\right){\vec{v}} = - \frac{\nabla P}{\rho} - \nabla\Phi + \frac{\nabla\cdot\tens{T}}{\rho} - \left[2\vec{\Omega}\times\vec{v}+\vec{\Omega}\times\left(\vec{\Omega}\times\vec{r}\right) \right]\, ,
  \label{eq:naviere_stokes}
\end{equation}
where the individual quantities stand for the volume density $\rho$, time $t$,
flow velocity vector $\vec{v}$ (with the radial, azimuthal and vertical components
$v_{r}$, $v_{\theta}$ and $v_{\phi}$), pressure $P$, gravitational potential 
$\Phi$ of the protostar and the protoplanet, viscous stress
tensor $\tens{T}$ \citep{Benitez-Llambay_Masset_2016ApJS..223...11B},
radius vector $\vec{r}$ and
angular velocity vector $\vec{\Omega}$ which is non-zero
because we work in the reference frame corotating with the protoplanet.

To account for the energy production, dissipation and transport in the disk,
we implement the two-temperature energy equations for the gas and the radiation field
according to the formulation of \cite{Bitsch_etal_2013A&A...549A.124B}:
\begin{equation}
  \frac{\partial E_{\mathrm{R}}}{\partial t} + \nabla\cdot\vec{F} = \rho\kappa_{\mathrm{P}}\left[4\sigma T^{4} - cE_{\mathrm{R}}  \right] \, ,
  \label{eq:e_rad}
\end{equation}
\begin{equation}
  \frac{\partial \epsilon}{\partial t} + \left(\vec{v}\cdot\nabla\right)\epsilon = -P\nabla\cdot\vec{v} - \rho\kappa_{\mathrm{P}}\left[ 4\sigma T^{4} - cE_{\mathrm{R}} \right] + Q_{\mathrm{visc}} + Q_{\mathrm{art}} + Q_{\mathrm{acc}} \, ,
  \label{eq:e_int}
\end{equation}
where $E_{\mathrm{R}}$ is the radiative energy,
$\vec{F}$ the radiation flux,
$\kappa_{\mathrm{P}}$ the Planck opacity, $\sigma$ the Stefan-Boltzmann constant,
$T$ the gas temperature,
$c$ the speed of light, $\epsilon$ the internal energy of the gas,
$Q_{\mathrm{visc}}$ the viscous heating term \citep{Mihalas_WeibelMihalas_1984frh..book.....M},
$Q_{\mathrm{art}}$ describes the heating due to the artificial viscosity \citep{Stone_Norman_1992ApJS...80..753S}
and $Q_{\mathrm{acc}}$ the heat released when the protoplanet is accreting.
Stellar irradiation is neglected in this paper for simplicity
although our code is capable of including it as well.

The state equation of the ideal gas is used:
\begin{equation}
  P = (\gamma-1)\epsilon = (\gamma-1)\rho c_{V} T \, ,
  \label{eq:eos}
\end{equation}
where $\gamma$ is the adiabatic index and $c_{V}$ is the
specific heat at constant volume, which
can be expressed as $c_{V}=R/(\mu(\gamma-1))$,
where $R$ is the universal gas constant and $\mu$ is the mean
molecular weight.

The flux-limited diffusion approximation \citep[FLD;][]{Levermoe_Pomraning_1981ApJ...248..321L}
is adopted to obtain a closure relation for the radiation flux
\begin{equation}
  \vec{F} = - \lambda_{\mathrm{lim}}\frac{c}{\rho\kappa_{\mathrm{R}}}\nabla E_{\mathrm{R}} \, ,
  \label{eq:fld}
\end{equation}
where $\kappa_{\mathrm{R}}$ is the Rosseland opacity
and $\lambda_{\mathrm{lim}}$ is the flux limiter of \cite{Kley_1989A&A...208...98K}.
For the opacities, we assume that the Planck and Rosseland means
are similar enough to be replaced with a single value $\kappa$.
This is a valid assumption in the cold regions of protoplanetary
disks that we aim to study \citep{Bitsch_etal_2013A&A...549A.124B}.
The detailed opacity law will be specified later in Sect.~\ref{sec:opacity}.

The accretion luminosity of the protoplanet is given by
\begin{equation}
  L = \frac{GM_{\mathrm{p}}\dot{M}_{\mathrm{p}}}{R_{\mathrm{p}}} = \frac{GM_{\mathrm{p}}^{2}}{R_{\mathrm{p}}\tau} \, ,
  \label{eq:luminosity}
\end{equation}
where $G$ is the gravitational constant, $M_{\mathrm{p}}$ is the mass
of the protoplanet, $\dot{M}_{\mathrm{p}}$ is its accretion rate,
and $R_{\mathrm{p}}$ is the protoplanet's radius. In writing the
second equality, we introduce the mass doubling time $\tau=M_{\mathrm{p}}/\dot{M}_{\mathrm{p}}$,
which is a free parameter that controls the accretion rate in our model (see Sect.~\ref{sec:inicond}).

We assume that the radiation flux from the protoplanet
is completely absorbed by the optically thick gas in the eight grid cells
enclosing the protoplanet
\citep{Benitez-Llambay_etal_2015Natur.520...63B,Eklund_Masset_2017MNRAS.469..206E,Lambrechts_Lega_2017A&A...606A.146L}.
The accretion heating, which is non-zero only
within these cells, is then simply
\begin{equation}
  Q_{\mathrm{acc}} = \frac{L}{V} \, ,
  \label{eq:q_acc}
\end{equation}
where $V$ is the total volume of the heated cells.

The disk evolves in the combined gravitational 
potential of the central protostar and the embedded
protoplanet:
\begin{equation}
  \Phi = - \frac{GM_{\star}}{r} - \frac{GM_{\mathrm{p}}}{d}f_{\mathrm{sm}} \, ,
  \label{eq:potential}
\end{equation}
where $M_{\star}$ is the mass of the protostar and $d$ is the cell-protoplanet
distance. The planetary potential
is smoothed to avoid numerical divergence at the protoplanet location ($d=0$) using
the tapering cubic-spline function $f_{\mathrm{sm}}$ of \cite{Klahr_Kley_2006A&A...445..747K}:
\begin{equation}
  f_{\mathrm{sm}} =
  \begin{dcases*}
    1 \,, &
    $\left(d > r_{\mathrm{sm}} \right)$ , \\
    \left[\left( \frac{d}{r_{\mathrm{sm}}} \right)^{4} -2\left( \frac{d}{r_{\mathrm{sm}}} \right)^{3} + 2\frac{d}{r_{\mathrm{sm}}}  \right] \,, &
    $\left(d \leq r_{\mathrm{sm}} \right)$ 
  \end{dcases*}
  \label{eq:taper}
\end{equation}
where the smoothing length $r_{\mathrm{sm}}$ is
a fraction of the protoplanet's Hill sphere radius (see Sect.~\ref{sec:inicond}).
Orbital evolution of the protoplanet is tracked using
the \textsc{ias15} integrator \citep{Rein_Spiegel_2015MNRAS.446.1424R}
from the \textsc{rebound}\footnote{\raggedright Public version of the code is available at \url{https://rebound.readthedocs.io/en/latest/}.}
package \citep{Rein_Liu_2012A&A...537A.128R},
which we interfaced with \textsc{fargo3d}.

Although \textsc{fargo3d} is an explicit hydrodynamic
solver, implementing Eqs.~(\ref{eq:e_rad}) and (\ref{eq:e_int})
in an explicit form would lead to a very
restrictive Courant condition on the largest allowed time step length.
To avoid such a time step limitation, it is advantageous to solve
the energy equations in an implicit form. We thus follow the discretisation
and linearisation proposed by \cite{Bitsch_etal_2013A&A...549A.124B},
with a minor modification introduced in Appendix~\ref{sec:numerics}.

The solution of the implicit problem is obtained iteratively,
by minimizing the residual $\vec{r}\equiv\mathbf{A}\vec{x}-\vec{b}$, where $\mathbf{A}$ is the matrix of the 
linear system, $\vec{x}$ is the solution vector and $\vec{b}$ is the
right-hand side vector. Our iterative solver uses the improved bi-conjugate stabilised
method \citep[IBiCGStab;][]{Yang_Brent_2002} with the Jacobi
preconditioning.
The convergence criterion is $\|{\vec{r}}\|/\|{\vec{b}}\|<10^{-4}$,
where the norms are calculated in the $\mathrm{L}_{2}$ space.

\subsection{Initial conditions and parameters}
\label{sec:inicond}

\begin{figure}[]
  \centering
  \includegraphics[width=8.8cm]{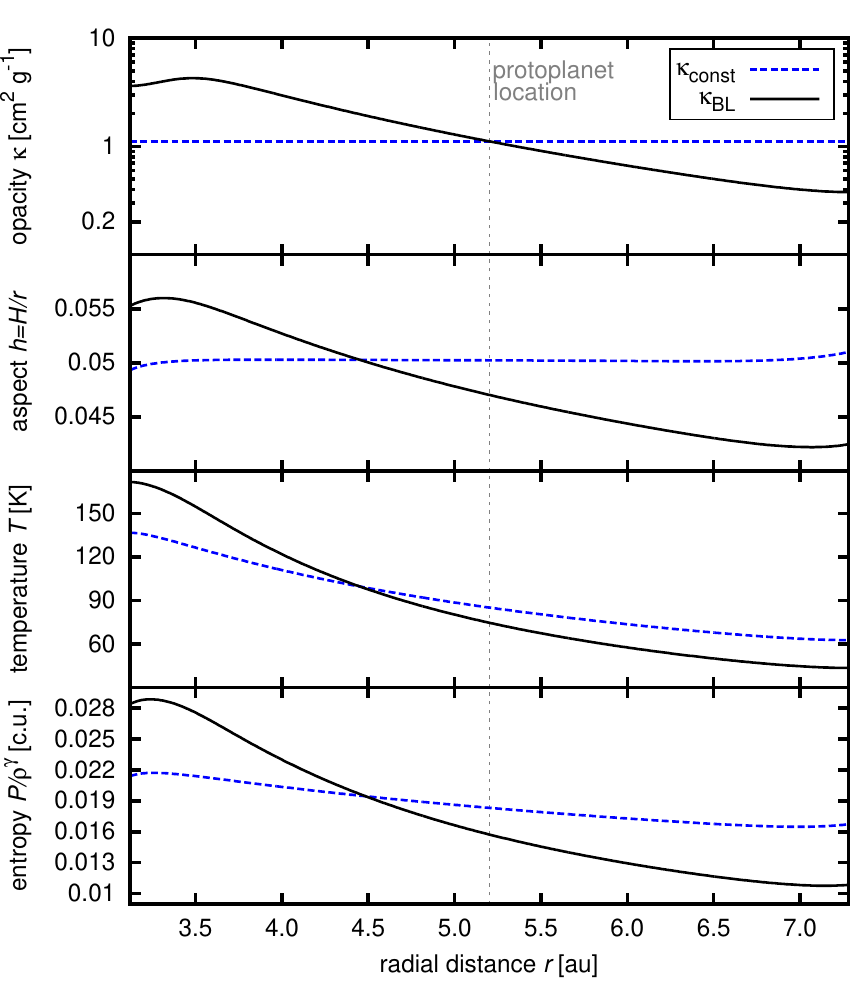}
  \caption{Radial profiles of the characteristic quantities
  in the \KC-disk (dashed blue curves) and the \KBL-disk
  (solid black curves) after their numerical relaxation over $t=100\,P_{\mathrm{orb}}$.
  From top to bottom, the panels show the opacity $\kappa$, aspect ratio $h=H/r$,
  temperature $T$ and entropy $S=P/\rho^{\gamma}$. Midplane values are displayed (or used to derive $h$).
  The dotted vertical line indicates the protoplanet's orbital distance $a_{\mathrm{p}}=5.2\,\mathrm{au}$.}
  \label{fig:profiles}
\end{figure}

Initial conditions describing the disk are adopted from \cite{Kley_etal_2009A&A...506..971K}
and \cite{Lega_etal_2014MNRAS.440..683L}.
The surface density follows the power-law function $\Sigma=484(r/(1\,\mathrm{au}))^{-0.5}\,\mathrm{g\,cm^{-2}}$.
The initial $\Sigma$ is converted into the initial $\rho$
assuming the disk is vertically isothermal.
The velocity field is given by balancing the
gravitational acceleration from the protostar,
the acceleration due to the pressure gradient
and the centrifugal acceleration.
We assign constant kinematic viscosity $\nu=10^{15}\,\mathrm{cm^{2}\,s^{-1}}$
to the gas to mimic the angular momentum transport in the disk driven by physical effects
outside the scope of this study (e.g. the turbulent eddy viscosity).
The initial aspect ratio is $h=H/r=0.05$, where $H$ is the
pressure scale height. The disk is therefore initially non-flaring
but we point our that $h$ evolves during the simulations.
The adiabatic index is $\gamma=1.43$ and the mean molecular weight
is $\mu=2.3\,\mathrm{g\,mol^{-1}}$, corresponding to the solar mixture
of the hydrogen and helium.

We will assume a single embedded super-Earth
with the mass $M_{\mathrm{p}}=3\,M_{\oplus}$,
orbiting on a circular orbit at the Jupiter's heliocentric distance
$a_{\mathrm{p}}\equiv a_{\mathrm{J}}=5.2\,\mathrm{au}$.
Unless otherwise specified, the orbit is held fixed in our simulations, i.e.
the protoplanet is not allowed to migrate and the
torque we measure is the static torque.
The simulations are performed in the reference frame corotating
with the protoplanet at the Keplerian angular frequency
$\Omega_{\mathrm{p}}\equiv\Omega_{\mathrm{J}}=\sqrt{GM_{\star}/a_{\mathrm{J}}^{3}}$
and the smoothing length for the tapering function of 
the gravitational potential $f_{\mathrm{sm}}$ is $r_{\mathrm{sm}}=0.5R_{\mathrm{H}}$,
where $R_{\mathrm{H}}=a_{\mathrm{p}}(M_{\mathrm{p}}/(3M_{\star}))^{1/3}$
is the Hill sphere radius.
We will study both cases without and with planetary accretion,
corresponding to the cold- and hot-protoplanet limit, respectively.
In the simulations with the accretion,
we assume the mass doubling time $\tau=100\,\mathrm{kyr}$, which 
is a value within the range of
the expected pebble accretion rates
\citep{Lambrechts_Johansen_2014A&A...572A.107L,Chrenko_etal_2017A&A...606A.114C}.
The resulting luminosity of the protoplanet is $L\simeq4.2\times10^{27}\,\mathrm{erg}\,\mathrm{s}^{-1}$.

\subsection{Opacities}
\label{sec:opacity}

Using the initial disk parameters,
we setup two fiducial disk models that only differ
in the material opacity function. The opacity is either constant or follows
a slightly modified (see below) prescription of \cite{Bell_Lin_1994ApJ...427..987B}.
To distinguish between the models, we use the abbreviations \KC-disk
and \KBL-disk, respectively.
The exact value of the opacity in the \KC-disk is tuned
to be the same as in the \KBL-disk
at the location of the protoplanet $a_{\mathrm{p}}$,
leading to $\kappa_{\mathrm{const}}\equiv\kappa_{\mathrm{BL}}(r=a_{\mathrm{p}})=1.11\,\mathrm{cm^{2}\,g^{-1}}$.

The unmodified opacity from \cite{Bell_Lin_1994ApJ...427..987B},
which we denote $\kappa_{\mathrm{BL}}^{\mathrm{full}}(\rho(r,\theta,\phi),T(r,\theta,\phi))$,
is a fitting law that sets $\kappa$ as a function
of the local gas density and temperature.
The table spans several regimes corresponding to the presence
(or absence) of dust or molecular species dominant in protoplanetary disks.
Using $\kappa_{\mathrm{BL}}^{\mathrm{full}}$ in our simulations with the accretion
heating of the protoplanet
could cause strong local opacity gradients,
because we expect the temperature perturbations to reach $\sim$$10^{1}\,\mathrm{K}$
at distances of one cell size from the centre of the protoplanet.

In practice, we apply the Bell \& Lin's opacity law
in a simplified way,
using $\kappa_{\mathrm{BL}}(\bar{\rho}(r,\phi),\bar{T}(r,\phi))$, where the bared
quantities are azimuthally computed arithmetic means and the dependence
on the $\theta$-coordinate is therefore dropped.
This helps us to distinguish the effects caused by the global structure
of the disk from those related to the local $\kappa$-$T$-$\rho$ feedback.
To justify our approach, we verify in Appendix~\ref{sec:supporting_simulations}
that the unmodified opacity table $\kappa_{\mathrm{BL}}^{\mathrm{full}}$
of \cite{Bell_Lin_1994ApJ...427..987B} does not change our conclusions.

The fiducial \KC- and \KBL-disks are discussed throughout the majority of the
paper, with the exception of Sect.~\ref{sec:parametric} where we study the dependence
of the heating torque on the opacity gradient within the disk.

\subsection{Grid resolution and boundary conditions}

Migration of low-mass protoplanets critically depends
on the grid resolution,
we therefore combine the well-established disk extent
and resolution from \cite{Lega_etal_2014MNRAS.440..683L}
(in the azimuthal and vertical direction) 
and \cite{Eklund_Masset_2017MNRAS.469..206E} (in the radial direction).
The disk radially stretches from $r_{\mathrm{min}}=3.12\,\mathrm{au}$
to $r_{\mathrm{max}}=7.28\,\mathrm{au}$ and is resolved by 512 rings.
We prevent any vertical motions of the protoplanet in our simulations
by assuming that the solution is symmetric with respect to the midplane.
One of the disk boundaries in the colatitude is therefore located
at the midplane ($\phi=\pi/2$); the vertical extent above the midplane is $7^{\circ}$.
The colatitude is resolved by 64 zones.
In the azimuth, we use only 1 sector in
our preparatory simulations without the protoplanet and 1382 
sectors in our simulations with the embedded protoplanet.
The resulting local resolution is 8 cells per $R_{\mathrm{H}}$ in the $r$- and $\phi$-directions
and 3 cells per $R_{\mathrm{H}}$ in the $\theta$-direction (see also Appendix~\ref{sec:increased_resol} for
a simple resolution test).

The azimuthal boundary conditions are periodic for all primitive quantities.
The radial boundary conditions are symmetric for $\rho$, $\epsilon$, and $v_{\phi}$
and reflecting for $v_{r}$. $E_{\mathrm{R}}$ is set to a zero gradient
and $v_{\theta}$ is extrapolated using the same radial dependence as for
the Keplerian rotation velocity. The boundary conditions
in colatitude are symmetric for $\rho$, $\epsilon$, $v_{r}$ and $v_{\theta}$,
and reflecting for $v_{\phi}$. $E_{\mathrm{R}}$ is symmetrised at the midplane
boundary and set to $a_{\mathrm{R}}T_{0}^{4}$ at the remaining boundary in the colatitude,
where $a_{\mathrm{R}}$ is the radiation constant and $T_{0}\equiv5\,\mathrm{K}$
is the ambient temperature that allows for vertical radiative cooling of the disk.
Additionally, wave-damping conditions of \cite{deValBorro_etal_2006MNRAS.370..529D}
are imposed near the radial edges and also near the disk surface.

\section{Simulations}
\label{sec:simulations}

\subsection{Equilibrium disks}
\label{sec:profiles}

Since we use a non-isothermal equation of state and we also account for
the energy production and transfer, the parametric setup
of the disk, which was discussed so far, is not stationary.
Therefore, before the simulations with the embedded protoplanet are conducted,
we let both disks (\KC-disk and \KBL-disk) numerically evolve
over the time scale $100\,P_{\mathrm{orb}}$,
where $P_{\mathrm{orb}}=2\pi/\Omega_{\mathrm{J}}$.

The equilibrium state after the relaxation is characterized
by Fig.~\ref{fig:profiles}.
We plot the radial profiles of the midplane temperature $T$,
opacity $\kappa$, entropy $S=P/\rho^{\gamma}$ and aspect ratio $h=H/r$,
where the pressure scale height is determined as $H=c_{\mathrm{s}}/\Omega_{\mathrm{K}}$
and the sound speed as $c_{\mathrm{s}}=\sqrt{\gamma P/\rho}$.

The constant opacity of the \KC-disk makes all
the remaining radial dependencies rather shallow.
The aspect ratio is almost flat, only slightly increasing with the 
radial distance.
In the \KBL-disk, on the other hand, the opacity has a peak near $3.5\,\mathrm{au}$,
where the water ice evaporates for the given setup, and decreases
at larger radii.
Therefore, the efficiency of the disk cooling increases
and simultaneously,  the efficiency of the viscous heating
diminishes with the dropping relative velocity of the shearing layers.
Consequently, the aspect ratio radially decreases
as the energy budget is not sufficient to puff up the disk.
In such a disk, $T$ and $S$ profiles radially decrease more steeply
compared to the \KC-disk.

\subsection{Torque evolution}

\begin{figure}[!t]
  \centering
  \includegraphics[width=8.8cm]{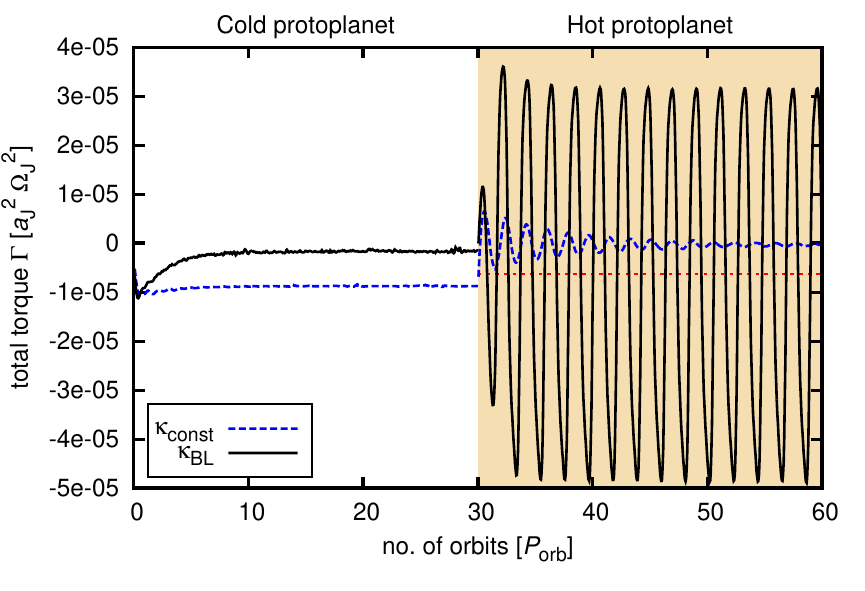}
  \caption{Temporal evolution of the total torque exerted on
    the protoplanet ($M_{\mathrm{p}}=3\,M_{\oplus}$) by the \KC-disk (dashed blue curve) and
    the \KBL-disk (solid black curve). For $t\leq30\,P_{\mathrm{orb}}$, the
    protoplanet is not accreting.
    Its accretion and the respective heating
    are activated for $t>30\,P_{\mathrm{orb}}$, as also highlighted by a yellow background.
    While the blue curve evolves as expected, i.e. gains a positive boost
    when the accretion heating is initiated
    \citep{Benitez-Llambay_etal_2015Natur.520...63B}, 
    the black curve does not converge and exhibits strong oscillations
    between positive and negative values instead.
    The red horizontal dotted line shows the mean value of the oscillating black curve
    and demonstrates that the heating torque makes the total torque
    more negative in this case.
   }
  \label{fig:torque}
\end{figure}

\begin{figure*}[!hpt]
  \centering
  \begin{tabular}{cc}
    \hline
    \multicolumn{2}{c}{cold non-luminous protoplanet (\KC-disk simulation, $t=30\,P_{\mathrm{orb}}$)} \\
     perturbed density & temperature \\
     \includegraphics[width=6.6cm]{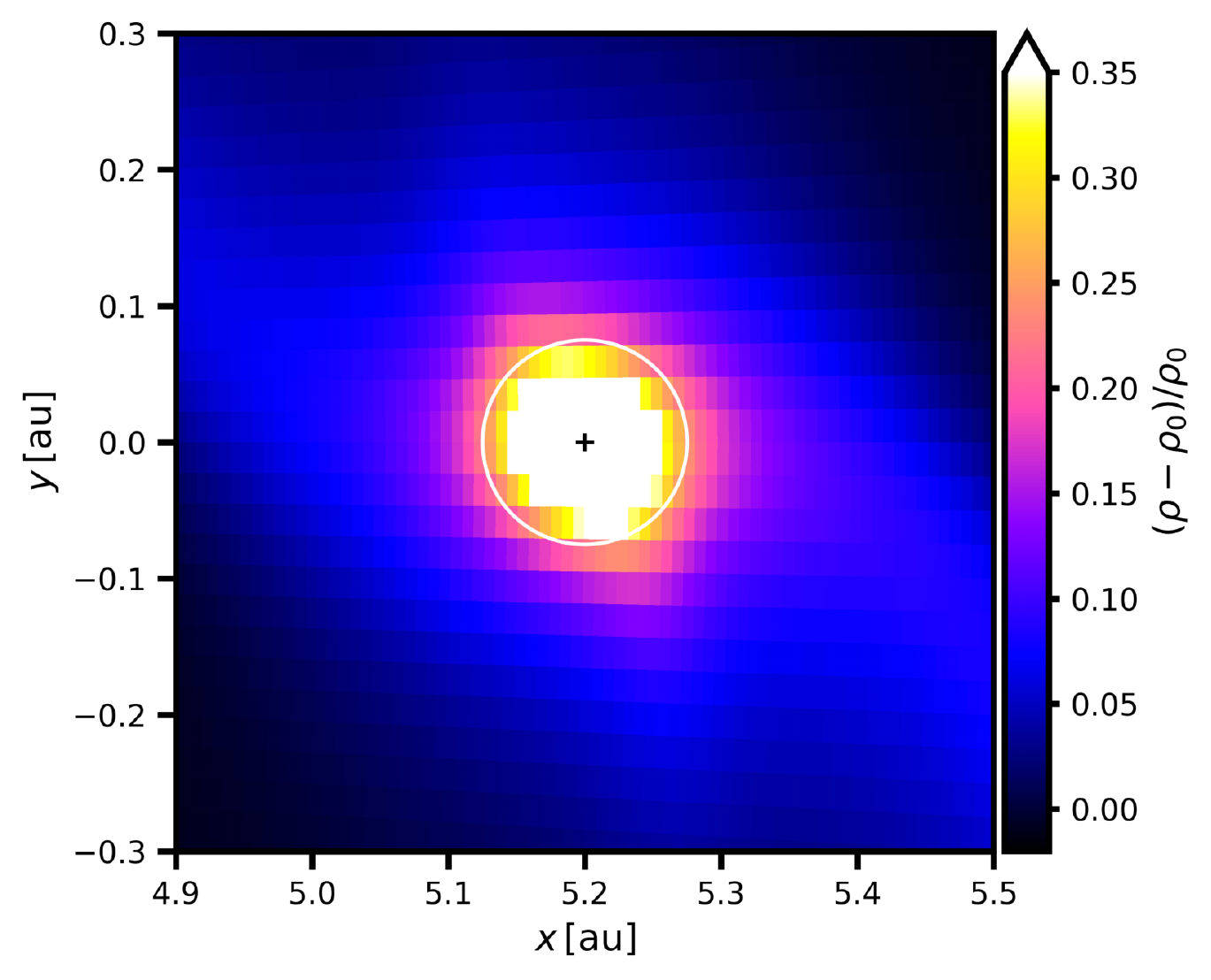} & \includegraphics[width=6.6cm]{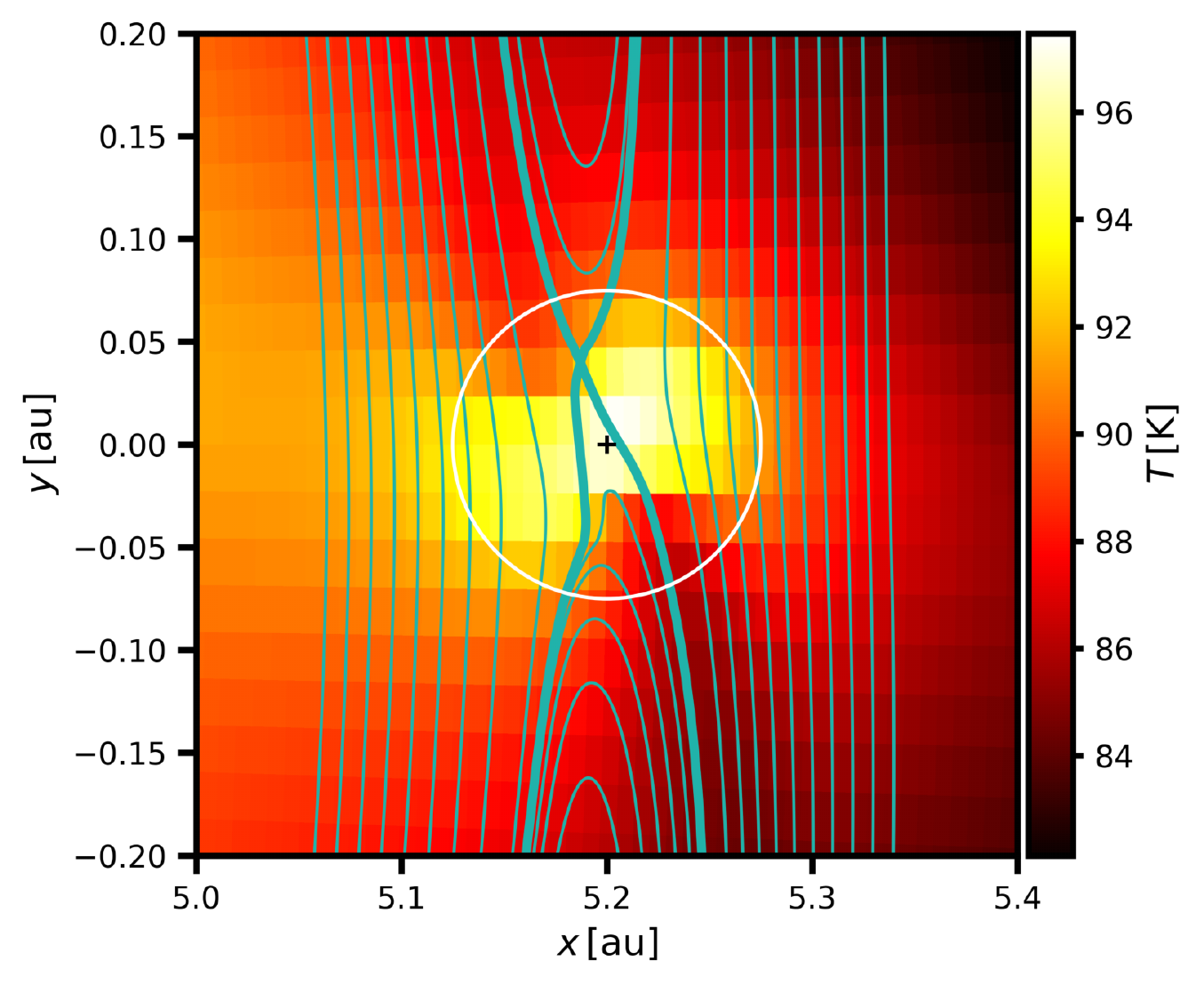} \\
     \hline
     \multicolumn{2}{c}{hot luminous protoplanet (\KC-disk simulation, $t=60\,P_{\mathrm{orb}}$)} \\
     perturbed density & temperature excess \\
     \includegraphics[width=6.6cm]{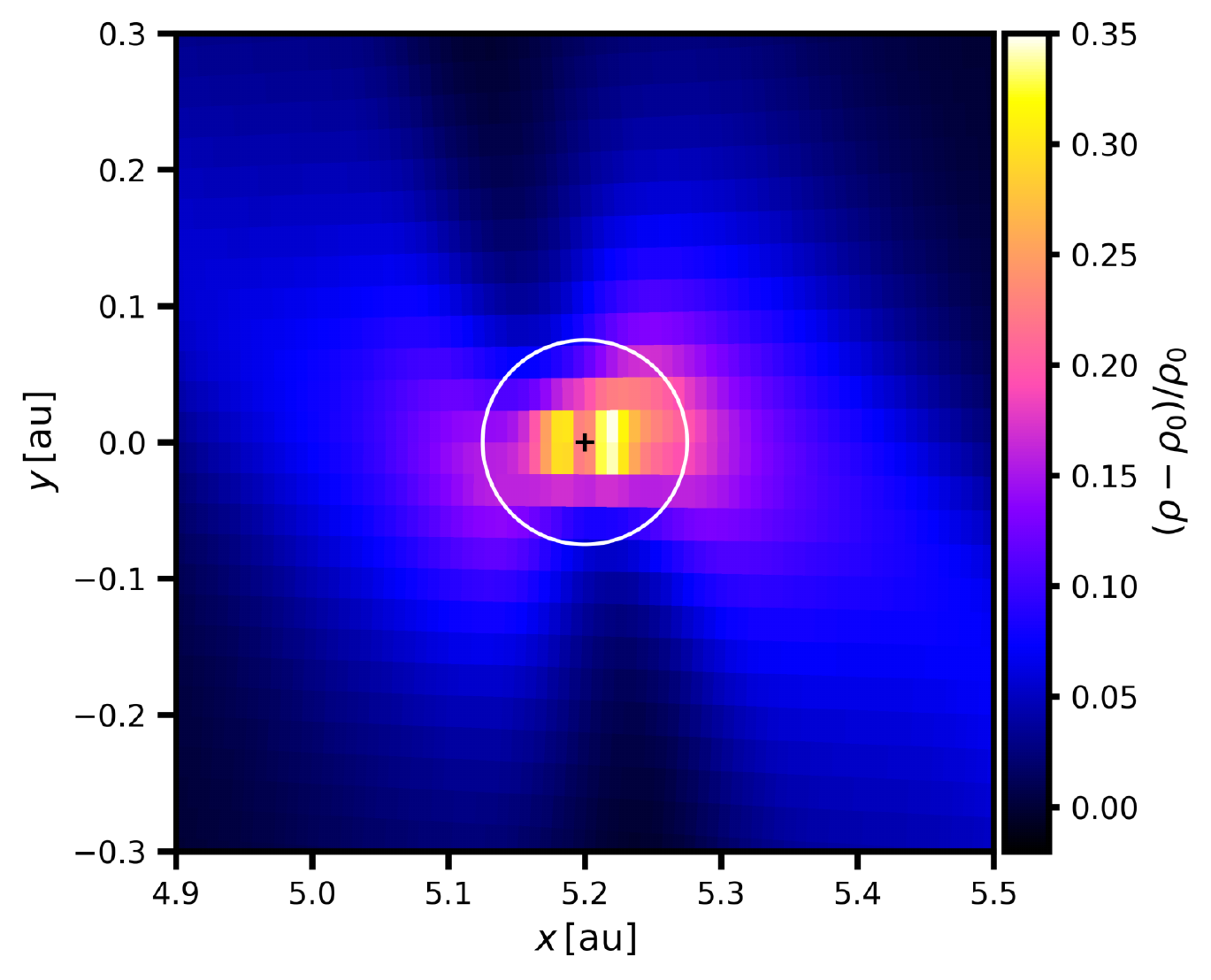} & \includegraphics[width=6.6cm]{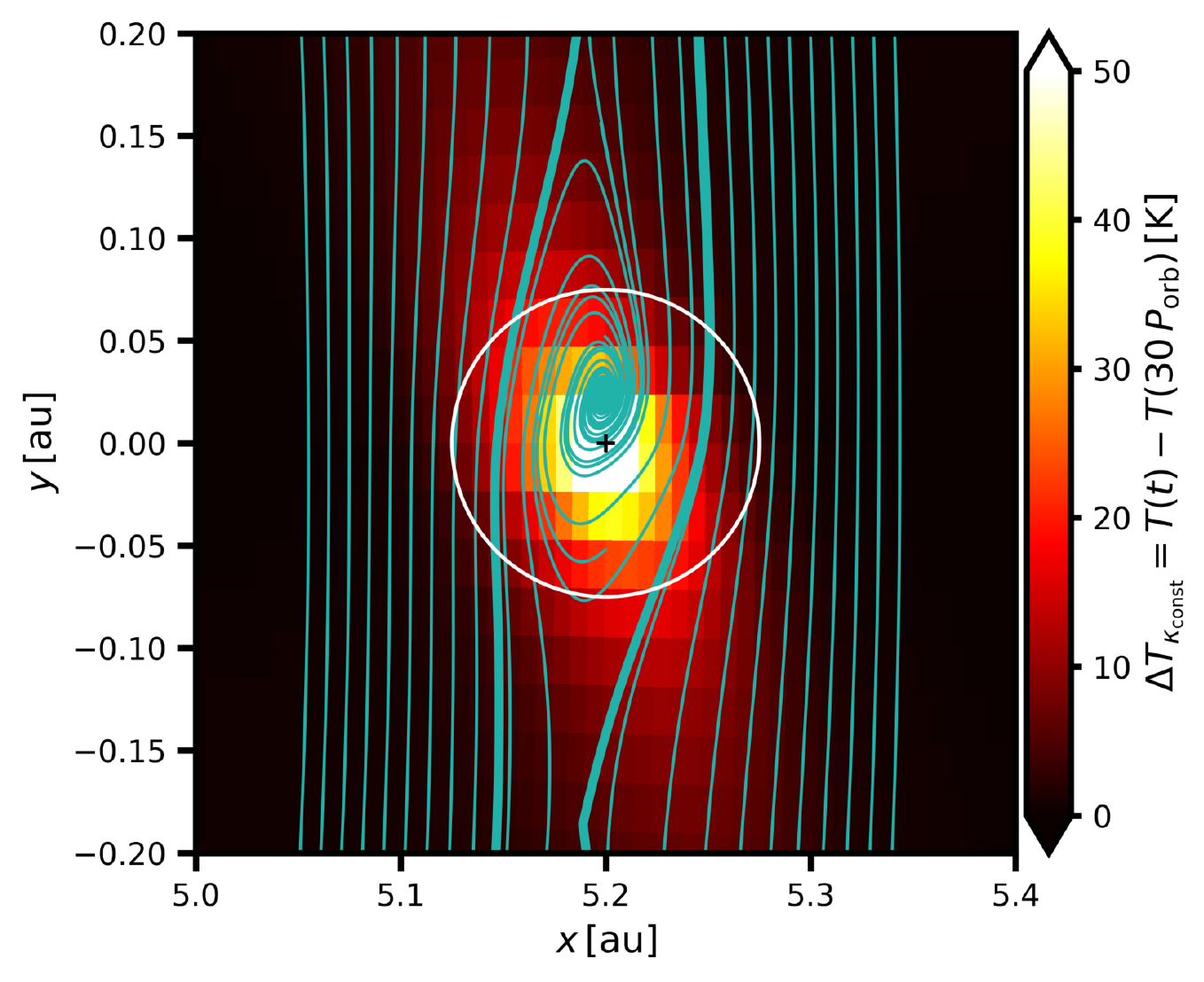} \\
  \end{tabular}
  \caption{Hydrodynamic quantities in the midplane of the \KC-disk close to the protoplanet.
    Top row corresponds to the cold protoplanet right before the accretion heating
    is initiated (at $t=30\,P_{\mathrm{orb}}$), bottom row shows the steady state 
    of gas around the hot protoplanet (at $t=60\,P_{\mathrm{orb}}$). 
    The figure is constructed as a Cartesian projection
    of the spherical grid. The density maps (right)
    display the perturbation $(\rho-\rho_{0})/\rho_{0}$ relative
    to the equilibrium disk ($t=0\,P_{\mathrm{orb}}$), the temperature
    map for the cold protoplanet (top right) shows the absolute values,
    and the temperature map for the hot protoplanet (bottom right) shows the excess
    with respect to the cold protoplanet (by subtracting $T(t=30\,P_{\mathrm{orb}})$ from $T(t=60\,P_{\mathrm{orb}})$).
    The position of the protoplanet is marked with the cross,
    the extent of its Hill sphere is bordered by the circle.
    The green curves (right) show the topology of
    streamlines in the frame corotating with the protoplanet. 
    In the inertial frame, the protoplanet would orbit counterclockwise.
    The streamlines outwards from its orbit thus depict the flow directed from $y>0$
    to $y<0$, the inward streamlines are oriented in the opposite direction.
    A detailed view of the streamlines is provided in Fig.~\ref{fig:stream_kc}
    where we also sort them according to their type.
  }
  \label{fig:hydro_kc}
\end{figure*}

Starting from the relaxed state of the disks, 
we copy the hydrodynamic quantities in the azimuthal direction
to expand the resolution from a single sector to the desired
1382 sectors. The protoplanet is inserted and 
we simulate $30\,P_{\mathrm{orb}}$ of evolution
while neglecting any accretion and accretion heating of the protoplanet.
The aim of this part of the simulation is to let the disk adjust to
the presence of a gravitational perturber and to acquire
a converged value of the disk torque in the absence of the accretion heating,
corresponding to the cold-protoplanet limit.
For the \KBL-disk, this part of the simulation is
similar to the experiments in \cite{Lega_etal_2014MNRAS.440..683L}.
Subsequently, we continue the simulation for another $30\,P_{\mathrm{orb}}$
during which we let the planetary mass grow 
while releasing the accretion heat into the gas disk 
according to Eqs.~(\ref{eq:luminosity}) and (\ref{eq:q_acc}).

Fig.~\ref{fig:torque} shows the temporal evolution of the torque
exerted on the protoplanet by the \KC-disk and \KBL-disk.
During the phase without accretion heating, the torque converges to
a stationary value during $10\,P_{\mathrm{orb}}$. The torque value 
in the \KBL-disk is more positive compared to the \KC-disk, which is because
the steeper radial decline of the entropy in the \KBL-disk
enhances the positive entropy-driven part of the corotation torque \citep{Paardekooper_Mellema_2006A&A...459L..17P,Baruteau_Masset_2008ApJ...672.1054B}.

When the accretion heating is activated in the \KC-disk,
a positive contribution is added to the torque,
in accordance with the results
of \cite{Benitez-Llambay_etal_2015Natur.520...63B}.
The torque slightly oscillates at first, but the amplitude of the
oscillations decreases in time and becomes negligible 
at $t=50\,P_{\mathrm{orb}}$.

On contrary, when the accretion heating is activated in the \KBL-disk,
the outcome of the heating torque becomes very different.
Strong oscillations of the disk torque
are excited almost immediately and they do not vanish in time;
instead, their amplitude remains the same. The arithmetic
mean of the torque measured in the time interval $30$--$60\,P_{\mathrm{orb}}$
is $\bar{\Gamma}\simeq-6.3\times10^{-6}\,a_{\mathrm{J}}^{2}\,\Omega_{\mathrm{J}}^{2}$,
implying that
the torque is more negative compared to the situation without accretion heating
(the mean torque in the interval $20$--$30\,P_{\mathrm{orb}}$ is
$\bar{\Gamma}\simeq-1.6\times10^{-6}\,a_{\mathrm{J}}^{2}\,\Omega_{\mathrm{J}}^{2}$).
The amplitude of the variations with respect to the mean value 
is $\simeq$$\pm3.9\times10^{-5}\,a_{\mathrm{J}}^{2}\,\Omega_{\mathrm{J}}^{2}$
and the oscillation period is $\simeq$$2.1\,P_{\mathrm{orb}}$.

The torque oscillations are unexpected and
we thus focus throughout the rest of the paper
on finding the physical mechanism that excites them.
The occurrence of oscillations suggests that
the gas distribution around the protoplanet is changing during the simulation,
as we can demonstrate using the integral expression for the disk torque
\begin{equation}
  \Gamma = \int\limits_{\mathrm{disk}}\left( \vec{r}_{\mathrm{p}}\times\vec{F}_{\mathrm{g}} \right)_{\perp}\mathrm{d}V \, ,
  \label{eq:torque}
\end{equation}
where $\vec{r}_{\mathrm{p}}$ is the protoplanet's radius vector,
$\vec{F}_{\mathrm{g}}$ is the gravitational force of a disk element,
the vertical component of the cross product is considered and we integrate
over the disk volume.
Only a non-zero azimuthal component of $\vec{F}_{\mathrm{g}}$ can
lead to a non-vanishing cross product in the integral, thus any oscillations
of $\Gamma$ must be related to a variation of $F_{\mathrm{g},\theta}$.
In other words, there must be an azimuthal asymmetry
in the gas distribution with respect to the protoplanet 
for the torque to be non-zero and only a temporal redistribution
of the asymmetry can cause a torque oscillation.

Our strategy throughout the rest of Sect.~\ref{sec:simulations}
is the following: First, we focus on the \KC-disk simulation in Sect.~\ref{sec:steady_state}.
Although similar simulations were analysed by \cite{Benitez-Llambay_etal_2015Natur.520...63B},
our aim is to focus on the 3D gas flow that has not been described yet.
Our findings are then expanded for the
\KBL-disk simulation in Sect.~\ref{sec:instability} where we relate the gas
redistribution to the oscillatory behaviour of the torque.
Sect.~\ref{sec:processes} is devoted to identifying the key
physical mechanisms that affect the gas flow.
In Sect.~\ref{sec:parametric},
we vary the disk opacity
gradient and we study its impact on the torque oscillation.
Finally, we relax the assumption
of a fixed orbit and explore how the protoplanet migrates in Sect.~\ref{sec:migration}.

\subsection{Steady state of the heated gas}
\label{sec:steady_state}

In this section, the \KC-disk simulation is analysed.

\subsubsection{Midplane}

Fig.~\ref{fig:hydro_kc} compares the midplane density and temperature 
distribution around the cold and hot protoplanet.
We display the state of the simulation at $t=30$
(i.e. at the final stage of the phase without accretion heating)
and $60\,P_{\mathrm{orb}}$ (i.e. at the final stage of the phase
with accretion heating). The latter represents the steady state
of gas around the accreting protoplanet and we verified that
such a distribution is achieved early (at $t\simeq31\,P_{\mathrm{orb}}$)
and does not greatly evolve since then.

For the non-luminous protoplanet, the gas state is 
in agreement with \cite{Lega_etal_2014MNRAS.440..683L} (see
their Fig.~10 for a comparison).
The density distribution is not spherically symmetric
with respect to the protoplanet but there are two
patches of slightly overdense gas along the outflow
from the Hill sphere known as the cold fingers (explained in Sect.~\ref{sec:intro}).
The temperature drop inside the fingers is clearly apparent
from the top right panel of Fig.~\ref{fig:hydro_kc}.

When the protoplanet becomes luminous, the gas distribution
is modified and the heating torque arises.
Following \cite{Benitez-Llambay_etal_2015Natur.520...63B}
and \cite{Masset_2017MNRAS.472.4204M}, one can imagine
the response of the gas to the heating from the protoplanet as follows:
First, an underdense disturbance appears close to the protoplanet,
with a characteristic scale length given by the
linear perturbation model of \cite{Masset_2017MNRAS.472.4204M}:
\begin{equation}
    \lambda_{c} = \sqrt{\frac{\chi}{q\Omega_{\mathrm{p}}\gamma}} \, ,
      \label{eq:length_masset}
\end{equation}
where $\chi$ is the thermal diffusivity and $q$ is a dimensionless
measure of the disk shear ($q=3/2$ for a Keplerian disk).
Second, the low-density gas is distorted by the shear motions.
The rotation of the inner disk with respect to the protoplanet
is faster and the low-density gas thus propagates ahead of the protoplanet.
The motion of the outer disk lags behind the protoplanet  
and so does the hot perturbation.
As a result, two hot lobes with decreased density
are formed along the streamlines outflowing from the Hill sphere,
as described by \cite{Benitez-Llambay_etal_2015Natur.520...63B}.
The size of the lobes is inherently asymmetric because the corotation between
the protoplanet and the sub-Keplerian gas is radially shifted inwards,
therefore the outer rear lobe is usually larger and the heating
torque should be positive.

Such an advection-diffusion interplay is indeed observed
in Fig.~\ref{fig:hydro_kc}, where we
find the typical two-lobed distribution of hot underdense gas around the luminous
protoplanet and the positive boost of the total torque (Fig.~\ref{fig:torque})
confirms that the outer lobe is slightly more pronounced.
The bottom right panel reveals the magnitude and spatial extent
of the temperature excess, as well as its skewed shape in the direction of the disk shear.

However, we make a new observation here concerning the streamlines
of the flow that are overlaid in the temperature maps.
It is obvious that the hot perturbation significantly changes the topology of the flow
with respect to the cold-protoplanet case.
U-turn streamlines no longer appear in the depicted part of the
disk, the direction of the circulating streamlines changes as they pass the protoplanet
and a new set of spiral-like retrograde streamlines appears.

\subsubsection{Vertical plane}

\begin{figure}[!thp]
  \centering
  \begin{tabular}{c}
    \hline
    cold protoplanet (\KC-disk simulation, $t=30\,P_{\mathrm{orb}}$) \\
    \includegraphics[width=8.8cm]{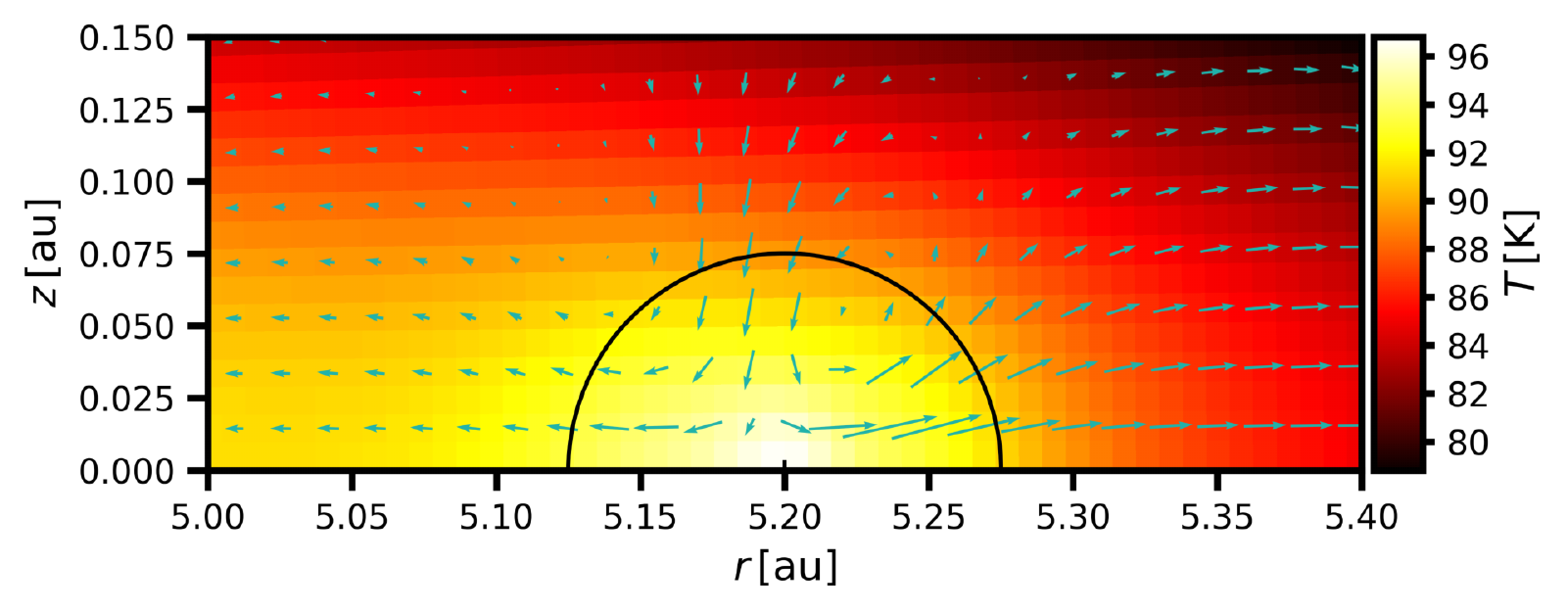} \\
    \hline
    hot protoplanet (\KC-disk simulation, $t=60\,P_{\mathrm{orb}}$) \\
    \includegraphics[width=8.8cm]{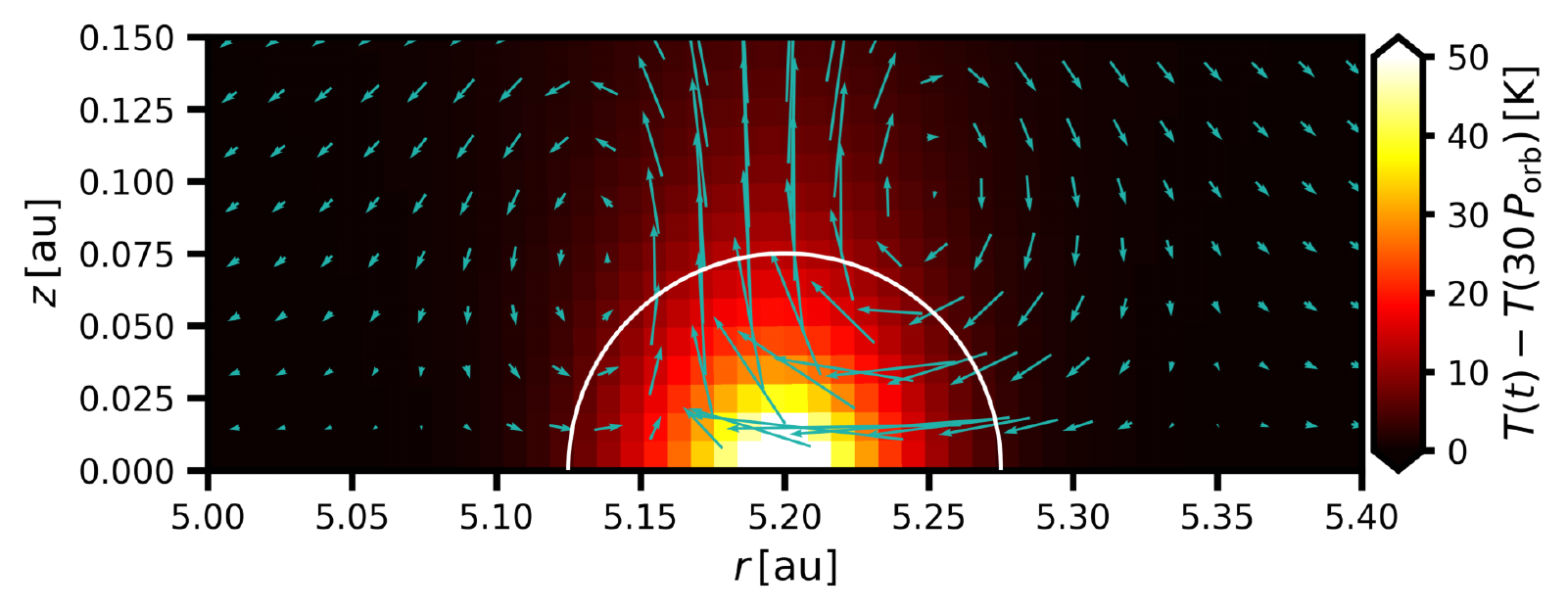} \\
  \end{tabular}
  \caption{Temperature map around the cold protoplanet at $t=30\,P_{\mathrm{orb}}$ (top)
    and temperature excess around the hot protoplanet at $t=60\,P_{\mathrm{orb}}$
    (bottom) in the \KC-disk.
    Vertical plane (perpendicular to the disk midplane) is displayed.
    The green arrows show the vertical velocity vector field.}
  \label{fig:vert_kc}
\end{figure}

It is known that vertical motions play an important role
for the structure of circumplanetary envelopes
\citep[e.g.][]{Tanigawa_etal_2012ApJ...747...47T,Fung_etal_2015ApJ...811..101F,Ormel_etal_2015MNRAS.447.3512O,Cimerman_etal_2017MNRAS.471.4662C,Lambrechts_Lega_2017A&A...606A.146L,Kurokawa_Tanigawa_2018MNRAS.479..635K,Popovas_etal_2019MNRAS.482L.107P}.
Since previous studies of the heating torque did not investigate the vertical perturbations,
we thus do so here.
Fig.~\ref{fig:vert_kc} shows the gas temperature and the
velocity field in the vertical plane intersecting the protoplanet's location.
When the protoplanet is cold, there is a vertical stream of gas 
descending towards the protoplanet and escaping as an outflow through the midplane,
in accordance with e.g. \cite{Lambrechts_Lega_2017A&A...606A.146L}.

In the presence of accretion heating, however, 
the direction of the gas flow above the protoplanet
is reverted. It forms an outflowing column while the midplane flow becomes directed towards
the protoplanet. Two overturning cells appear on each side of the vertical column
(although it is important to point out that no such cells are apparent
in the full 3D flow which is discussed later).
We notice that the hot perturbation is not spherically symmetric
but rather elongated in the direction of the column,
indicating that the envelope is not hydrostatic.

\subsubsection{2D and 3D streamline topology}

\begin{figure*}[!hpt]
  \centering
  \begin{tabular}{cc}
    \hline
    \multicolumn{2}{c}{cold protoplanet (\KC-disk simulation, $t=30\,P_{\mathrm{orb}}$)} \\
    \includegraphics[width=7.6cm]{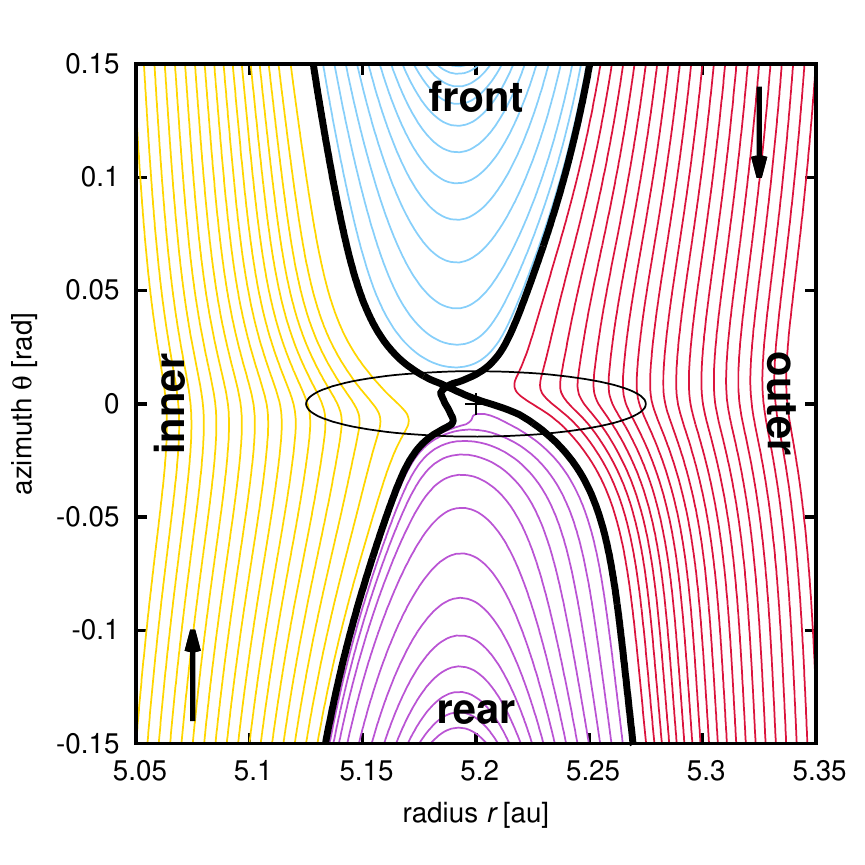} & \includegraphics[width=7.6cm]{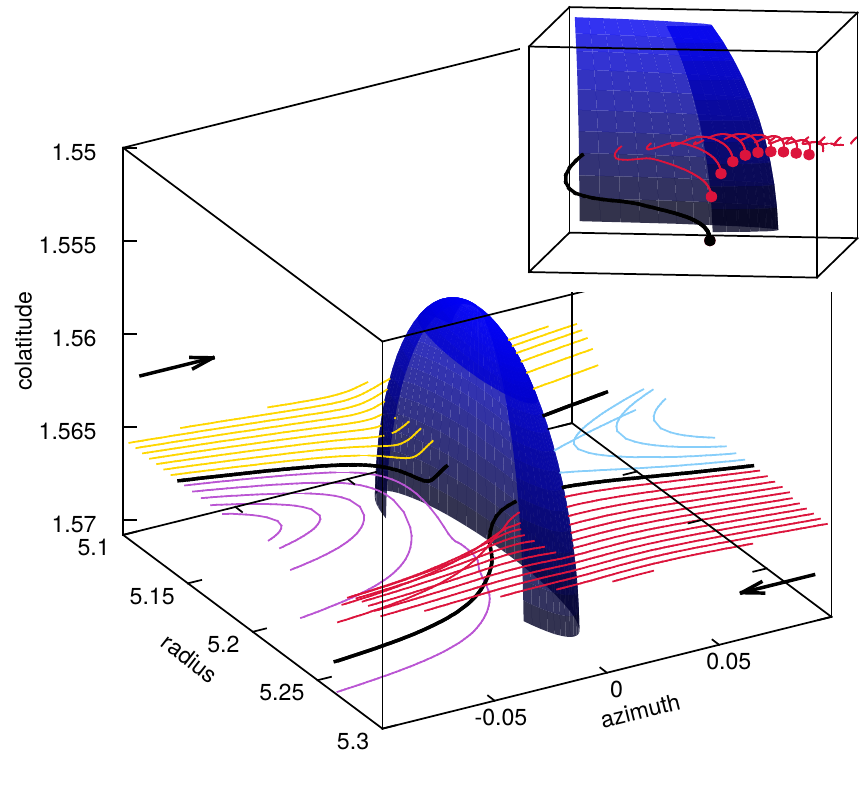} \\
    \hline
    \multicolumn{2}{c}{hot protoplanet (\KC-disk simulation, $t=60\,P_{\mathrm{orb}}$)} \\
    \includegraphics[width=7.6cm]{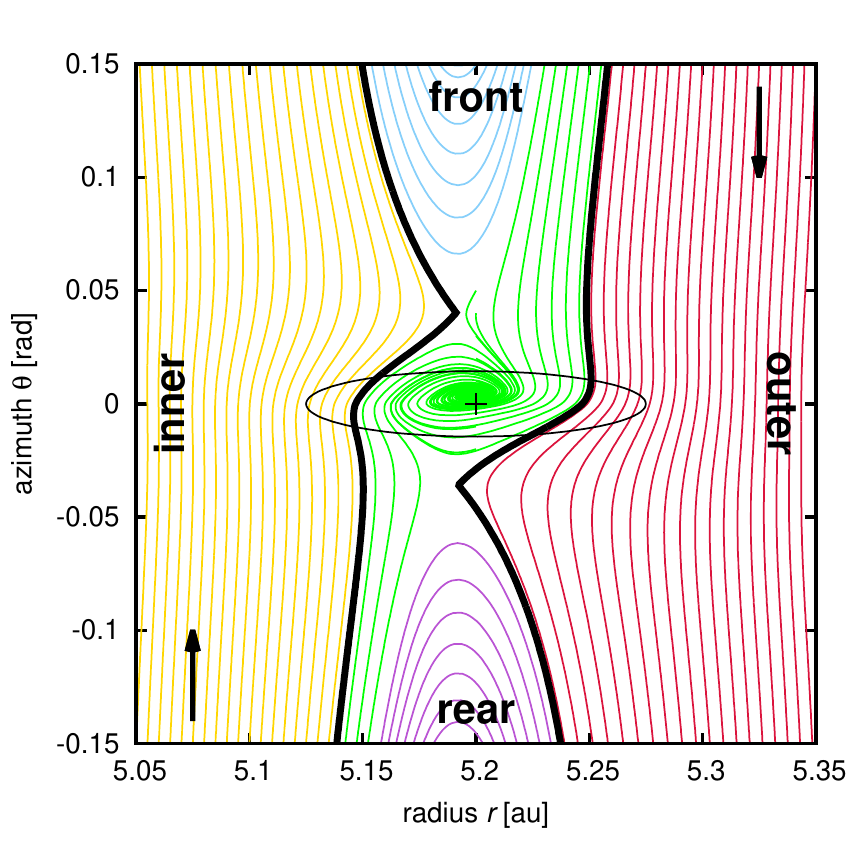} & \includegraphics[width=7.6cm]{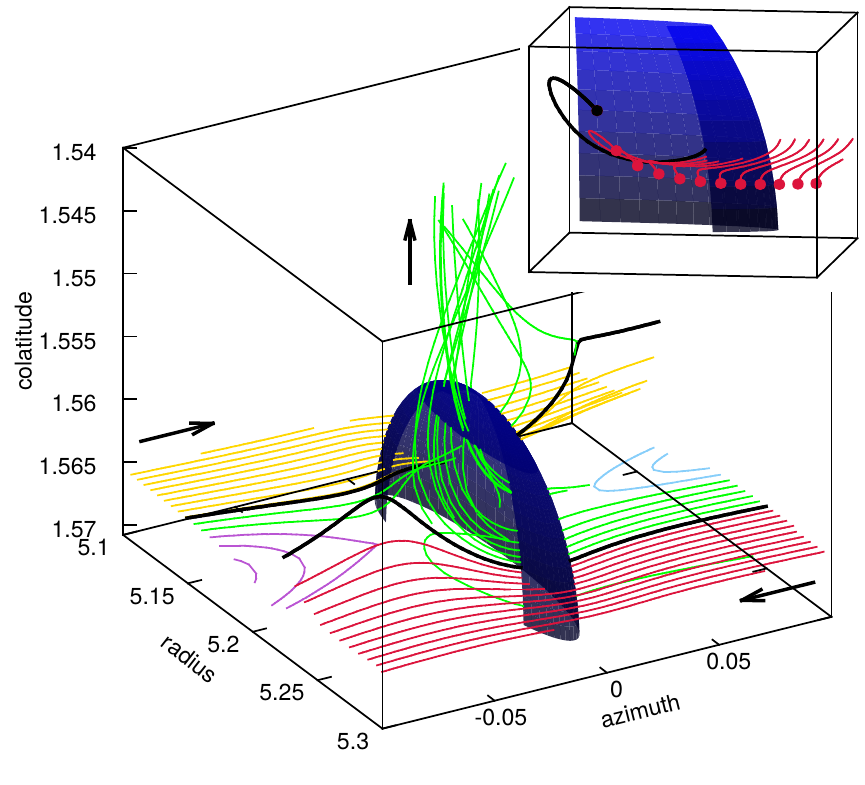} \\
  \end{tabular}
  \caption{
   Detailed streamline topology in the \KC-disk simulation.
   Top row corresponds to the cold protoplanet at $t=30\,P_{\mathrm{orb}}$
   and the bottom row to the hot protoplanet at $t=60\,P_{\mathrm{orb}}$.
   Rectangular projection in the spherical coordinates is used
   to display the disk midplane near the protoplanet (left)
   and the actual 3D flow (right). The colour of the curves
   distinguish individual sets of streamlines: inner circulating
   (yellow), outer circulating (red), front horseshoe (light blue),
   rear horseshoe (purple) and other (green). The thick black
   lines highlight the critical circulating streamlines closest to the protoplanet.
   The black cross and the ellipse mark the protoplanet's location
   and Hill sphere; the black arrows indicate the flow direction with respect to the protoplanet. In 3D figures (right),
   the dark blue hemisphere corresponds to the Hill sphere above the midplane.
   Additionally, the insets in the corners of the 3D figures 
   provide a close-up of the upstream outer circulating streamlines 
   viewed from a slightly different angle. The endpoints
   indicate where the flow exits the depicted part of the space
   and highlight if the initially coplanar streamlines
   descend towards the protoplanet (top) or rather rise vertically (bottom).
   We emphasise that the streamlines in the 3D figures
   are generated above the midplane and do not directly
   correspond to those in the 2D figures.
  }
  \label{fig:stream_kc}
\end{figure*}

\begin{figure*}[!hpt]
  \centering
  \begin{tabular}{c}
      \hline
      \multicolumn{1}{c}{hot protoplanet (\KBL-disk simulation)} \\
  \includegraphics[width=16cm]{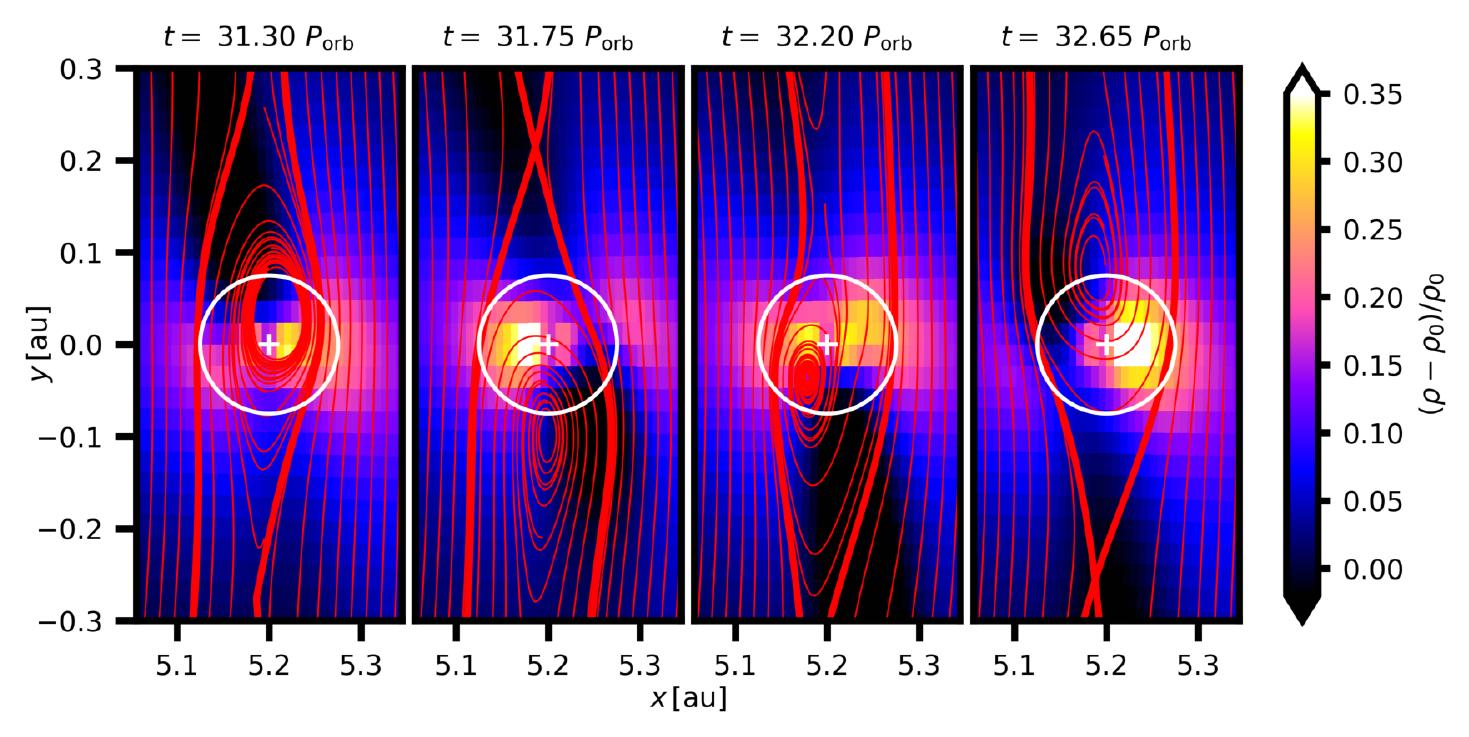}  \\
  \end{tabular}
  \caption{Evolution of the perturbed midplane gas density in the \KBL-disk
  simulation. The corresponding simulation time $t$ is given by labels.
  Individual snapshots represent the state of gas when the 
  total torque acting on the protoplanet is minimal (left), maximal (third)
  and oscillating in between (second and right). The streamlines
  are overplotted for reference. The figure is also available as an online movie 1, showing
    the temporal evolution from $t=30$ to $33\,P_{\mathrm{orb}}$.}
  \label{fig:hydro_avrBL}
\end{figure*}

So far, we described two new findings that
were not incorporated in the existing descriptions of the heating torque
\citep{Benitez-Llambay_etal_2015Natur.520...63B,Masset_2017MNRAS.472.4204M}:
The distortion of the streamline topology and the reversal of the vertical
motions.
The flow direction is directly linked to the heating torque
because it determines the redistribution of the hot gas by advection
and thus contributes to the shape of the underdense regions.
Therefore we focus on the streamline topology in this section.

The streamlines are calculated using the explicit first order Euler
integrator and the trilinear interpolation of the velocity field.
The interpolation allows us to obtain the velocity vector 
at an arbitrary location within the spherical grid.
The size of the integration step is chosen so that the
length integrated during a single propagation does not exceed $0.1$
of the shortest cell dimension.

We construct 2D and 3D projections of the streamline topologies.
For the 2D projections, the streamlines are generated 
exactly at the midplane where $v_{\phi}=0$ and
although they provide a useful visualisation,
we emphasise that they -- by construction -- carry
zero information about the adjacent vertical flows.
For the 3D projections, the streamlines are generated
slightly above the midplane (at $\phi=\pi/2-0.005\,\mathrm{rad}$)
to take into account non-zero $v_{\phi}$.

Fig.~\ref{fig:stream_kc} shows 2D and 3D streamlines in the \KC-disk,
again for the exact same simulation stages as in Figs.~\ref{fig:hydro_kc}
and \ref{fig:vert_kc}. In the plots, we distinguish the following types
of streamline: First, there are circulating streamlines that
do not cross the protoplanet's corotation. To imagine
the direction of the relative motion, we point out that the gas on inner circulating
streamlines moves faster than the protoplanet whereas
the gas on outer circulating streamlines lags behind the protoplanet.
Second, there are horseshoe streamlines that make a single U-turn
and cross the corotation once at each side of the protoplanet.
Such streamlines  ahead of the protoplanet's orbital motion
form the front horseshoe region, those located behind the protoplanet
belong to the rear horseshoe region.
To outline the separatrices between the horseshoe and circulating regions,
we highlight the critical inner and outer circulating streamlines 
that are located closest to the protoplanet.
Finally, some of our plots contain streamlines that do not fall
in any of the aforementioned categories.

When the protoplanet is non-luminous, the 2D midplane
streamlines in Fig.~\ref{fig:stream_kc} do not exhibit any unexpected
features. The stagnation point (X-point) of the flow
is located within the Hill sphere, which is also intersected by
both horseshoe and circulating streamlines. In 3D, we notice
that upon making their U-turn, the horseshoe streamlines
vertically descend towards the midplane, as already pointed out
by \cite{Fung_etal_2015ApJ...811..101F} or \cite{Lambrechts_Lega_2017A&A...606A.146L}.
The descend is also exhibited by some of the circulating
streamlines closest to the protoplanet.

For the hot protoplanet, we now obtain a clear picture
of the streamline distortion. In the midplane,
the following changes appear:
\begin{itemize}
  \item Circulating streamlines cross a smaller portion
    of the Hill sphere. When passing the protoplanet, they are bent towards
    it (unlike near the non-luminous protoplanet where
    they are rather deflected away),
    which is especially apparent for the critical circulating streamlines.
  \item The classical horseshoe streamlines
    detach from the Hill sphere and 
    make their U-turns at greater azimuthal separations.
  \item Part of the streamlines originating downstream
    the horseshoe regions is captured inside the Hill
    sphere where it rotates around the protoplanet in
    a retrograde fashion.
\end{itemize}

In 3D, the distortion has the following additional features:
\begin{itemize}
  \item The `captured' streamlines are uplifted 
    and form a spiral-like vertical column, outflowing
    and escaping from the Hill sphere.
  \item When the circulating streamlines pass the protoplanet
    and become perturbed, they
    are also uplifted.
    This behaviour is exactly the opposite to the
    cold-protoplanet situation.
\end{itemize}

\subsection{Instability of the heated gas}
\label{sec:instability}

We now return to the \KBL-disk simulation
for which we discovered the strong oscillations
of the heating torque (Fig.~\ref{fig:torque}).

\subsubsection{Evolving underdense lobes}

Investigating the evolution of the gas density,
we find that the position and size of the underdense
lobes never becomes stationary, as
shown in Fig.~\ref{fig:hydro_avrBL} (see also the online movie 1).
The figure shows a sequence of snapshots
corresponding to
$t=31.3$, $31.75$, $32.20$ and $32.65\,P_{\mathrm{orb}}$.

The first panel depicts the state when
the total torque reaches its first minimum during the
beginning of the oscillatory phase. There is a dominant underdense
lobe located ahead of the protoplanet while the rear
lobe almost disappears. Such a distribution can be easily
related to the strong negative torque: The
excess of the gas mass behind the protoplanet 
(and the paucity of mass ahead of it) leads to
an azimuthal pull acting against the orbital motion,
imposing a negative torque.

When the torque is reversing from negative to positive (second panel),
both lobes are similarly pronounced. The rear one
seems to be located closer to the protoplanet.
The third panel corresponds to the torque maximum.
The rear lobe is dominant and thus the overabundance of the
gas ahead of the protoplanet makes the torque positive.
The final panel shows the state when the oscillating
torque is approximately half-way from positive to negative.
The gas distribution indeed looks like a counterpart to
the second panel since both lobes are again apparent 
but the front one is now closer to the protoplanet.

\begin{figure*}[!hpt]
  \centering
  \begin{tabular}{ccc}
    \hline
    \multicolumn{3}{c}{hot protoplanet (\KBL-disk simulation)} \\
    (a) $t=31.3\,P_{\mathrm{orb}}$ & (b) $t=31.35\,P_{\mathrm{orb}}$ & (c) $t=31.75\,P_{\mathrm{orb}}$ \\
    \includegraphics[width=5.3cm]{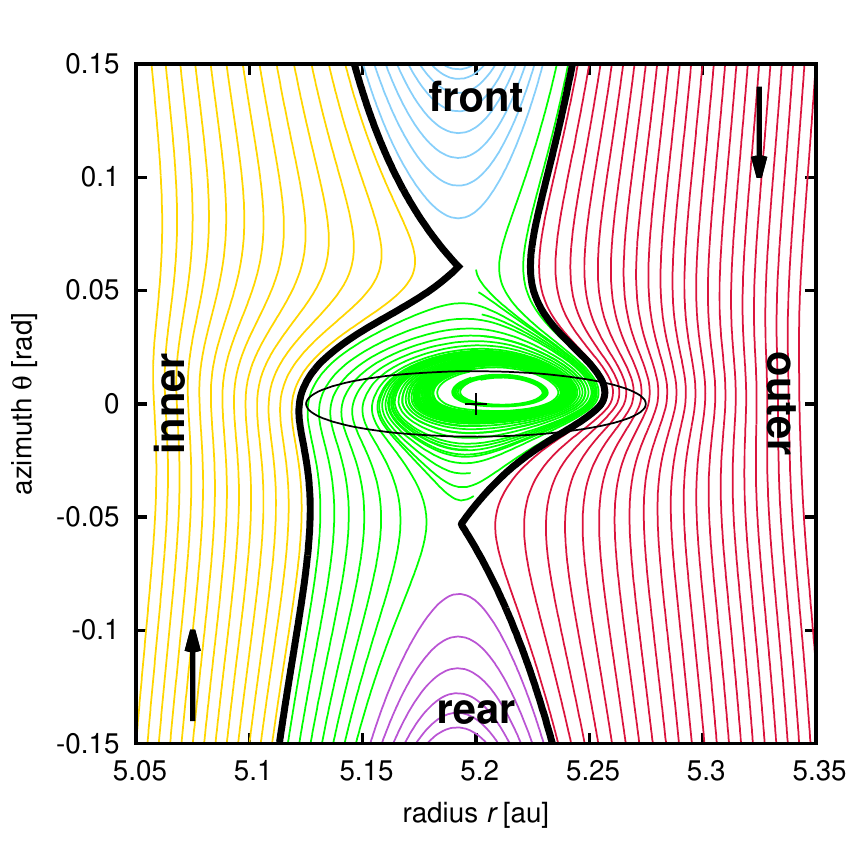} & \includegraphics[width=5.3cm]{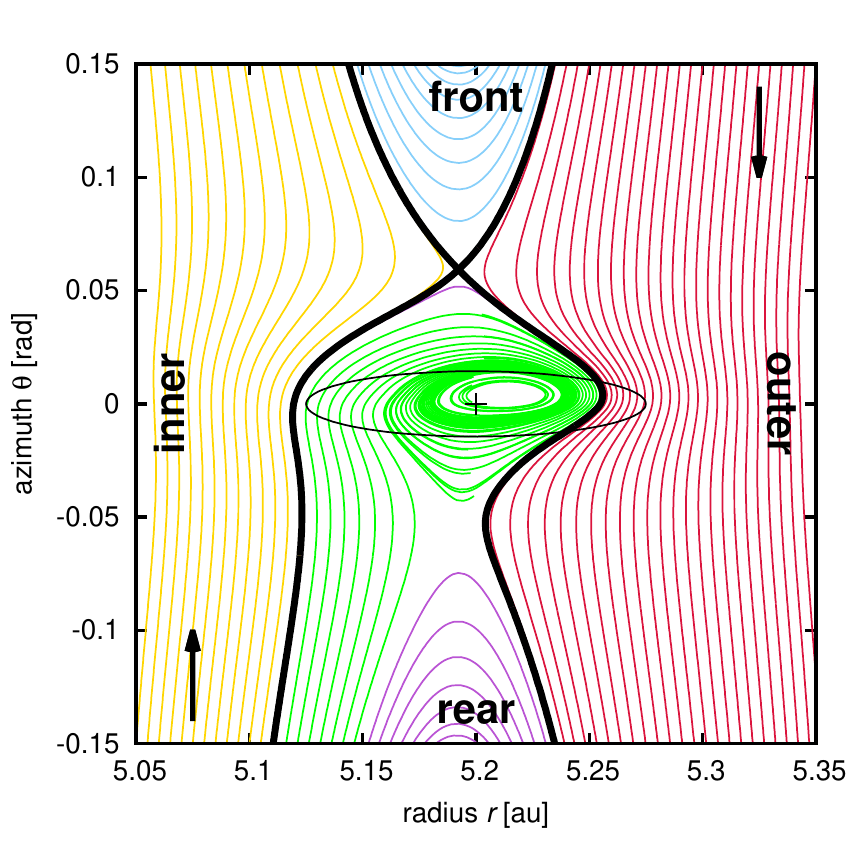} & \includegraphics[width=5.3cm]{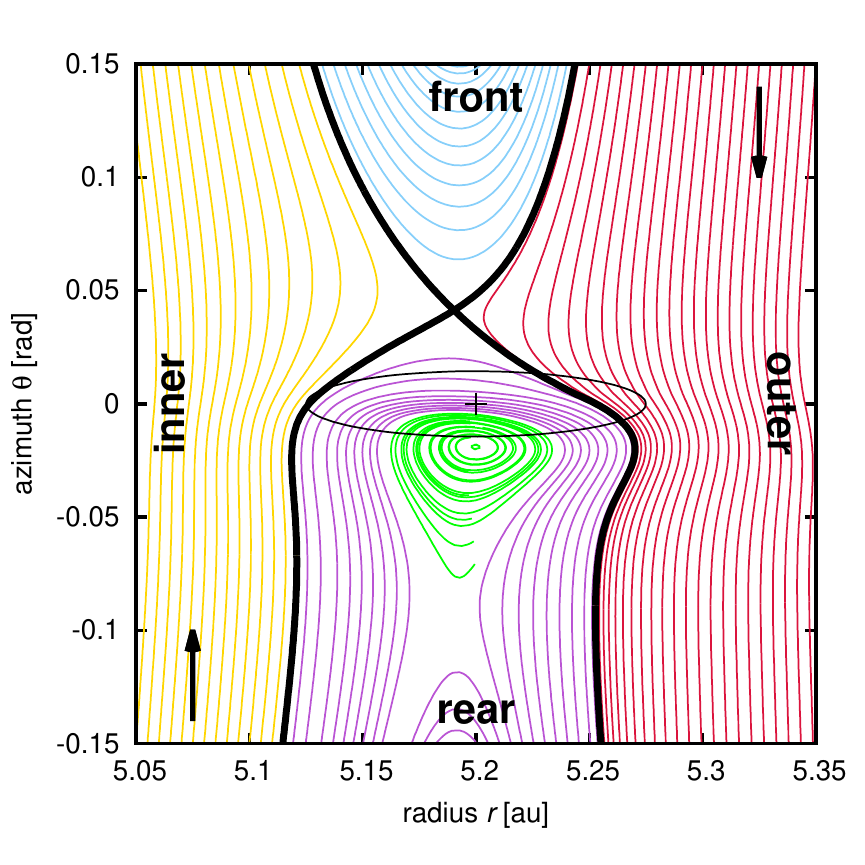}\\
    (d) $t=31.95\,P_{\mathrm{orb}}$ & (e) $t=32\,P_{\mathrm{orb}}$ & (f) $t=32.15\,P_{\mathrm{orb}}$ \\
    \includegraphics[width=5.3cm]{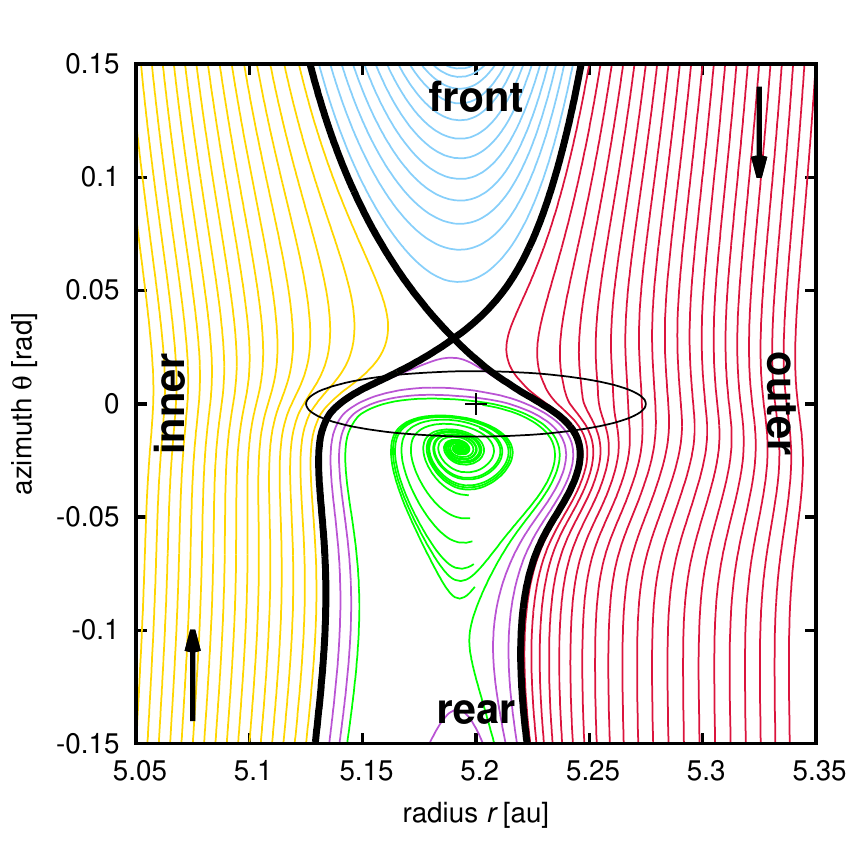} & \includegraphics[width=5.3cm]{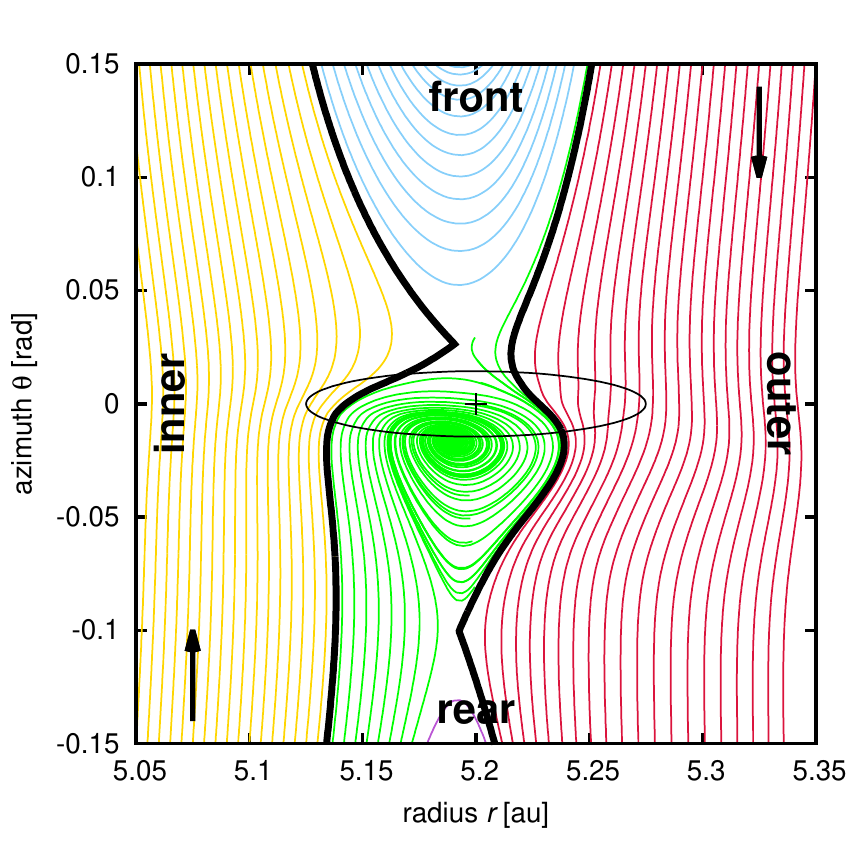} & \includegraphics[width=5.3cm]{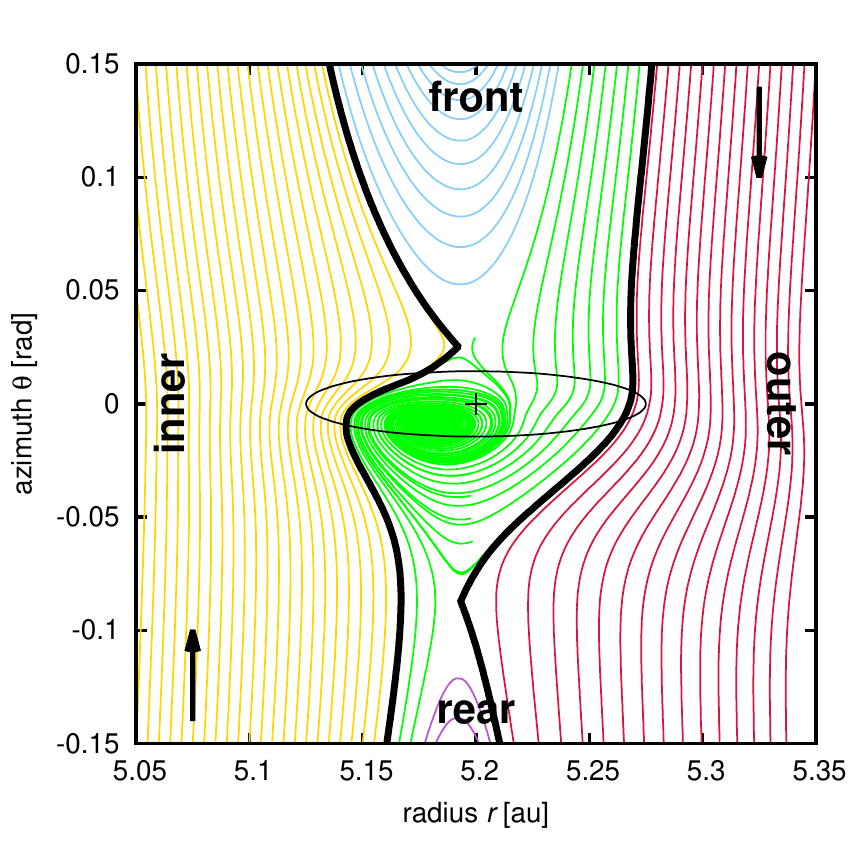}\\
  \end{tabular}
  \caption{Midplane streamline topology in the \KBL-disk simulation. The panels
  are labelled by the simulation time. The individual types of streamlines are
  the same as in Fig.~\ref{fig:stream_kc}. The sequence (a) to (f) represents the transition
  between the states corresponding to the minimum and maximum of the torque, respectively
  (see Fig.~\ref{fig:torque} to relate the panels to the
  torque evolution).}
  \label{fig:stream_2D_avrBL}
\end{figure*}

The gas redistribution is clearly related to the topology of
the streamlines and to the position of the spiral-like flow.
We notice that the centre of the captured streamlines
undergoes retrograde (`clockwise') rotation around the protoplanet.
One underdense lobe is always associated with this rotating flow.
Apparently, the redistribution of the hot gas by advection
tends to favour the lobe which is intersected by the majority
of the captured streamlines at a given time.
In the first panel of Fig.~\ref{fig:hydro_avrBL}, for example,
the hot gas is transported more efficiently into
the front lobe, creating a strong front-rear asymmetry
between the lobes. In the third panel, the situation
is exactly the opposite and the rear lobe is more pronounced.

We point out that the continuous variations of the hot lobes
and their alternating dominance
are unexpected features of the heating torque
which was previously thought to be strictly
positive \citep{Benitez-Llambay_etal_2015Natur.520...63B}.

\subsubsection{Evolving 2D and 3D streamline topology}

We again explore the streamline topology and its
changes related to the redistribution of the hot gas.
Fig.~\ref{fig:stream_2D_avrBL} shows the midplane
streamlines near the protoplanet for a selection
of simulation times between $t=31.3$ and $32.15\,P_{\mathrm{orb}}$.
The former corresponds to the torque minimum,
the latter to the torque maximum.
The time intervals between the individual panels are not uniform
but rather selected to highlight the most interesting transitions.

The sequence reveals the following features:
\begin{itemize}
  \item In panel (a), the streamlines are similar to the 
    steady-state \KC-disk case (bottom of Fig.~\ref{fig:stream_kc}) in several ways,
    mostly in the detachment of the classical horseshoe streamlines and in the
    existence of the retrograde streamlines captured from downstream horseshoe regions.
    But there are also important differences: There are none inner circulating
    streamlines intersecting the Hill sphere and moreover, a larger part of 
    the captured streamlines originate in the rear horseshoe region
    while the front horseshoe region is almost disconnected.
  \item In panel (b), the front horseshoe region becomes entirely disconnected.
    The captured streamlines originate exclusively in the rear horseshoe region.
  \item In panel (c), also the outer circulating streamlines stop crossing the Hill sphere.
    Some of the streamlines that were captured in (b) now overshoot the protoplanet
    and make a U-turn ahead of it.
    The centre around which the captured streamlines enclose becomes shifted
    behind the protoplanet.
  \item In panel (d), the front X-point moves closer to the Hill sphere
    and so do the front horseshoe streamlines.
    The rear horseshoe region becomes radially narrower and the 
    number of U-turn streamlines overshooting the protoplanet diminishes.
  \item In panel (e), the front horseshoe region reconnects with the
    captured streamlines.
  \item In panel (f), the captured streamlines originate mostly
    in the front horseshoe region while the rear horseshoe region
    is evolving towards its disconnection, similarly to what we 
    saw for the front horseshoe region in panels (a) and (b).
    The centre of the captured streamlines moves inwards from the protoplanet
    (and will continue to propagate ahead).
\end{itemize}

\begin{figure}[!hpt]
  \centering
  \begin{tabular}{c}
      \hline
     \multicolumn{1}{c}{hot protoplanet (\KBL-disk simulation)} \\
  $t=31.3\,P_{\mathrm{orb}}$ \\
  \includegraphics[width=7.8cm]{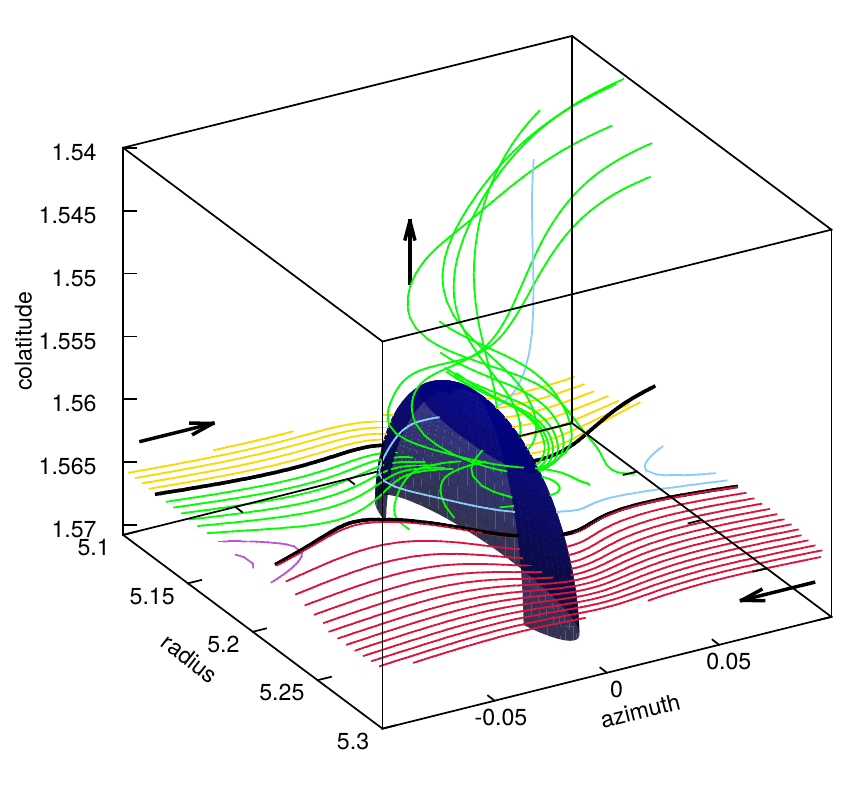} \\
  $t=31.75\,P_{\mathrm{orb}}$ \\
  \includegraphics[width=7.8cm]{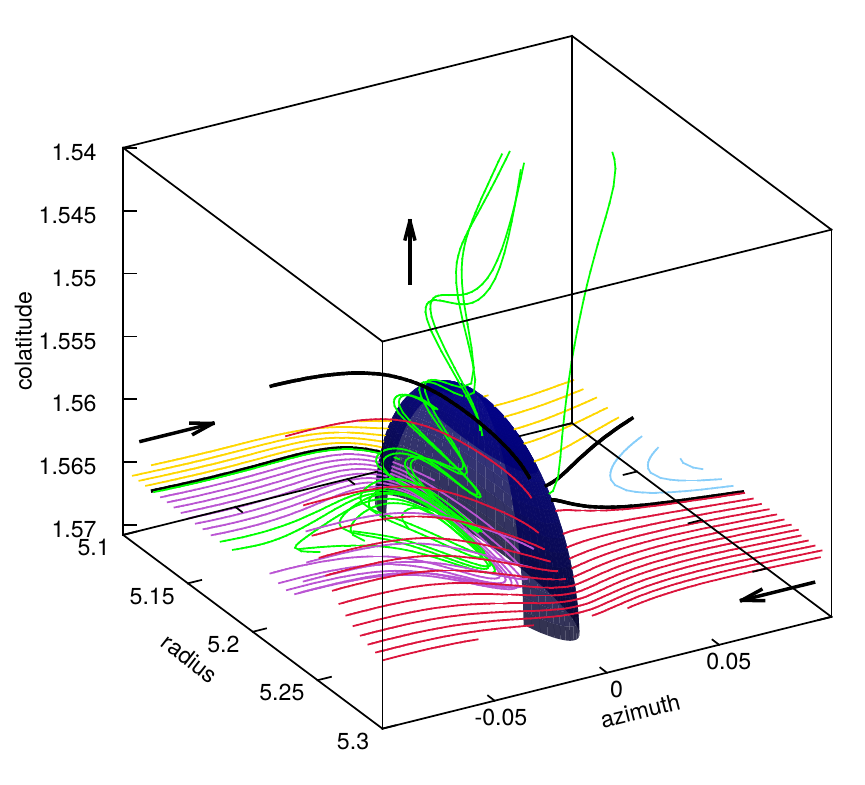} \\
  $t=32.15\,P_{\mathrm{orb}}$ \\
  \includegraphics[width=7.8cm]{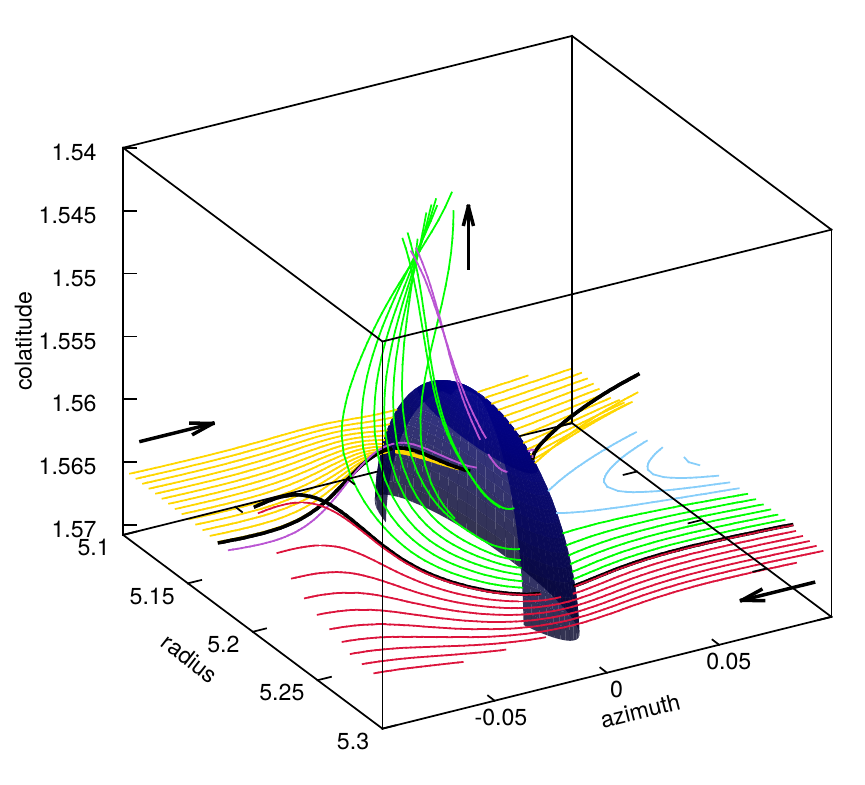} \\
  \end{tabular}
  \caption{3D streamlines in the \KBL-disk simulation.
  Each snapshot is labelled by the simulation time.
  The panels correspond to the minimum (top) and maximum (bottom) torque
  and to the state in between (middle).}
  \label{fig:stream_3D_avrBL}
\end{figure}

Fig.~\ref{fig:stream_3D_avrBL} shows three selected snapshots of 3D streamlines
($t=31.3$, $31.75$ and $32.15\,P_{\mathrm{orb}}$)
when the oscillating torque is at its minimum (top),
grows halfway towards the maximum (middle) and reaches it (bottom).
Clearly, the reshaping of the streamline topology 
that we described for the midplane propagates
in a complicated way into the vertical direction as well:

\begin{itemize}
  \item In the first panel, the spiraling streamlines of the vertical
    column are more or less centred above the protoplanet and 
    as they rise above the midplane they penetrate the
    majority of the Hill sphere.
    
  \item In the second panel, we see the overshooting rear horseshoe streamlines
    making their U-turns within the Hill sphere. The vertical column
    of captured streamlines is displaced to the rear of the Hill sphere.
    As the captured streamlines spiral up, the column tilts towards the Hill sphere.
    The outer circulating streamlines are strongly uplifted
    towards colatitudes above the Hill sphere.
    
  \item In the third panel, some uplifted outer circulating streamlines
    penetrate into the vertical column and by this reconfiguration,
    the column reconnects with the front horseshoe region
    while the rear one starts to disconnect.
\end{itemize}

\subsection{Physical processes distorting the gas flow}
\label{sec:processes}

In previous sections, we revealed that the gas heating from
an accreting protoplanet changes the topology of the flow.
Perturbed streamlines have a tendency to bent towards
the protoplanet and also to rise vertically.
If the perturbations become strong enough, the
streamlines can form a vertical spiral.
In this section, we investigate the physical processes responsible
for such a streamline distortion.

In Sect.~\ref{sec:vorticity_evol}, we theorise that the streamline distortion
is a result of vorticity perturbations arising
because the vigorous accretion heating renders the circumplanetary
gas baroclinic.
A confirmation is provided
for the steady state of the \KC-disk simulation
in Sects.~\ref{sec:baroclinic_vorticity} and \ref{sec:baroclinic_region}.
Finally, Sect.~\ref{sec:vertical_convection}
demonstrates that the vertical temperature
gradient above the accreting protoplanet is superadiabatic
and we also highlight differences between the \KC-disk and \KBL-disk.

\begin{figure*}[!hpt]
  \centering
  \begin{tabular}{ccc}
    \hline
    \multicolumn{3}{c}{cold non-luminous protoplanet (\KC-disk simulation, $t=30\,P_{\mathrm{orb}}$)} \\
    azimuthal components & radial components & polar components \\
    \includegraphics[width=5.8cm]{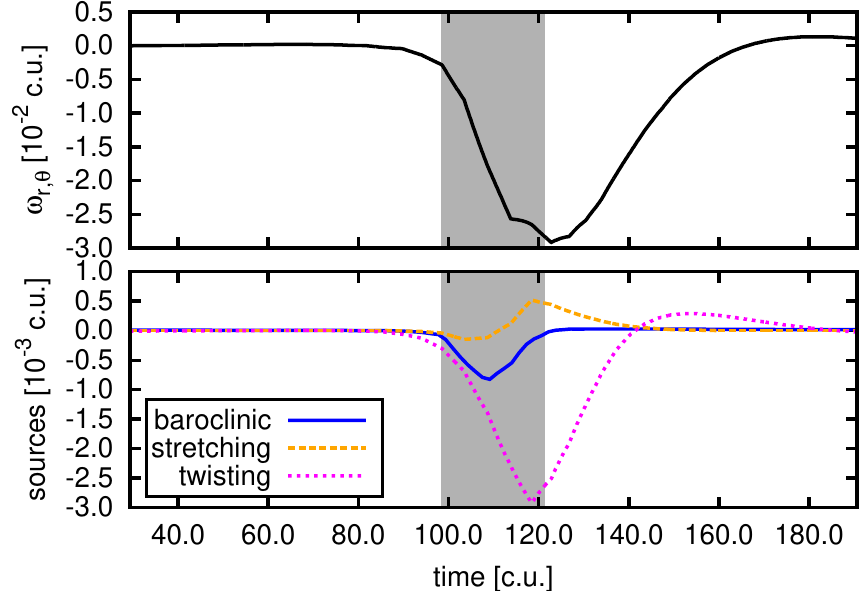} & \includegraphics[width=5.8cm]{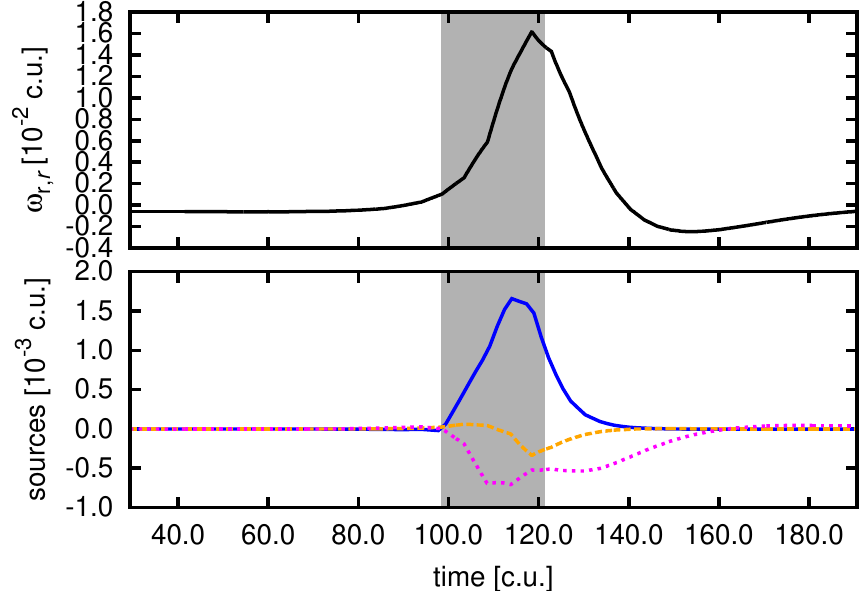} & \includegraphics[width=5.8cm]{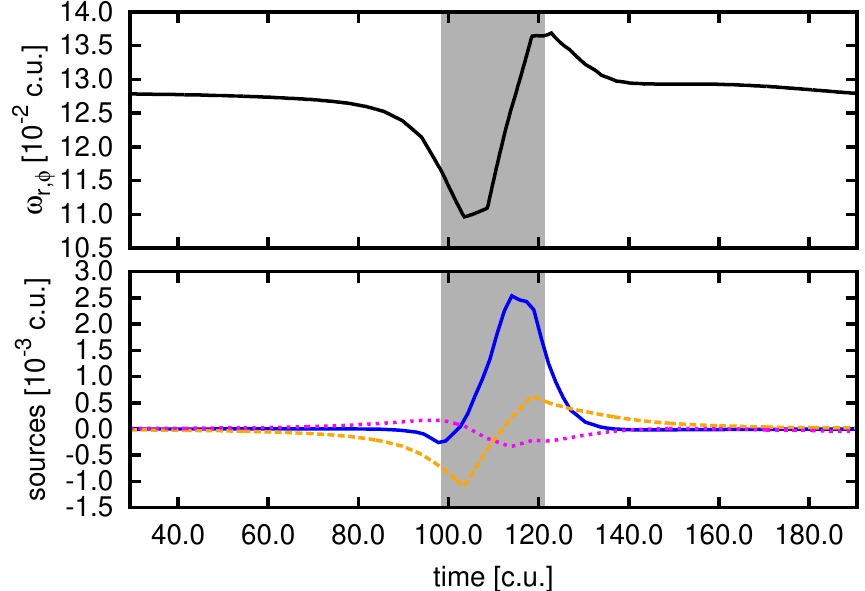} \\
    \hline
    \multicolumn{3}{c}{hot luminous protoplanet (\KC-disk simulation, $t=60\,P_{\mathrm{orb}}$)} \\
    azimuthal components & radial components & polar components \\
    \includegraphics[width=5.8cm]{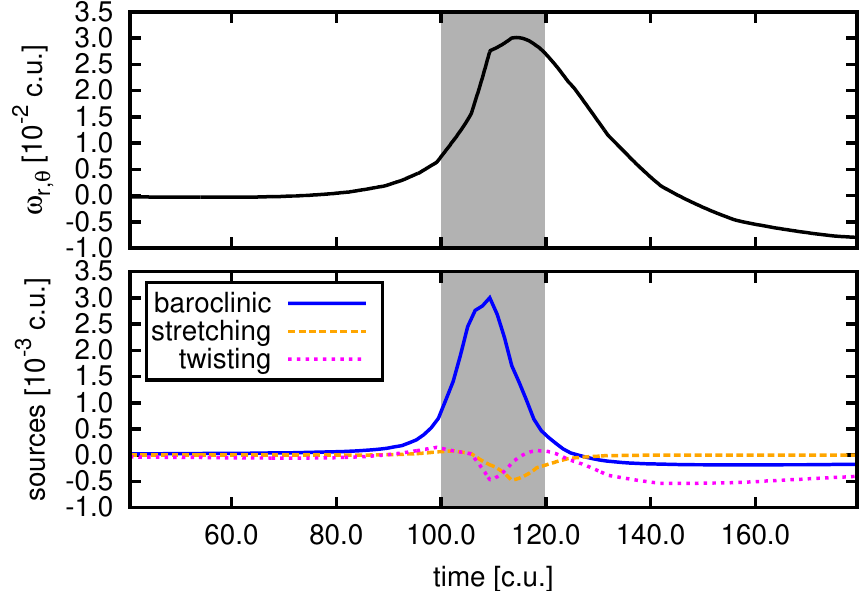} & \includegraphics[width=5.8cm]{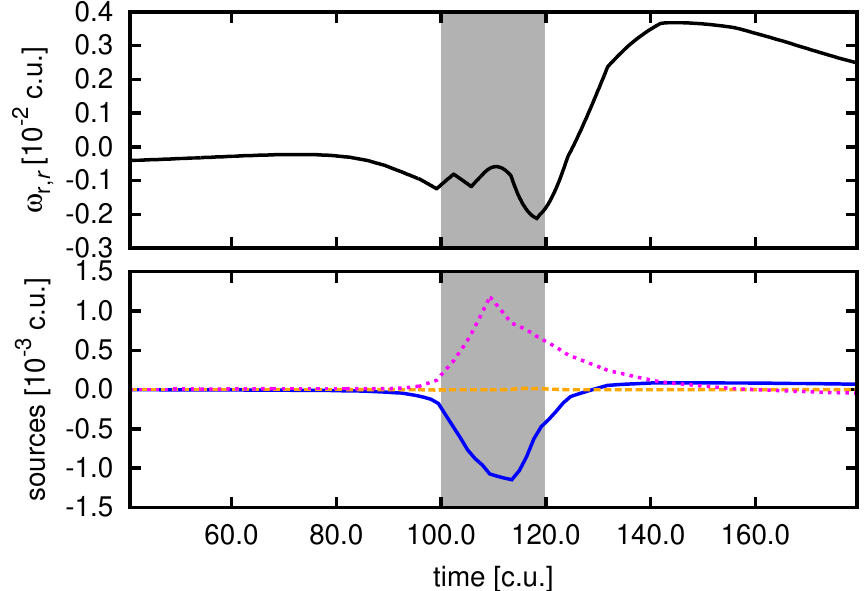} & \includegraphics[width=5.8cm]{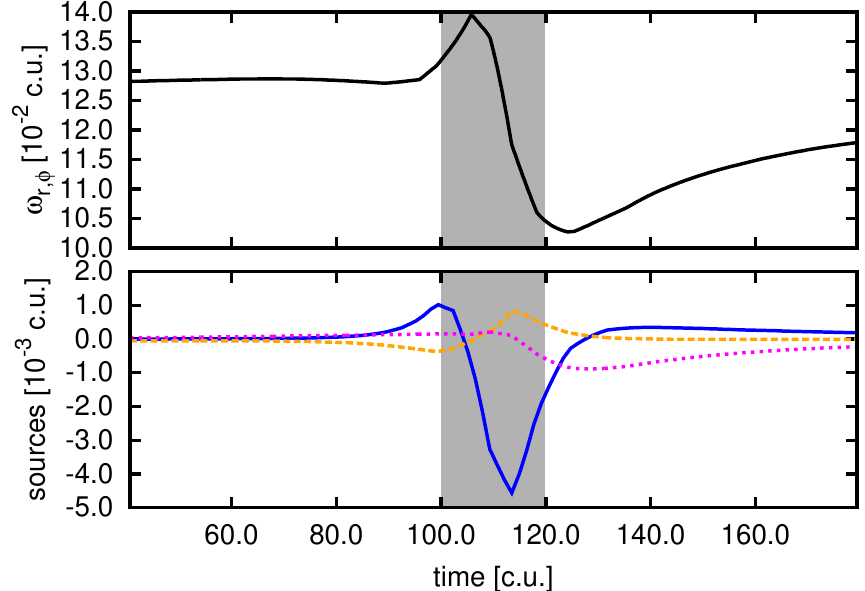} \\
  \end{tabular}
  \caption{Evolution of the relative vorticity (first and third row)
    and balance of the vorticity source terms (second and fourth row) along a
    single streamline in the \KC-disk simulation.
  The streamlines for these measurements are chosen from the 3D sets displayed
  in Fig.~\ref{fig:stream_kc}.
  For the hot protoplanet (bottom two rows),
  the extremal outer circulating streamline is selected
  and for the cold protoplanet (top two rows), we select an outer circulating streamline
  with a comparable Hill sphere crossing time.
  Azimuthal (left), radial (middle) and polar (right) components of the
  vorticity (black solid curve), baroclinic term (blue solid curve), stretching
  term (orange dashed curve) and twisting term (magenta dotted curve) are displayed
  using scaled code units. The grey rectangle marks the Hill sphere crossing.
  We point out that the source terms represent the rate of change of the vorticity
  and also that the vertical range is not kept fixed among individual panels.
  }
  \label{fig:vorticity_terms}
\end{figure*}

\subsubsection{Vorticity evolution}
\label{sec:vorticity_evol}

The spiral-like structure of the captured streamlines
and the bending of nearby circulating streamlines
suggests that the vorticity of the flow is modified when 
the protoplanet becomes hot.
The vorticity can be defined via the relation
\begin{equation}
  \vec{\omega_{\mathrm{a}}} \equiv \vec{\omega_{\mathrm{r}}} + 2\vec{\Omega} \equiv \nabla\times\vec{v} + 2\vec{\Omega} \, ,
  \label{eq:vorticity_def}
\end{equation}
where $\vec{\omega_{\mathrm{a}}}$ is the absolute vorticity and
$\vec{\omega_{\mathrm{r}}}$ is the relative vorticity in the
reference frame corotating with the protoplanet.

Evolution of $\vec{\omega_{\mathrm{r}}}$ is described
by the vorticity equation in the corotating frame (see Appendix~\ref{sec:vorticity_eq})
\begin{equation}
  \frac{\mathrm{D}\vec{\omega_{\mathrm{r}}}}{\mathrm{D}t} = \left(\vec{\omega_{\mathrm{a}}}\cdot\nabla  \right)\vec{v} - \vec{\omega_{\mathrm{a}}}\left( \nabla\cdot\vec{v} \right) + \frac{\nabla\rho\times\nabla P}{\rho^{2}} \, ,
  \label{eq:vorticity_eq}
\end{equation}
where $\mathrm{D}/\mathrm{D}t$ denotes the Lagrangian derivative.
In writing the equation, we neglected the effects of viscous diffusivity
(large Reynolds number limit) but otherwise the equation is general.

Regarding the right-hand side terms, the first one
describes the tendency of vortex tubes to become twisted
due to velocity field gradients. The second one characterises
the stretching or contraction of vortex tubes due to
flow expansion or compression. These first two terms,
usually called the twisting and stretching term,
are only important if there is non-zero absolute vorticity already
existing in the flow and they can cause its redistribution.

The remaining term on the right-hand side is the baroclinic term.
It vanishes in barotropic flows where the pressure and density
gradients are always parallel, but since our model is not barotropic,
$\nabla\rho$ and $\nabla P$ can be misaligned, leading to 
vorticity production or destruction since their cross product can
be non-zero. Because $\rho$ near the accreting protoplanet exhibits
asymmetric perturbations
while $P$ remains roughly spherically symmetric,
one can expect non-zero baroclinic perturbations to arise.
The respective non-zero vorticity then enhances circulation
around a given point of the continuum, twisting the streamlines
with respect to the situation unperturbed by accretion heating.

\subsubsection{Baroclinic vorticity generation}
\label{sec:baroclinic_vorticity}

It is not a priori evident which source term is the most
important for the vorticity evolution in our simulations.
In Fig.~\ref{fig:vorticity_terms}, we study the variations
of $\vec{\omega_{\mathrm{r}}}$ and the source terms of Eq.~(\ref{eq:vorticity_eq})
along a single outer circulating streamline. 
We compare the situation near the cold and hot protoplanet in the \KC-disk.

Downstream, before the streamline encounters the protoplanet,
the situation is similar for the compared cases: the azimuthal and radial
vorticity components $\omega_{\mathrm{r},\theta}$ and $\omega_{\mathrm{r},r}$
are negligible, while the polar component $\omega_{\mathrm{r},\phi}$
is non-zero and positive.
The positive value of $\omega_{\mathrm{r},\phi}$ corresponds
to the inherent vorticity in a flow with the Keplerian shear:
Taking $v_{r}=0$, $v_{\phi}=0$ and $v_{\theta}=\sqrt{GM/r} - \Omega r$, one obtains
$\omega_{\mathrm{r},\phi}=-0.5\Omega_{\mathrm{K}}\left( r \right) + 2\Omega$.
At the same time, the source terms
are zero because far from the protoplanet there are no strong
velocity gradients, no compression and $\nabla\rho$ and $\nabla P$ are aligned.

To relate the vorticity variation with the streamline distortion close to
the protoplanet, let us make a thought experiment, utilising
the fact that the vorticity describes the tendency of the flow to circulate
around some point in space.
First we focus on the hot-protoplanet case which is the most important for us.
Imagine an observer moving along the 
critical outer circulating streamline, corresponding to the outer thick black curve
in the bottom right panel of Fig.~\ref{fig:stream_kc}.
The observer moves with the flow, predominantly in the $-\theta$ direction.
For this experiment, we dub the directions $\theta$, $-\theta$, $r$, $-r$, $\phi$ and $-\phi$
as behind, ahead, outwards, inwards, down and up, respectively.
Considering only the streamlines of Fig.~\ref{fig:stream_kc} originating at $r>r_{\mathrm{p}}$,
they initially define a plane at constant $\phi=\pi/2-0.005\,\mathrm{rad}$
and the observer propagates through that plane. When
the flow reaches the Hill sphere, the observer studies the instantaneous
displacement of nearby gas parcels which will correspond to the 
deformation of the surface defined by neighbouring streamlines.

According to bottom panels of Fig.~\ref{fig:vorticity_terms},
the observer measures $\omega_{\mathrm{r},\theta}>0$ when crossing the Hill sphere.
Thus $\omega_{\mathrm{r},\theta}$ points against the direction of the observer's motion,
forcing the circulation in the local $(r,\phi)$ plane.
The nearby gas parcels must obey the following right-hand rule: When the thumb points
in the direction of $\omega_{\mathrm{r},\theta}$, wrapping fingers determine
the direction of circulation. From the observer's point of view, an outer gas parcel
will be falling downwards to the midplane and an inner gas parcel will be rising upwards.
This is exactly in accordance with Fig.~\ref{fig:stream_kc} where
streamlines passing the protoplanet are uplifted from 
the $\phi=\pi/2-0.005\,\mathrm{rad}$ plane and the kick gets stronger
with decreasing separation to the protoplanet.

$\omega_{\mathrm{r},r}$ only slightly oscillates during the Hill sphere
crossing but becomes positive (although relatively small) upstream.
Using again the same considerations as above, $\omega_{\mathrm{r},r}$
points outwards from the observer after crossing the Hill sphere,
promoting circulation in the $(\theta,\phi)$
plane. Using the right-hand rule, a gas parcel ahead of the observer
will be falling downwards and a gas parcel behind the observer will be rising
upwards. In Fig.~\ref{fig:stream_kc}, this is reflected by the streamline
topology when the red streamlines rising after the Hill sphere passage
suddenly start to fall back towards the midplane.

As for the remaining vorticity component
$\omega_{\mathrm{r},\phi}$,
it remains positive during the Hill sphere crossing, but it acquires
a positive boost at first and a more prominent negative perturbation afterwards.
The later diminishes the circulation related to shear
in the $(r,\theta)$ plane. This is only possible if
gas parcels near the observer become displaced towards trajectories
with smaller shear velocities.
In Fig.~\ref{fig:stream_kc}, the streamline topology indeed exhibits
such a behaviour because when the red streamlines (and similarly green and yellow ones)
pass the protoplanet, they are being bent towards it.

Looking at the source terms, it is obvious that the baroclinic term
is responsible for perturbing $\omega_{\mathrm{r},\theta}$ and $\omega_{\mathrm{r},\phi}$,
while counteracting the twisting term contributing to $\omega_{\mathrm{r},r}$.
The importance of the baroclinic term for the flow approaching the hot protoplanet
is thus confirmed. Moreover, the perturbation is indeed 3D 
as each of the studied components is important for the resulting streamline
topology.

Comparing the hot-protoplanet case to the cold-protoplanet case,
we notice that the evolution of $\omega_{\mathrm{r},\theta}$ and $\omega_{\mathrm{r},\phi}$
is roughly antisymmetric, as well as the evolution of the baroclinic source term.
This is consistent with our finding that streamlines near the cold protoplanet
are distorted in the opposite manner (they fall towards the midplane and
deflect away from the protoplanet).
In other words, the baroclinic behaviour of the protoplanet's vicinity is 
reverted between the cold- and hot-protoplanet case.

\subsubsection{Baroclinic region}
\label{sec:baroclinic_region}

\begin{figure}[!hpt]
  \centering
  \begin{tabular}{c}
    \hline
    cold protoplanet (\KC-disk simulation, $t=30\,P_{\mathrm{orb}}$) \\
    \includegraphics[width=7.8cm]{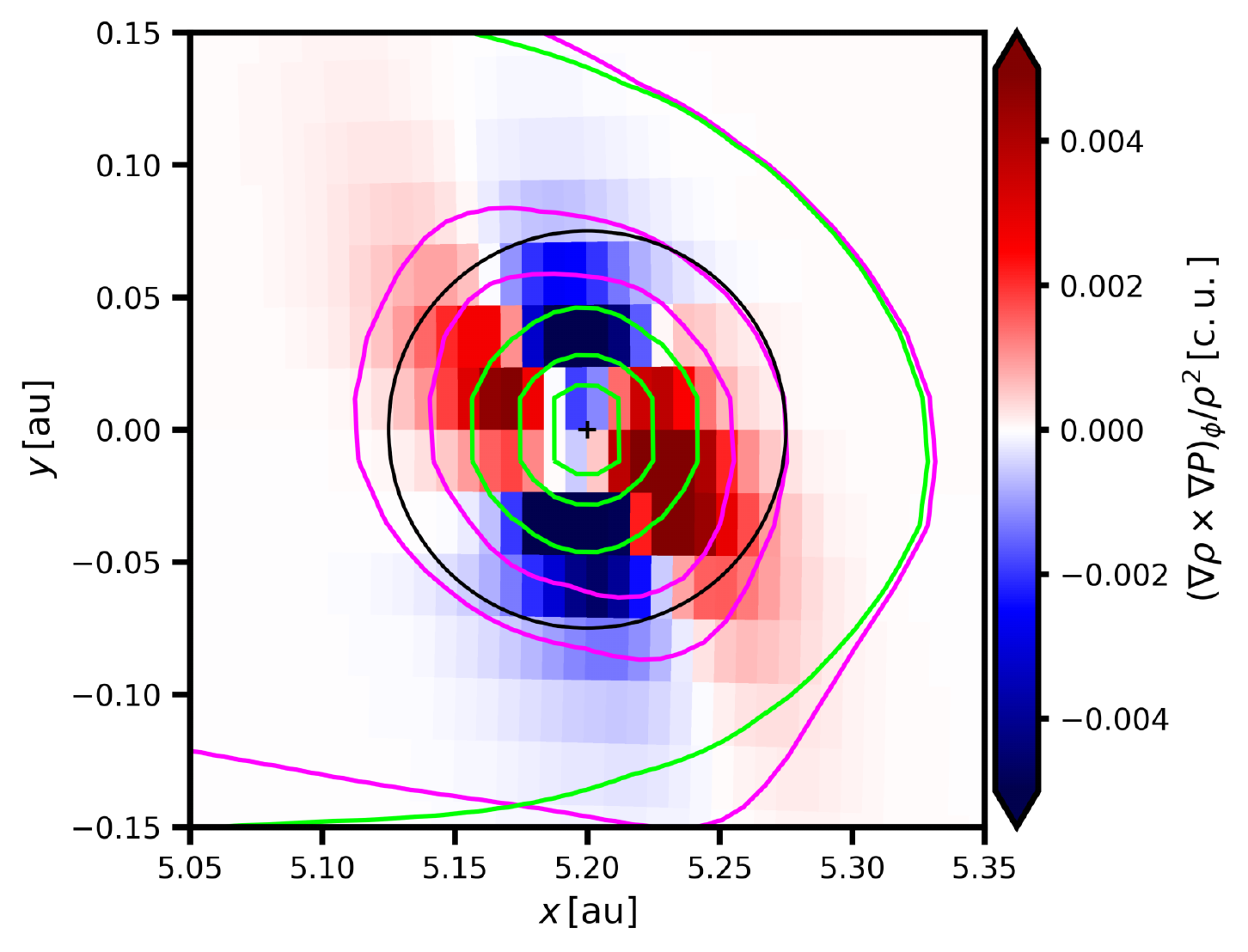} \\
    \hline
    hot protoplanet (\KC-disk simulation, $t=60\,P_{\mathrm{orb}}$) \\
    \includegraphics[width=7.8cm]{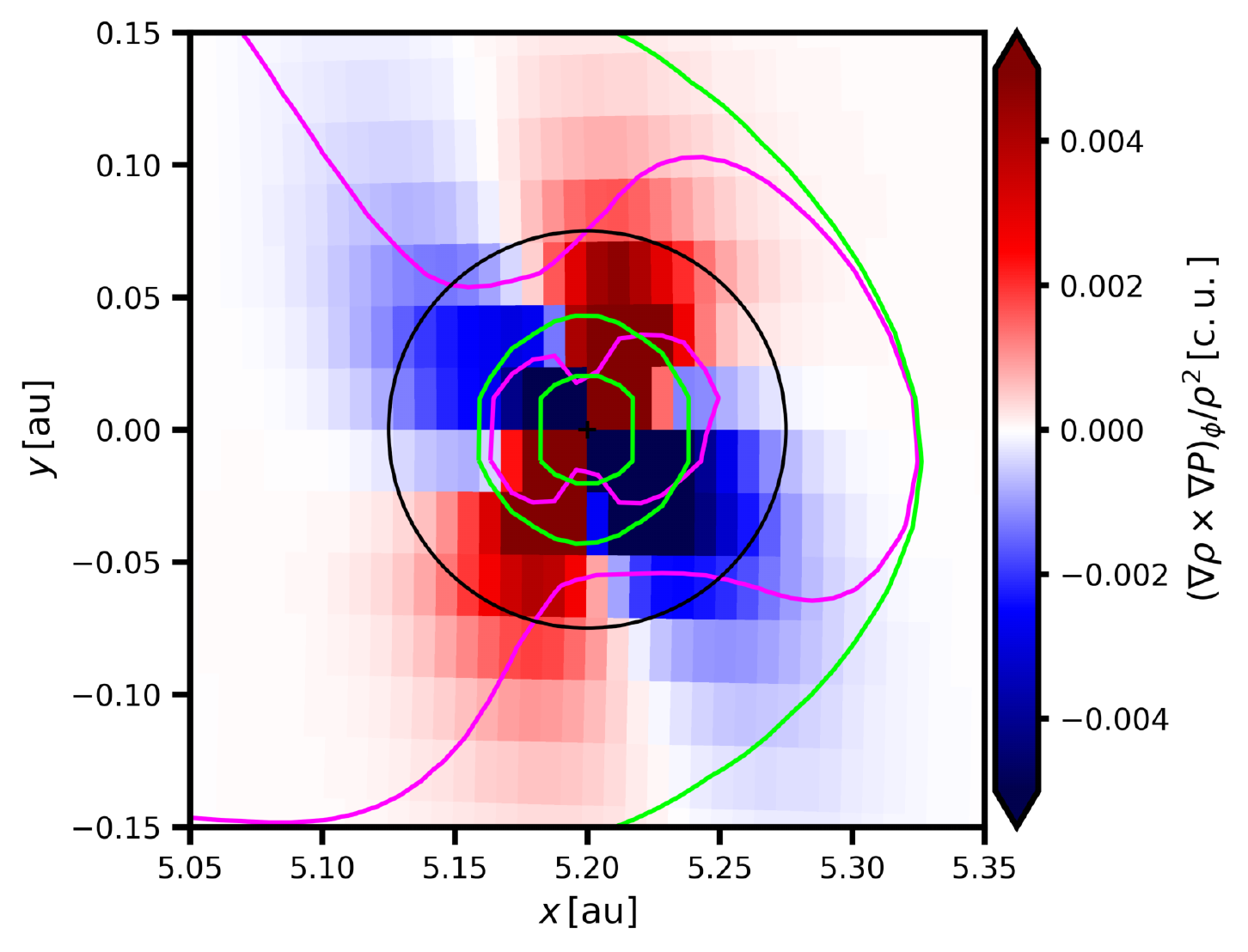} \\
  \end{tabular}
  \caption{Maps of the $\phi$-component of the baroclinic term in the \KC-disk
  simulation. The purple isocontours depict several levels of the constant
  volume density and the green isocontours correspond to the isobars. The levels
  of the contours are kept fixed between the panels.}
  \label{fig:baroclinic_kc}
\end{figure}

Although we do not repeat the vorticity analysis for the remaining sets of streamlines,
it is clear that features of the streamline distortion
can be explained by the baroclinic generation of the vorticity.
To further support this claim, we compare in Fig.~\ref{fig:baroclinic_kc}
the map of the baroclinic term near the cold and hot protoplanet in the \KC-disk. 
We plot the polar component of the baroclinic term $\left(\nabla\rho\times\nabla P\right)_{\phi}/\rho^{2}$
in the midplane.
Since the midplane flow is effectively 2D (because $v_{\phi}=0$ in midplane),
only the polar component of $\vec{\omega_{\mathrm{r}}}$ is non-zero and thus
also the polar component of the baroclinic term is the most important one.

Fig.~\ref{fig:baroclinic_kc} reveals that the gas is baroclinic
near both the cold and hot protoplanet,
but for the latter case, the map becomes approximately antisymmetric compared
to the cold-protoplanet case,
slightly rotated in the retrograde sense and the baroclinic region
is more extended.

The existence of the baroclinic region can be explained using the isocontours of constant 
volume density and isobars. For a barotropic gas ($\nabla\rho\parallel\nabla P$),
any nearby isocontours of constant $\rho$ and $P$ should have the same shape.
Wherever the isocontours depart from one another, it means
that the local gradients of $\rho$ and $P$ are misaligned and that
the baroclinic term is non-zero.

Looking at the cold-protoplanet case, the isobars and contours
of constant density are nearly spherically symmetric. However,
we notice that each density isocontour exhibits two bumps
and appears to be stretched in the direction where the
gas outflows from the Hill sphere. Clearly, these bumps
are associated with the cold-finger perturbation 
that appears in the same location (see Fig.~\ref{fig:hydro_kc}).
Since the cold fingers are filled with overdense gas,
they perturb the local density gradient, making
it to point towards them. At the same time, the
isobars remain approximately spherically symmetric.

When the protoplanet is hot, the cold fingers are replaced
with hot underdense perturbations. Therefore, the
local density gradient tends to point away
from them. We can see that this is indeed true
because the density isocontours do not exhibit
bumps, but rather concavities across the
overheated region (compare with Fig.~\ref{fig:hydro_kc}).
Clearly, this is the reason
why the baroclinic map for the hot protoplanet
looks reverted compared to the cold protoplanet.

Summarising these findings, we saw that the
thermal perturbations associated either with the
cold-finger effect or the heating torque make 
the circumplanetary region baroclinic. 
But the influence on the vorticity evolution
is the opposite when comparing the cold- and hot-protoplanet
case.

\subsubsection{Vertical convection}
\label{sec:vertical_convection}

Although the baroclinic distortion of the gas flow
is a robust mechanism, it does not provide a simple
explanation why the \KBL-disk simulation exhibits
gas instability whereas the \KC-disk simulation remains stable.
We now explore the vertical stability of both disks
against vertical convection, considering only
the hot-protoplanet limit.

To do so, we employ the Schwarzschild criterion
\begin{equation}
  %\Bigl|\frac{\mathrm{d}T}{\mathrm{d}\phi}\Bigr|_{\mathrm{rad}} > \Bigl|\frac{\mathrm{d}T}{\mathrm{d}\phi}\Bigr|_{\mathrm{ad}}
  |\nabla_{\mathrm{rad},\phi}| > |\nabla_{\mathrm{ad},\phi}|
  \,\Leftrightarrow\,\frac{|\nabla_{\mathrm{rad},\phi}|}{|\nabla_{\mathrm{ad},\phi}|}-1>0 \, ,
  \label{eq:schwarzschild}
\end{equation}
where the subscript `rad' denotes the vertical\footnote{We use the colatitude $\phi$ to study
the vertical gradients because the curvature
of spherical coordinates near the midplane does not
significantly depart from the true vertical direction.}
temperature 
gradient found in our simulations and the subscript `ad' denotes the 
temperature gradient that an adiabatic gas would establish.
The $\nabla$ symbol stands for the logarithmic gradient $\mathrm{d}\log{T}/\mathrm{d}\log{P}$,
yielding $|\nabla_{\mathrm{ad},\phi}|=(\gamma-1)/\gamma$ for the adiabatic case.

The Schwarzschild criterion is not necessarily a universal way to determine if the disk
is unstable to convection. The reason is that 
convective destabilisations in protoplanetary disks
are opposed by diffusive effects \citep[e.g.][]{Held_Latter_2018MNRAS.480.4797H}
and shear motions \citep{Rudiger_etal_2002A&A...391..781R}.
However, to our knowledge there are no convective criteria that would take
into account the disk perturbation by the protoplanet,
we thus use the Schwarzschild criterion for its simplicity,
keeping the limitations in mind.

\begin{figure}[!hpt]
  \centering
  \begin{tabular}{c}
  \hline
  \KC-disk, hot protoplanet ($t=60\,P_{\mathrm{orb}}$) \\
  \includegraphics[width=8.8cm]{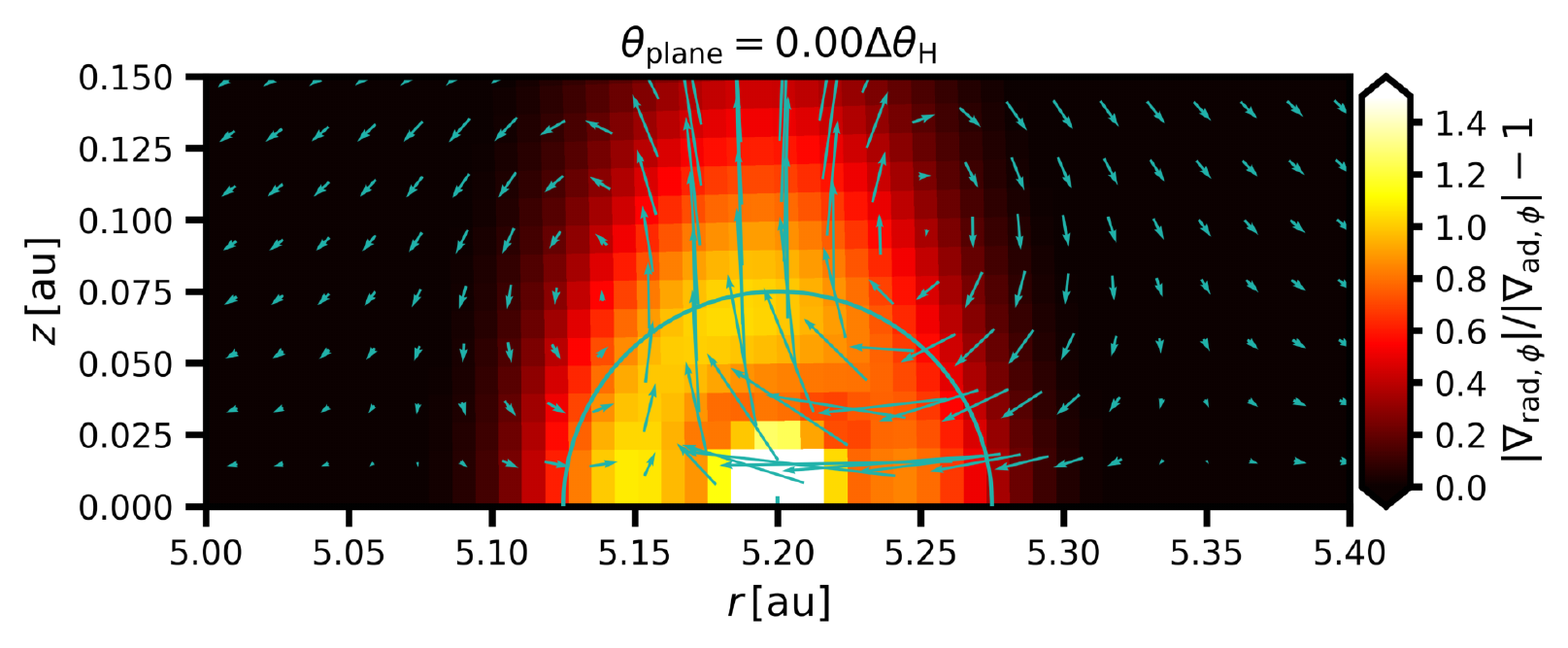} \\
  \hline
  \KBL-disk, hot protoplanet ($t=31.3$, $31.75$, $32.2$ and $32.65\,P_{\mathrm{orb}}$) \\
  \includegraphics[width=8.8cm]{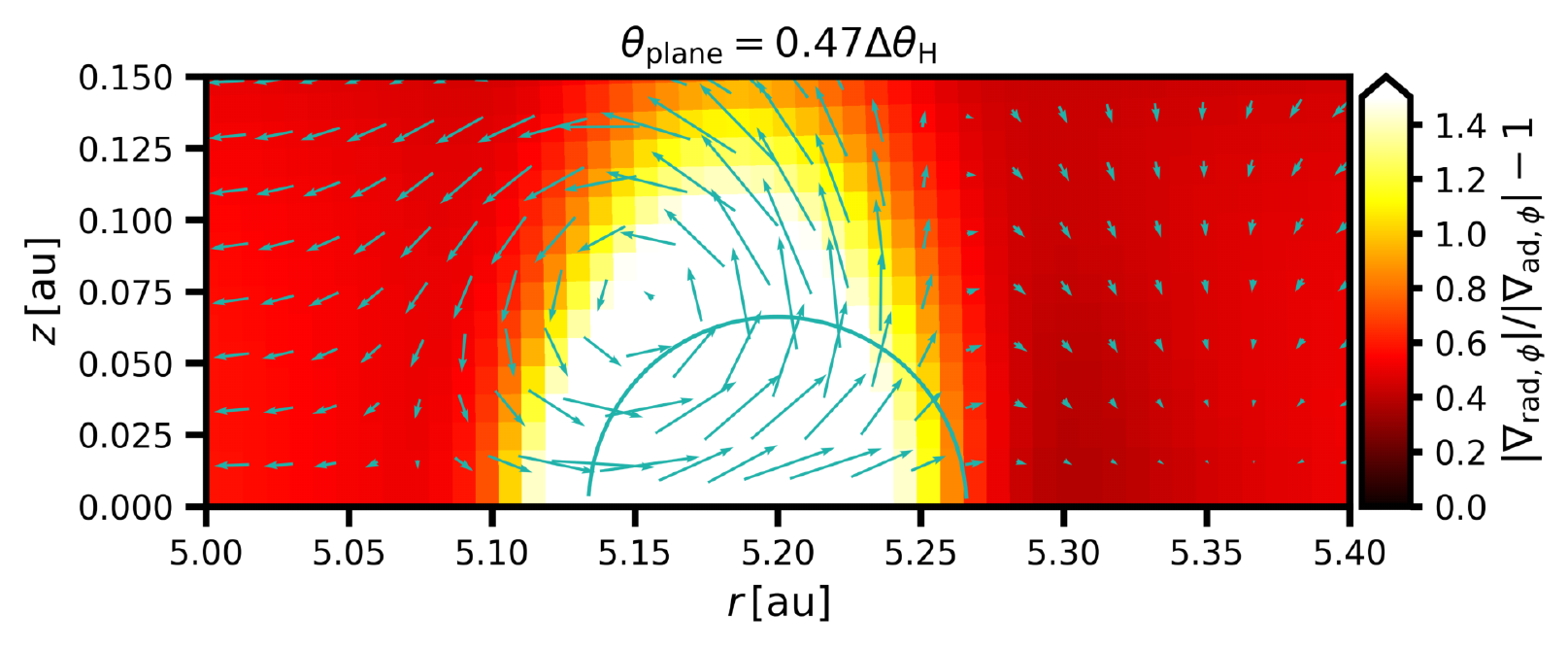} \\
  \includegraphics[width=8.8cm]{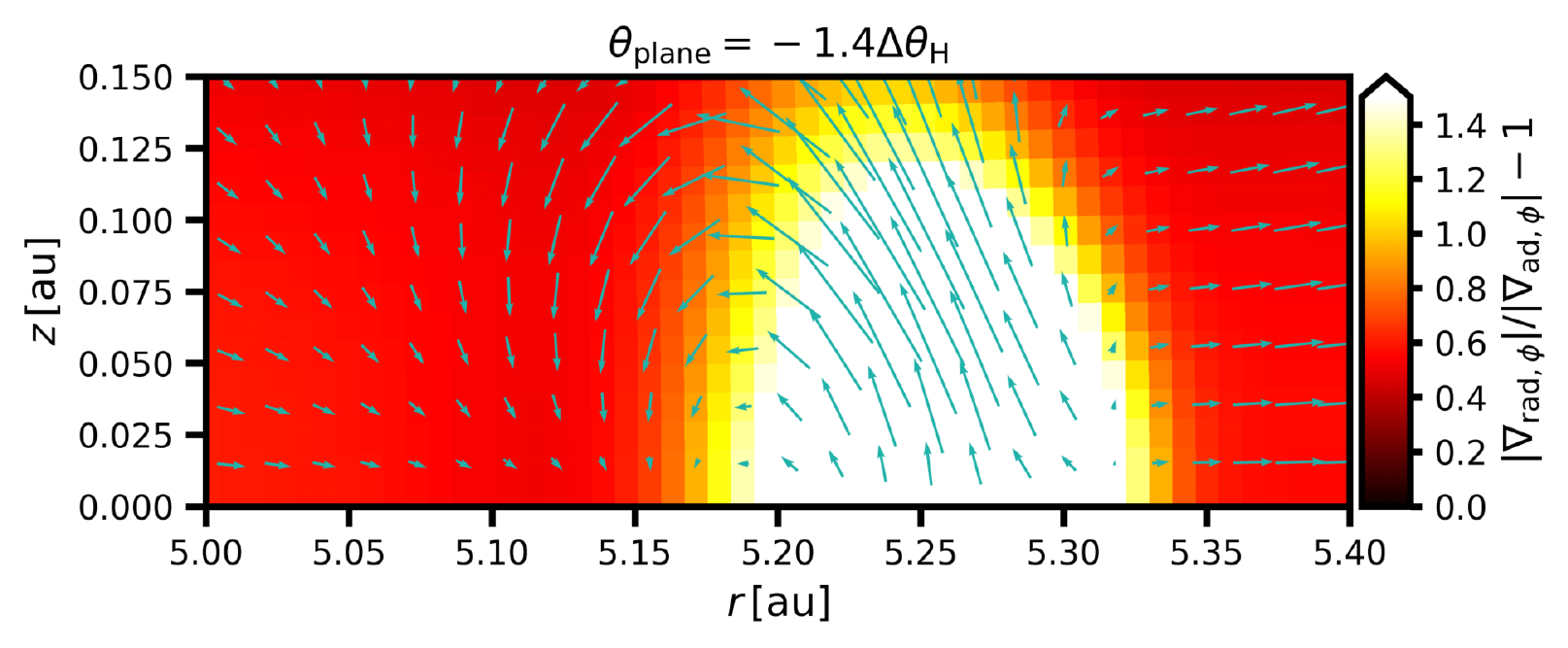} \\
  \includegraphics[width=8.8cm]{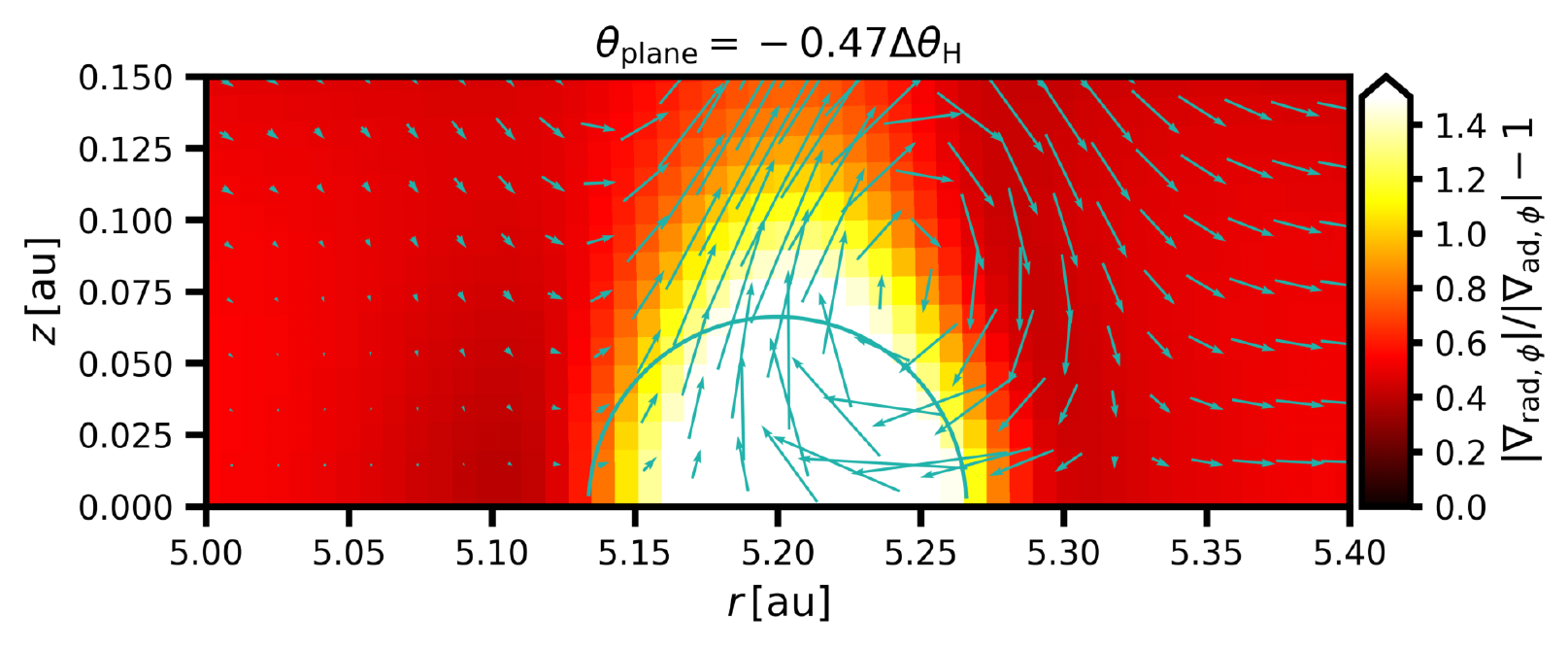} \\
  \includegraphics[width=8.8cm]{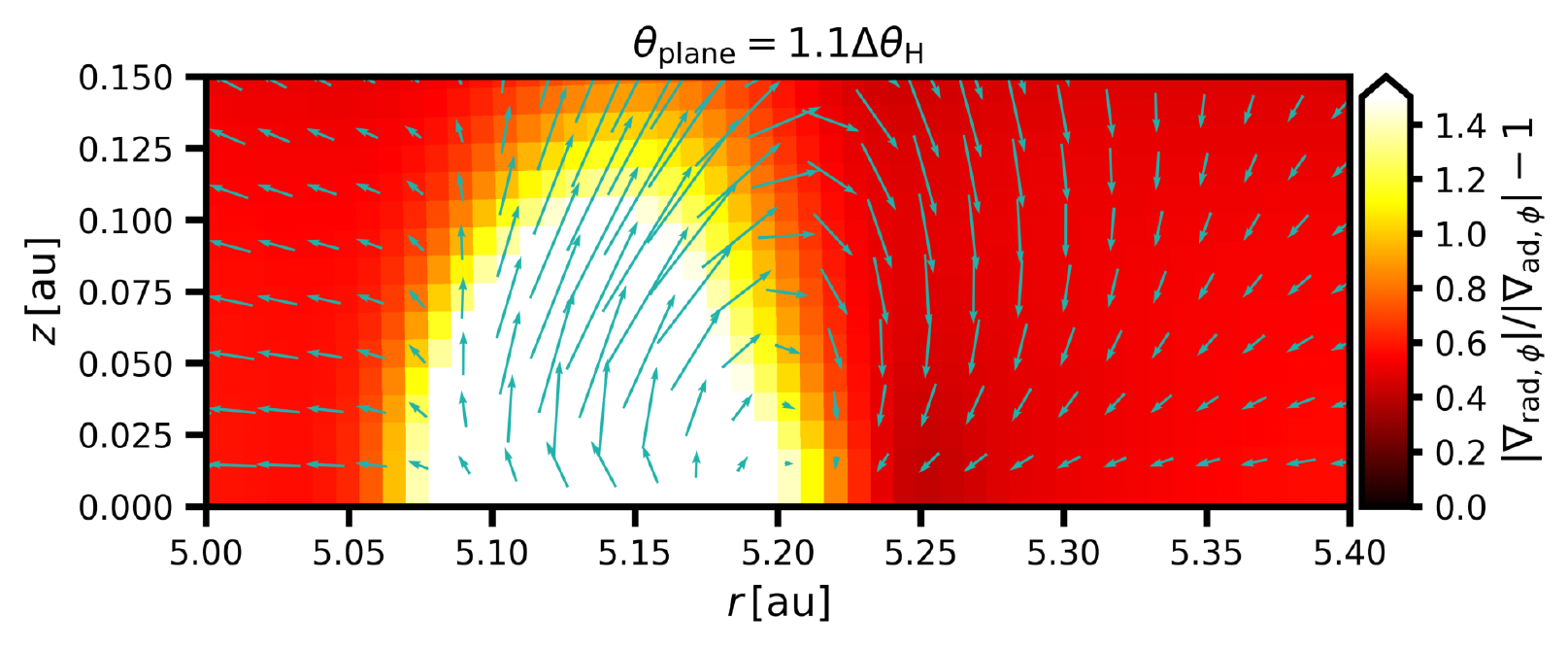} \\
  \end{tabular}
  \caption{Balance of the Schwarzschild criterion in the vertical planes
    of the \KC-disk (top) and the \KBL-disk (remaining panels).
    The individual vertical planes are chosen to track the 
    most prominent vertical flow at the given simulation time $t$ (see the main text)
    and their azimuthal separations from the protoplanet location 
    are given by panel labels (using multiples of the azimuthal
    span $\Delta\theta_{\mathrm{H}}$ of the Hill sphere radius).
    The colour maps evaluate Eq.~(\ref{eq:schwarzschild}).
    Positive values indicate superadiabatic vertical temperature gradients.
    The vertical velocity
    vector field is overlaid in the plots.
    The half-circles mark the overlap of a given plane with the Hill sphere
    (the planes in panels 3 and 5 do not overlap with the Hill sphere).
  }
  \label{fig:convecgrad}
\end{figure}

Fig.~\ref{fig:convecgrad} shows the vertical velocity field and balance 
of the Schwarzschild criterion in the vertical direction
of the \KC-disk and \KBL-disk. Each panel shows a different simulation time and vertical plane.
For the \KC-disk, we display $t=60\,P_{\mathrm{orb}}$ and
the vertical plane intersecting the protoplanet's location,
whereas for the \KBL-disk, each plane approximately intersects the centre around
which the captured streamlines of Fig.~\ref{fig:hydro_avrBL} circulate
at the given simulation time ($t=31.3$, $31.75$, $32.2$ and $32.65\,P_{\mathrm{orb}}$).
In other words, we choose the vertical
planes where we expect the most prominent vertical outflow.

The first thing we point out is that the background 
differs between the studied disks.
The background of the \KBL-disk is slightly superadiabatic,
contrary to the \KC-disk.
The difference arises as a result of the opacity laws.
As derived by \cite{Lin_Papaloizou_1980MNRAS.191...37L} and \cite{Ruden_Pollack_1991ApJ...375..740R},
one can make a qualitative estimate for optically
thick regions unperturbed by the protoplanet
\begin{equation}
  \left(\frac{|\nabla_{\mathrm{rad},\phi}|}{|\nabla_{\mathrm{ad},\phi}|}-1\right)_{\mathrm{background}}=\frac{1/\left( 4-\beta \right)}{\left( \gamma-1 \right)/\left( \gamma \right)}-1 \, ,
  \label{eq:convec_estimate}
\end{equation}
where $\beta$ is the power-law index of the
$\kappa \propto T^{\beta}$ dependence. In the \KC-disk,
$\beta=0$ and thus $|\nabla_{\mathrm{rad},\phi}|/|\nabla_{\mathrm{ad},\phi}|-1\simeq-0.17$.
In the \KBL-disk, the \cite{Bell_Lin_1994ApJ...427..987B} opacity
law in the given temperature range corresponds to water-ice
grains and exhibits $\beta=2$, therefore $|\nabla_{\mathrm{rad},\phi}|/|\nabla_{\mathrm{ad},\phi}|-1\simeq0.67$.
These estimated values are in a good agreement with the
background values of Fig.~\ref{fig:convecgrad}.

The background itself is not convective because vertical convection
is usually not self-sustainable
unless there is a strong heat deposition within the disk
\citep[e.g.][]{Cabot_1996ApJ...465..874C,Stone_Balbus_1996ApJ...464..364S,Klahr_etal_1999ApJ...514..325K,Lesur_Ogilvie_2010MNRAS.404L..64L}
which, however, can be provided by the hot accreting protoplanet
in our case. Indeed, looking at the \KC-disk in Fig.~\ref{fig:convecgrad} (top),
the region where we previously identified the most significant
temperature excess due to planetary luminosity (compare with Fig.~\ref{fig:vert_kc}) is superadiabatic
and the corresponding vertical outflow can be considered convective. 

In the \KBL-disk simulation, the temperature gradient departs
from the adiabatic one even more and additionally, the excess
is no longer centred above the protoplanet itself but rather
spans its vicinity. This can be seen
from the varying azimuthal coordinate of the displayed vertical planes
and also from the radial offset of the highly superadiabatic
region in panels 3 and 5.

We summarise the section by speculating that
the \KBL-disk simulation becomes destabilised because
the hot disturbance created by the accreting protoplanet
is subject to vertical buoyant forces
acting over a more extended region (compared to the \KC-disk).
The reason is that the hot disturbance is imposed
over an already-superadiabatic background.
The uplift of the material cannot be compensated
for in a stationary manner and eventually,
the vertical outflow becomes offset with respect to the protoplanet
and starts to change its position in a cyclic manner.
However, such a description is rather qualitative
and precise conditions for triggering
the instability should be explored in future works.

\subsection{Torque oscillation vs opacity gradient}
\label{sec:parametric}

We now perform a partial exploration of the
parametric space by varying the opacity gradient within
the disk. The aim is twofold: First, we would like to support
the claim of the previous Sect.~\ref{sec:vertical_convection}
about the importance of the vertical stratification for the torque oscillations.
Second, it is desirable to show that the appearance of torque oscillations in the \KBL-disk
is not coincidental and that it can be recovered
for a wider range of parameters.

We construct six additional disk models with
artificial opacity laws that i) conserve
the opacity value at the protoplanet location ($1.11\,\mathrm{cm^{2}\,g^{-1}}$)
and ii) lead to opacity gradients which are intermediate
between the \KC- and \KBL-disks. The latter property accounts
for the highly unconstrained size distribution
of solid particles in protoplanetary disks which
manifests itself in a large parametric freedom of the power-law
slope of the opacity profile \citep[e.g.][]{Piso_etal_2015ApJ...800...82P}.

The first additional set of disks utilises the opacity law which 
we dub $T$-dependent:
\begin{equation}
  \kappa\left(\bar{T}(r,\phi)\right) = \kappa_{0}\bar{T}^{\beta} \, .
  \label{eq:kappa_artif_T}
\end{equation}
Similarly to the \cite{Bell_Lin_1994ApJ...427..987B} opacity
in the water-ice regime, it is exclusively a function of temperature.
The temperature $\bar{T}$ is again azimuthally averaged to disentangle the 
influence of global opacity gradients (which we focus on) from
those related to accretion heating of the protoplanet.
We examine the values $\beta=1.5$, $1$ and $0.5$ to span
the range between $\beta=0$ (\KC-disk) and $2$ (\KBL-disk).
The constant of proportionality $\kappa_{0}$ is always chosen to recover
$\kappa=1.11\,\mathrm{cm^{2}\,g^{-1}}$ at $r=a_{\mathrm{p}}$ and $\phi=\pi/2$
in an equilibrium disk.
We find $\kappa_{0}\simeq0.0016$, $0.014$ and $0.122\,\mathrm{cm^{2}\,g^{-1}}$
for the respective values of $\beta$.

The opacity law used for the second additional set of disks, referred
to as $r$-dependent, is
\begin{equation}
  \kappa(r) = 1.11\left(\frac{r}{a_{\mathrm{p}}}\right)^{-\delta}\,\mathrm{cm^{2}\,g^{-1}} \, ,
  \label{eq:kappa_artif_r}
\end{equation}
and we choose $\delta=3$, $2$ and $1$ (because
$\delta=4$ leads to an opacity profile similar
to the \KBL-disk). The motivation for choosing
this purely radially dependent opacity law is to
distinguish between effects caused by radial and 
vertical opacity gradients. The latter
will not appear when $\kappa=\kappa(r)$.

\begin{figure}[!hpt]
  \centering
  \includegraphics[width=8.8cm]{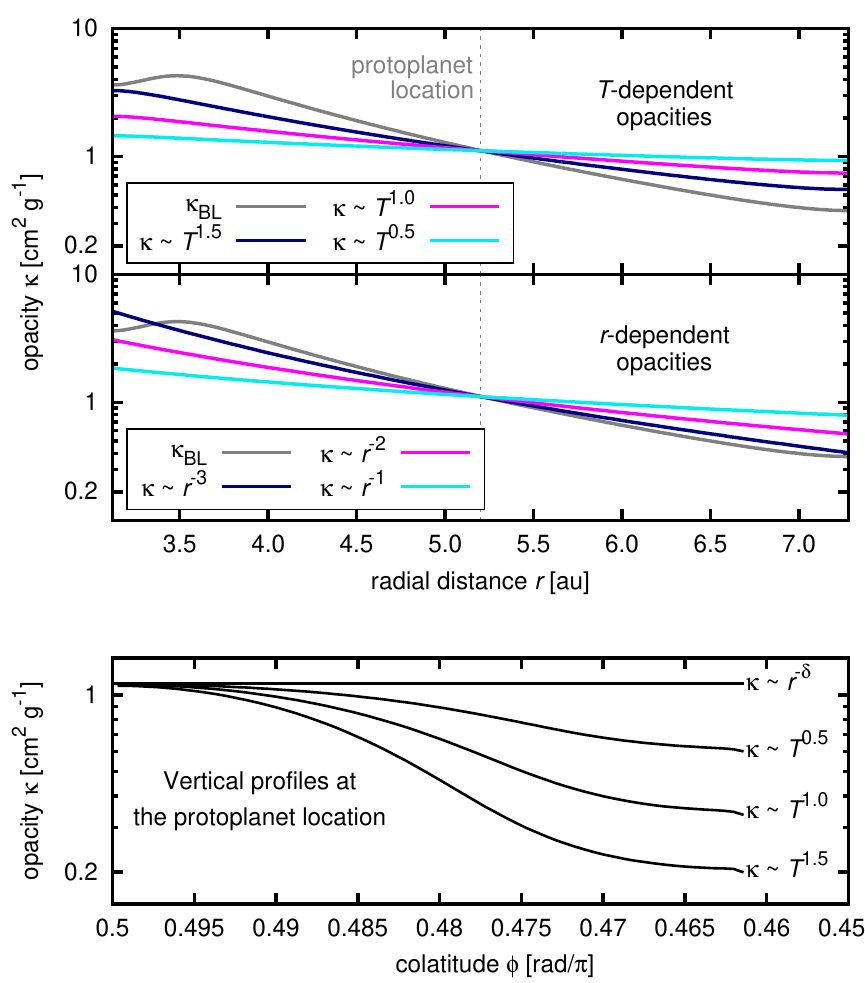}
  \caption{Radial (top two panels) and vertical (bottom) opacity profiles
  in equilibrium disks which we use to study the torque dependence on the opacity
  gradients. In top panels, the individual cases are distinguished by colour
  and labelled in the legend.
  The profile of the \KBL-disk (solid grey curve) is plotted for comparison.
  The bottom panel corresponds to the protoplanet location and demonstrates
  that only the disks with $T$-dependent opacities develop a vertical opacity gradient
  (which is not allowed for $r$-dependent opacities by construction).}
  \label{fig:kappa_artif}
\end{figure}

The opacity profiles of these disks in radiative equilibrium
are summarised in Fig.~\ref{fig:kappa_artif}
which reveals that all radial opacity gradients (top two panels)
are indeed shallower compared to the \KBL-disk.
However, all disks with $r$-dependent opacities
have zero vertical opacity gradient by construction (bottom panel),
unlike disks with $T$-dependent opacities which vertically decrease.

\begin{figure}[!hpt]
  \centering
  \includegraphics{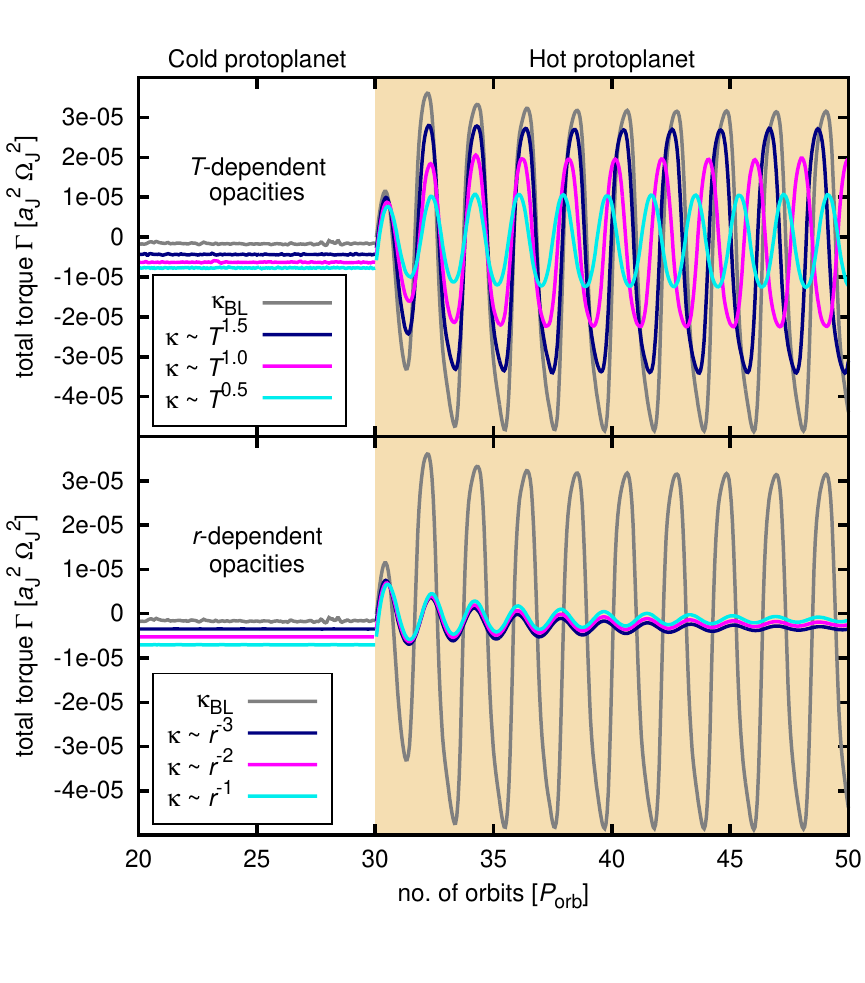}
  \caption{Torque evolution in disks described in Fig.~\ref{fig:kappa_artif}.
   The top panel corresponds to disks with $T$-dependent opacities and the bottom panel
   to those with $r$-dependent opacities. The individual cases are distinguished
   by colour and labelled in the legend. The evolution from the \KBL-disk
   simulation (solid grey curve) is given for reference. In the top panel,
   the torque amplitude diminishes with the power-law index of the opacity law,
   yet the oscillations appear in all studied cases.
   In the bottom panel, we find that oscillations are rapidly damped.}
  \label{fig:tqwk_artif}
\end{figure}

The torque measurements are done the same way as in our
previous experiments and the results are given in Fig.~\ref{fig:tqwk_artif}.
For disks with $T$-dependent opacities (top), we find that the torque
oscillations appear for all investigated values of $\beta$. The oscillation amplitude,
however, linearly decreases with $\beta$ and the period becomes slightly shorter as well.
In case of $r$-dependent opacities (bottom), the torque evolution
does not strongly depend on $\delta$ and exhibits
marginal and vanishing oscillations, as in the \KC-disk case.

Since the only qualitative difference between
the disks with $T$-dependent and $r$-dependent opacities
is in the vertical opacity gradient, our torque measurements
confirm the importance of the vertical structure for 
the torque oscillations.
The torque oscillates
wherever the vertical
opacity profile favours superadiabatic vertical stratification 
and the oscillation amplitude scales with the strength of the opacity-temperature
coupling.
In the absence of the vertical opacity gradient,
the oscillations are not established.

\subsection{Evolution of a migrating protoplanet}
\label{sec:migration}

The previous simulations were conducted with the assumption
of the fixed orbit of the protoplanet and the static
torque was examined.
Here we explore whether our findings
can be readily applied to a dynamical case when the protoplanet
is allowed to radially migrate and its semimajor axis evolves.
In this section, we focus only on the \KBL-disk in which we found
the flow instability.

\begin{figure}[!hpt]
  \centering
  \begin{tabular}{c}
    \includegraphics[width=7.8cm]{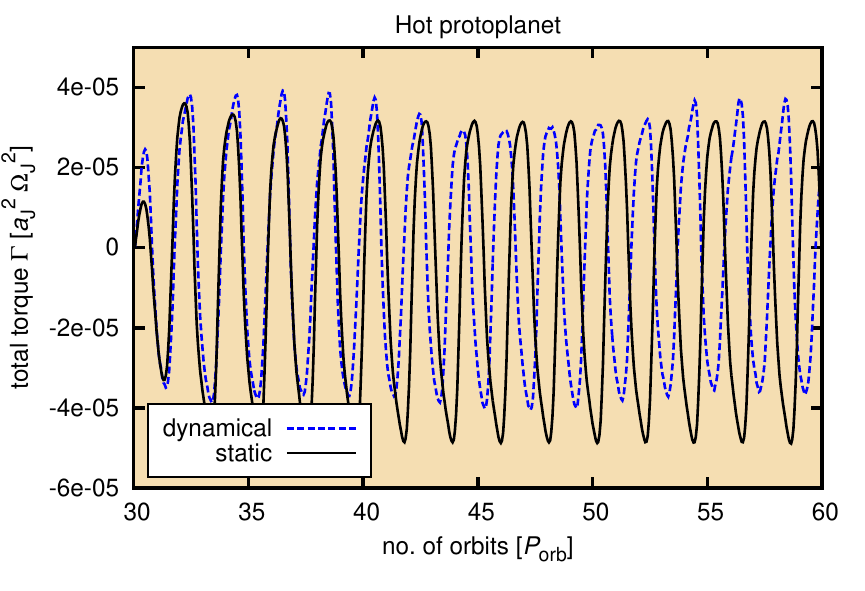} \\
    \includegraphics[width=7.8cm]{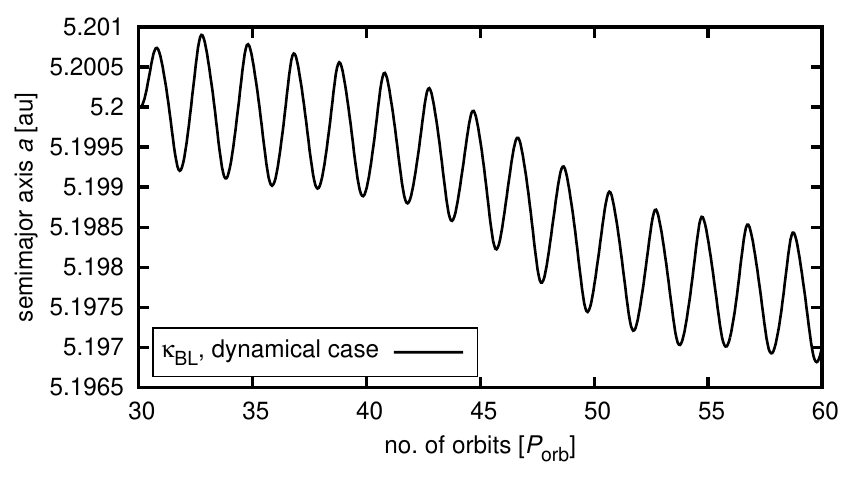} \\
  \end{tabular}
  \caption{
    Top: Comparison of the static (solid black curve)
    and dynamical torque (dashed blue curve) acting on the hot protoplanet
    in the \KBL-disk.
    Bottom: Evolution of the semimajor axis in the \KBL-disk
    when the protoplanet is allowed to migrate.
    The migration is inward and oscillatory.
  }
  \label{fig:dynamical}
\end{figure}

Starting from $t=30\,P_{\mathrm{orb}}$, we release
the protoplanet and run the simulation until $t=60\,P_{\mathrm{orb}}$.
Fig.~\ref{fig:dynamical} compares the obtained dynamical torque
with the previous result of our static experiments. 
It is obvious that the oscillating character of the torque
is retained and therefore the instability of the flow
operates near a moving protoplanet as well. 
There are differences both in the amplitude and phase of the torque oscillations,
but the mean value of the torque over the simulated period of time
is $\bar{\Gamma}\simeq-1.3\times10^{-6}\,a_{\mathrm{J}}^{2}\,\Omega_{\mathrm{J}}^{2}$.
Although this value is slightly more positive than
the static heating torque ($\bar{\Gamma}\simeq-6.3\times10^{-6}\,a_{\mathrm{J}}^{2}\,\Omega_{\mathrm{J}}^{2}$),
it is almost the same as the torque acting on the cold protoplanet
($\bar{\Gamma}\simeq-1.6\times10^{-6}\,a_{\mathrm{J}}^{2}\,\Omega_{\mathrm{J}}^{2}$).
We thus confirm that in the \KBL-disk, the heating torque does not add
any considerable positive contribution to the mean torque but makes it strongly oscillating instead.

Bottom panel of Fig.~\ref{fig:dynamical} shows the actual evolution
of the protoplanet's semimajor axis.
On average, the protoplanet slowly migrates inwards
but this drift is not smooth. The protoplanet exhibits fast periodic inward and outward
excursions on an orbital time scale. The migration rate
of these individual excursions (not to be confused with the mean
migration rate stated above) is $\dot{a}\sim(10^{-3}\,\mathrm{au})/P_{\mathrm{orb}}$.

Finally, we note that the mean migration rate is
not constant, which also corresponds to the varying offset of the dynamical
torque with respect to the static torque in Fig.~\ref{fig:dynamical}.
Probably, the unstable gas distribution around the protoplanet
is further affected by the protoplanet's radial drift.

\section{Discussion}
\label{sec:discussion}

This section discusses the applicability of our results,
draws links to some previous studies
and speculates about possible implications
for planet formation. Additionally,
we outline possibilities for future work.
Although most of them are beyond the scope
of this paper, we at least present several
additional simulations in Appendix~\ref{sec:supporting_simulations}
to verify the correctness of our model (we study the impact
of the luminosity increase, grid resolution, opacity treatment
and computational algorithm on the evolution of torque oscillations).

\subsection{Relation to previous works}
\label{sec:relation}

Our results revealed unexpected perturbations
of the gas flow in the vicinity of the protoplanet
undergoing strong accretion heating,
one of them being the vertical outflow.
The resolution of our simulations was originally tailored
for studying the torque and may lack some accuracy 
close to the protoplanet where the outflow occurs.
Follow-up studies should thus utilise 
simulations with high resolution
of the protoplanetary envelope,
similarly to e.g.
\cite{Tanigawa_etal_2012ApJ...747...47T,Fung_etal_2015ApJ...811..101F,Ormel_etal_2015MNRAS.447.3512O,Lambrechts_Lega_2017A&A...606A.146L}.

It is likely that a critical combination
of the opacity $\kappa$ and luminosity $L$
for a given planetary mass $M_{\mathrm{p}}$
exists for which the vertical outflow is triggered, similarly to the
dependencies of the heating torque explored by
\cite{Benitez-Llambay_etal_2015Natur.520...63B}.
\cite{Popovas_etal_2018MNRAS.479.5136P,Popovas_etal_2019MNRAS.482L.107P}
studied the stability
of the circumplanetary envelope during pebble accretion
and found that the gas within the Bondi sphere exhibits 3D convective motions,
assuming\footnote{Greater number of parameters were
  discussed in \cite{Popovas_etal_2018MNRAS.479.5136P,Popovas_etal_2019MNRAS.482L.107P} but we quote those closest to this paper.}
$L\simeq1.4\times10^{26}\,\mathrm{erg}\,\mathrm{s}^{-1}$, $M_{\mathrm{p}}=0.95\,M_{\oplus}$
and $\kappa=1\,\mathrm{cm}^{2}\,\mathrm{g}^{-1}$.
\cite{Lambrechts_Lega_2017A&A...606A.146L} also explored
a set of ($L$, $\kappa$, $M_{\mathrm{p}}$) parameters
and its impact on the
structure of the circumplanetary envelope.
They found the inner region of the envelope
to depart from hydrostatic equilibrium
when the luminosity exceeds $L=10^{27}\,\mathrm{erg}\,\mathrm{s}^{-1}$
around a $M_{\mathrm{p}}=5\,M_{\oplus}$ core within a $\kappa=1\,\mathrm{cm}^{2}\,\mathrm{g}^{-1}$ environment, but they did not identify any vertical outflow.
The outflow in our simulations 
appears for higher $L$ and smaller $M_{\mathrm{p}}$ compared
to \cite{Lambrechts_Lega_2017A&A...606A.146L},
we can therefore assume that our parameters cross the critical ones.

Regarding the baroclinic perturbations, although they are known to produce
vortical instabilities in protoplanetary disks
\citep{Klahr_Bodenheimer_2003ApJ...582..869K,Petersen_etal_2007ApJ...658.1236P,Petersen_etal_2007ApJ...658.1252P,Lesur_Papaloizou_2010A&A...513A..60L,Raettig_etal_2013ApJ...765..115R,Barge_etal_2016A&A...592A.136B},
they have been rarely considered in relation to
hot protoplanets. For example, \cite{Owen_Kollmeier_2017MNRAS.467.3379O}
claim that that hot protoplanets can excite large-scale
baroclinic vortices but we do not identify any of those in our simulations.
Instead, we find baroclinic perturbations to be
responsible for 3D distortion of the gas flow near the protoplanet.

\subsection{Implications for the formation of planetary systems}

Although the heating torque has been previously thought
to be strictly positive and also efficient in high-opacity
locations of protoplanetary disks, our paper
shows that it can exhibit more complicated
behaviour if the temperature dependence of the disk opacity is taken
into account.
We identified an oscillatory mode of the heating
torque in the disk region
with $\kappa\propto T^{2}$
and we demonstrated that it can operate
  even for shallower dependencies such as $\kappa\propto T^{0.5}$,
albeit with a decreased amplitude. This behaviour resembles
the nature of baroclinic and convective disk
instabilities which usually operate in the most opaque regions 
with the steepest entropy gradients 
but become less effective elsewhere \citep[e.g.][]{Pfeil_Klahr_2019ApJ...871..150P}.

It is worth noting that our simulations neglected the effect
of stellar irradiation. Stellar-irradiated disks
tend to have vertical temperature
profiles that are closer to being isothermal
\citep[e.g.][]{Flock_etal_2013A&A...560A..43F},
unlike disk models used in this work which have
rather adiabatic or slightly superadiabatic vertical temperature gradients.
On the other hand, even the irradiated disks often contain shadowed regions
protected against stellar irradiation,
for example behind the puffed-up inner rim
\citep[e.g.][]{Dullemond_etal_2001ApJ...560..957D}
or between the viscously
heated inner disk and a flared irradiated outer disk
\citep[e.g.][]{Bitsch_etal_2013A&A...549A.124B}.
In such regions, the oscillatory migration could
still operate.

For the aforementioned reasons,
it is likely that transition zones might exist
in protoplanetary disks, separating regions
where protoplanets migrate under the influence
of the standard positive heating torque
and where they undergo oscillatory migration.
Migration at the edges of such zones
could be convergent, leading to a pileup of
protoplanets.

\section{Conclusions}
\label{sec:conclusions}

By means of 3D radiation-hydrodynamic
simulations, we investigated the heating torque
\citep{Benitez-Llambay_etal_2015Natur.520...63B}
acting on a luminous $3\,M_{\oplus}$ protoplanet heated by
accretion of solids.
The aim was to compare the torque evolution and physics
in a disk with non-uniform opacities \citep{Bell_Lin_1994ApJ...427..987B}
with the outcome of a constant-opacity simulation.

We discovered that the gas flow near the protoplanet
is perturbed by two mechanisms:
\begin{enumerate}
  \item The gas advected past the protoplanet becomes
    hot and underdense. Consequently, a misalignment is created
    between the gradients of density and pressure within the protoplanet's
    Hill sphere. The baroclinic term of the
    vorticity equation ($\sim$$\nabla\rho\times\nabla P$)
    then becomes non-zero and modifies the vorticity of the flow.
  \item The efficient heat deposition in the midplane makes
    the vertical temperature gradient superadiabatic,
    thus positively
    enhancing vertical gas displacements.
\end{enumerate}
The streamline topology exhibits
a complex 3D distortion.
The most important feature are
spiral-like streamlines rising vertically
above the hot protoplanet,
forming an outflow column of gas escaping 
the Hill sphere.

In the constant-opacity disk, the vertical outflow is centralised
above the protoplanet, it temporarily captures streamlines from both horseshoe
regions and such a state is found to be stationary over the simulation
time scale. The distribution of the hot gas then remains
in accordance with findings of \cite{Benitez-Llambay_etal_2015Natur.520...63B}, having
a two-lobed structure, and so does the
resulting positive heating torque.

In the disk with non-uniform opacity
$\kappa\propto T^{2}$ (typically outside the water-ice line),
we find the superadiabatic temperature gradient
to be steeper and the distorted gas flow to be unstable.
The vertical spiral flow becomes offset with respect to the protoplanet
and periodically changes its position, spanning the edge
of the Hill sphere in a retrograde fashion.
Its motion is followed by the underdense gas and the resulting
heating torque strongly oscillates in time.
The interplay can be characterised by the following sequence:
\begin{enumerate}
  \item A stage when most of the captured streamlines
    originate in the rear horseshoe region
    and their spiral-like structure is offset ahead of the protoplanet.
    Therefore the hot gas cumulates ahead of the protoplanet,
    a dominant underdense lobe is formed there and the torque becomes negative,
    reaching the minimum of its oscillation.
  \item A stage when the front horseshoe region becomes completely isolated
    from the captured streamlines.
    Some of the rear horseshoe streamlines start to overshoot
    the protoplanet and make U-turns ahead of it. At the same time, the spiral-like
    structure recedes behind the protoplanet. The lobe from stage~1
    starts to decay while a rear lobe starts to grow and the torque changes
    from negative to positive.
  \item Antisymmetric situation to stage~1, when most of the captured streamlines
    originate in the front horseshoe region, the dominant lobe
    trails the protoplanet and the torque is positive, reaching the maximum
    of its oscillation.
  \item Antisymmetric situation to stage~2, when the torque decreases
    from positive to negative and the cycle repeats.
\end{enumerate}
Such an advective redistribution of the hot underdense gas
is sustained over the simulation time scale.

We also studied the dependence of the torque oscillations
on the opacity gradient and found that they can appear
even for $\kappa\propto T^{0.5}$, although their amplitude
linearly decreases with the power-law slope of the $\kappa(T)$ dependence.
We also demonstrated that the oscillations would vanish
in a disk with zero vertical opacity gradient.

If the protoplanet is allowed
to migrate, its mean migration rate is nearly unaffected
but the radial drift is not smooth. It is rather oscillatory, consisting
of brief inward and outward radial excursions with
a characteristic rate $\dot{a}\sim 10^{-4}\,\mathrm{au}\,\mathrm{yr}^{-1}$.

We discussed possible implications of the oscillating heating torque
for planet formation and pointed out
that it can affect the global evolution
of hot migrating low-mass protoplanets.
During their migration through disk regions with varying opacities
it might be possible for the protoplanets to
switch between the standard positive heating
torque of \cite{Benitez-Llambay_etal_2015Natur.520...63B}
and the positive/negative torque oscillations
discovered in this paper.

\begin{acknowledgements}

The work of OC has been supported 
by Charles University (research program no. UNCE/SCI/023;
project GA UK no. 624119; project SVV-260441),
by the Grant Agency of the Czech Republic (grant No. 18-06083S)
and by The Ministry of Education, Youth and Sports from the Large Infrastructures
for Research, Experimental Development and Innovations
project IT4Innovations National Supercomputing Centre
LM2015070.
Access to computing and storage facilities owned by
parties and projects contributing to the National Grid
Infrastructure MetaCentrum, provided under the programme
“Projects of Large Research, Development, and Innovations
Infrastructures” (CESNET LM2015042), is greatly appreciated.
OC would like to thank M.~Bro{\v z} and
B.~Bitsch for useful
discussions and also to E.~Lega
who kindly provided her numerical
code \textsc{fargoca} for comparison simulations.
The authors are grateful to an anonymous referee
whose insightful ideas helped to improve the
manuscript.

\end{acknowledgements}

\bibliographystyle{aa}
\bibliography{references}

\begin{appendix}

\section{Modifications of the numerical scheme}
\label{sec:numerics}

We implemented Eqs.~(\ref{eq:e_rad}) and (\ref{eq:e_int})
into \textsc{fargo3d} using their discrete form
derived by \cite{Bitsch_etal_2013A&A...549A.124B}
(see their Appendix~B). We introduce a minor
modification of the numerical scheme,
which allows to account for all the source terms
in a single substep. Using the same notation
as in \cite{Bitsch_etal_2013A&A...549A.124B},
the relation between the temperature
and radiative energy at $t+\Delta t$ is
\begin{equation}
  T^{n+1} = \eta_{1}+\eta_{2}E_{\mathrm{R}}^{n+1} \, ,
  \label{eq:T_ER_relation}
\end{equation}
and we redefine
\begin{equation}
  \eta_{1} = \frac{T^{n}+12\Delta t\frac{\kappa_{\mathrm{P}}}{c_{V}}\sigma\left(T^{n}\right)^{4}+\Delta t\frac{Q_{\epsilon{\mathrm{-indep}}}}{\rho c_{V}}}{1+16\Delta t\frac{\kappa_{\mathrm{P}}}{c_{V}}\sigma\left( T^{n} \right)^{3}+\Delta t\left( \gamma-1 \right)\nabla\cdot\vec{v}} \, ,
  \label{eq:eta1}
\end{equation}
\begin{equation}
  \eta_{2} = \frac{\Delta t\frac{\kappa_{\mathrm{P}}}{c_{V}}c}{1+16\Delta t\frac{\kappa_{\mathrm{P}}}{c_{V}}\sigma\left( T^{n} \right)^{3}+\Delta t\left( \gamma-1 \right)\nabla\cdot\vec{v}} \, .
  \label{eq:eta2}
\end{equation}

There are two changes with respect to \cite{Bitsch_etal_2013A&A...549A.124B}.
First, the compressional heating is included via the last term in the denominator of Eqs.~(\ref{eq:eta1}) and (\ref{eq:eta2}). Second, the $Q_{\epsilon\mathrm{-indep}}$
term is a sum of all heat sources that do not depend on $\epsilon$ (or $T$);
in our case $Q_{\epsilon\mathrm{-indep}}=Q_{\mathrm{visc}}+Q_{\mathrm{art}}+Q_{\mathrm{acc}}$.
We note that when necessary, the stellar irradiation term can also be easily
included in $Q_{\epsilon\mathrm{-indep}}$ but it is neglected
in this work.

\section{Supporting simulations}
\label{sec:supporting_simulations}

In this appendix, we summarize
several additional simulations designed
to confirm the robustness of our
conclusions.

\subsection{Simulation with a smoothly increasing luminosity of the protoplanet}
\label{sec:smooth_luminosity}

\begin{figure}[]
  \centering
  \includegraphics[width=8.8cm]{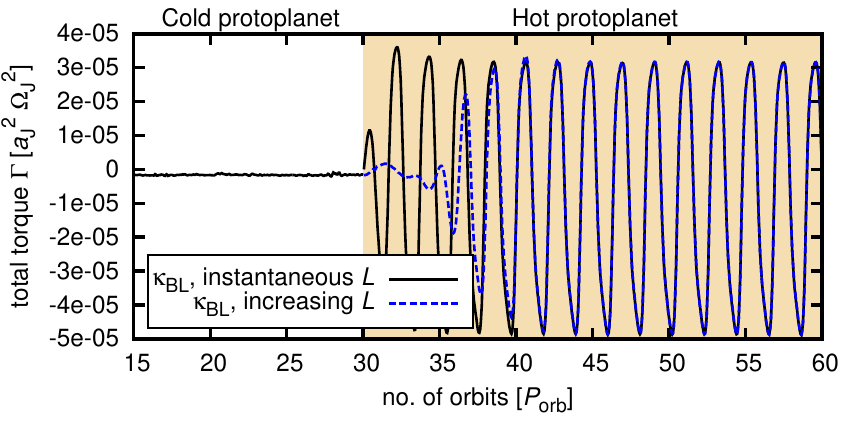}
  \caption{
    Torque evolution in the \KBL-disk
    with instantaneously increased luminosity of the protoplanet $L$
    (solid black curve; same as in Fig.~\ref{fig:torque})
    compared to the \KBL-disk with smoothly increasing $L$
    (dashed blue curve). Even before $L$ reaches its maximum value
    in the latter case, the curves start to overlap.
    After $t=40\,P_{\mathrm{orb}}$, the agreement is almost exact.
  }
  \label{fig:tqwk_smoothL}
\end{figure}

In our main simulations, we usually start the phase
with accretion heating abruptly, by instantaneously
increasing the luminosity of the protoplanet from $L=0$
to the value corresponding to the mass doubling time $\tau=100\,\mathrm{kyr}$.
Such a sudden appearance of a strong heat source might
produce an unexpected behaviour and instabilities
by itself. In order to exclude any undesirable 
behaviour, we repeat the \KBL-disk simulation with 
accretion heating of the protoplanet, but now we linearly
increase $L$ from zero at $t=30\,P_{\mathrm{orb}}$
to its maximal value over the time interval of $10\,P_{\mathrm{orb}}$.

Fig.~\ref{fig:tqwk_smoothL} shows the measured torque evolution.
Clearly, the gradual increase of $L$ has no impact on the
final character of the torque, the oscillations related to the flow
reconfigurations inevitably appear.

\subsection{Simulation with an increased azimuthal resolution}
\label{sec:increased_resol}

\begin{figure}[]
  \centering
  \includegraphics[width=8.8cm]{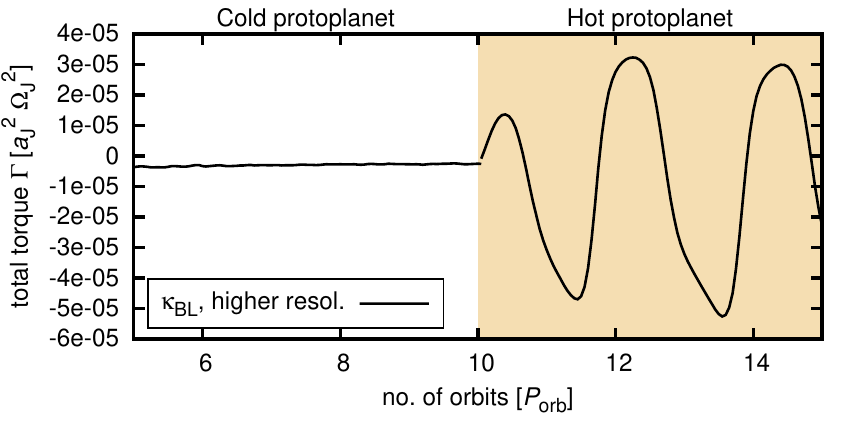}
  \caption{
    Torque evolution obtained in the \KBL-disk simulation
    with an increased azimuthal resolution of $2764$ cells.
    The time span of the individual phases is shortened
    to save computing time. The instability in the presence
    of the accretion heating is, however, recovered again.
  }
  \label{fig:tqwk_highres}
\end{figure}

The resolution in our simulations is motivated by works
of \cite{Lega_etal_2014MNRAS.440..683L} and \cite{Eklund_Masset_2017MNRAS.469..206E}.
Although we do not perform extended convergence tests of our own,
the resolution should be sufficient to recover a realistic
torque value and also a realistic advection-diffusion redistribution
of the hot gas near the protoplanet.

However, once the circumplanetary flow becomes unstable, it is no
longer clear if the chosen resolution is sufficient.
For example, one might argue that the coverage of the Hill
sphere by the grid cells in the azimuthal direction is too poor.
Here we present an experiment in which we double the 
number of the grid cells in the azimuthal direction in order to have
the same coverage of the Hill sphere in all directions.
We perform the \KBL-disk simulation again, however, we
shorten the phase without accretion heating to $10\,P_{\mathrm{orb}}$
and the phase with accretion heating to $5\,P_{\mathrm{orb}}$.

The result is shown in Fig.~\ref{fig:tqwk_highres} and
demonstrates that the toque oscillations are recovered
even when the increased resolution is used.
However, the amplitude of the torque
oscillations slightly changes, implying that the resolution dependence
should be explored more carefully in the future.

\subsection{Simulation with the unmodified Bell \& Lin opacity table}
\label{sec:unmodified_bl}

\begin{figure}[]
  \centering
  \includegraphics[width=8.8cm]{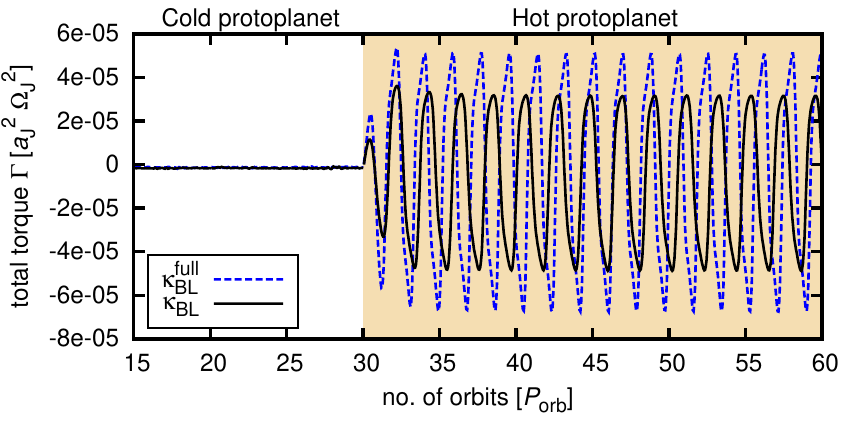}
  \caption{
    Comparison of the torque evolutions obtained in the 
    \KBL-disk (solid black curve; same as in Fig.~\ref{fig:torque}) and
    the $\kappa_{\mathrm{BL}}^{\mathrm{full}}$-disk (dashed blue curve).
    The first model inputs azimuthally-averaged values of $\rho$ and $T$
    to the opacity function of \cite{Bell_Lin_1994ApJ...427..987B},
    while the latter uses local values of $\rho$ and $T$.
    Both cases lead to the instability of the circumplanetary
    flow, but they differ in the amplitude of the torque oscillations.
  }
  \label{fig:tqwk_BLvsBL}
\end{figure}

The simulations of the \KBL-disk presented
in this paper are performed with a simplified
opacity law of \cite{Bell_Lin_1994ApJ...427..987B}
(explained in Sect.~\ref{sec:inicond}).
Here we test how the results change if
the unmodified opacity law $\kappa_{\mathrm{BL}}^{\mathrm{full}}$
is used and the dependence
on the local values of $T$ and $\rho$ is retained.

Fig.~\ref{fig:tqwk_BLvsBL} compares the torque evolution
in our standard \KBL-disk with that 
in a $\kappa_{\mathrm{BL}}^{\mathrm{full}}$-disk.
Clearly, the instability occurs in both disks, regardless
of whether or not the input values for the opacity
function are azimuthally averaged.
The only difference is in the torque amplitude
which is larger in the $\kappa_{\mathrm{BL}}^{\mathrm{full}}$-disk.

The increased amplitude occurs because 
if $T$ locally rises, so does the material opacity
($\kappa_{\mathrm{BL}}^{\mathrm{full}}\propto T^{2}$ 
in the given disk region).
Subsequently, the radiative cooling
of the hot gas becomes less efficient
and $T$ rises even more. As a result,
the underdense perturbations related to 
any temperature excess are more pronounced, leading to the
larger amplitude of the torque oscillations.

\subsection{Code comparison}
\label{sec:verif}

\begin{figure}[]
  \centering
  \includegraphics[width=8.8cm]{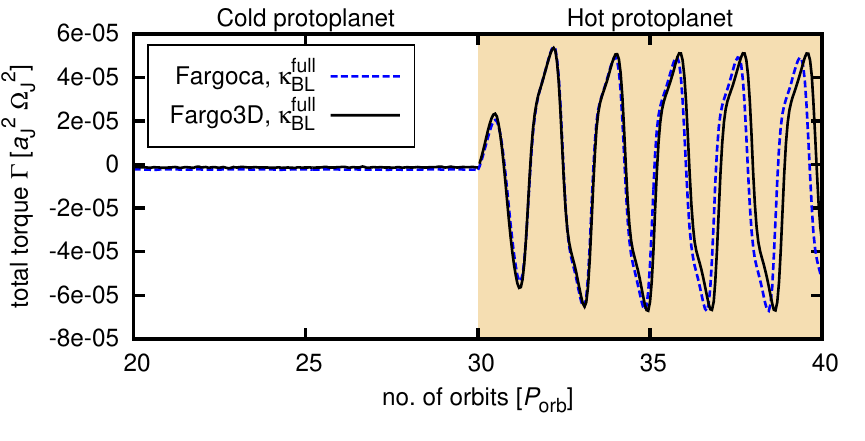}
  \caption{
    Comparison of the torque evolutions obtained with two independent
    codes \textsc{fargo3d} (solid black curve) and \textsc{fargoca} (dashed blue curve).
    The unmodified $\kappa_{\mathrm{BL}}^{\mathrm{full}}$ opacity table of \cite{Bell_Lin_1994ApJ...427..987B}
    was used in this case.}
  \label{fig:verif}
\end{figure}

With the aim to confirm that our implementation
of the energy equations in \textsc{fargo3d} is correct
and also that the instability of the circumplanetary
flow does not arise due to numerical artefacts,
we present a comparison simulation obtained
with an independent and well-tested
radiation hydrodynamic code \textsc{fargoca}
\citep{Lega_etal_2014MNRAS.440..683L}.
The simulation is performed
with the unmodified $\kappa_{\mathrm{BL}}^{\mathrm{full}}$ 
opacity law of \cite{Bell_Lin_1994ApJ...427..987B}.

Fig.~\ref{fig:verif} compares the torque evolution
found using our code with the one obtained with \textsc{fargoca}.
We can see that the converged torque for the cold protoplanet
is in a satisfactory agreement.

When the accretion heating is initiated, the same oscillatory trend
is observed with both codes. The curves overlap at first;
later they start to depart in terms of the oscillation phase.
However, the amplitude remains the same.

Since the converged torques are in agreement and the instability 
is recovered, we conclude that
the differences that we identify in Fig.~\ref{fig:verif}
arise only because our radiation module in \textsc{fargo3d}
relies on a slightly different numerical scheme (see Appendix~\ref{sec:numerics})
compared to \textsc{fargoca}.

\section{Vorticity equation in the corotating frame}
\label{sec:vorticity_eq}

For reader's convenience, we provide a step-by-step
derivation of the vorticity equation. Starting with Eq.~(\ref{eq:naviere_stokes}),
we apply the curl on both sides.
In the following, we neglect the viscous
term (large Reynolds number limit).

The following identities of the vector calculus will be utilised:
\begin{equation}
  \left(\vec{a}\cdot\nabla\right)\vec{a} = \frac{1}{2}\nabla\left( \vec{a}\cdot\vec{a} \right) + \left( \nabla\times\vec{a} \right)\times\vec{a} \, ,
  \label{eq:identity_curl_dirderiv}
\end{equation}
\begin{equation}
  \nabla\times\left( \vec{a}\times\vec{b} \right) = \vec{a}\left( \nabla\cdot\vec{b} \right) - \vec{b}\left( \nabla\cdot\vec{a} \right)
  + \left( \vec{b}\cdot\nabla \right)\vec{a} - \left( \vec{a}\cdot\nabla \right)\vec{b}\, ,
  \label{eq:identity_curl_cross}
\end{equation}
\begin{equation}
  \nabla\cdot\left(\nabla\times\vec{a} \right) = 0 \, ,
  \label{eq:identity_div_curl}
\end{equation}
\begin{equation}
  \nabla\times\left(\nabla f\right) = \vec{0} \, ,
  \label{eq:identity_curl_grad}
\end{equation}
where the last identity holds for
scalar functions that are at least twice continuously differentiable.

The curl of the advection term yields
\begin{equation}
  \nabla\times\left[\left(\vec{v}\cdot\nabla\right)\vec{v}\right] = 
    \nabla\times\left[\frac{1}{2}\nabla\left( \vec{v}\cdot\vec{v} \right) + \left( \nabla\times\vec{v} \right)\times\vec{v} \right]= 
    \nabla\times\left[\vec{\omega_{\mathrm{r}}}\times\vec{v} \right] \, ,
  \label{eq:advec_term_1}
\end{equation}
where we used Eq.~(\ref{eq:identity_curl_dirderiv}) in writing the first equality,
Eq.~(\ref{eq:identity_curl_grad}) to remove the $\sim$$\vec{v}\cdot\vec{v}$ term
and we defined the relative vorticity in the corotating frame $\vec{\omega_{\mathrm{r}}}=\nabla\times\vec{v}$.
Using Eqs.~(\ref{eq:identity_curl_cross}) and (\ref{eq:identity_div_curl}), we further obtain
\begin{equation}
  \nabla\times\left[\left(\vec{v}\cdot\nabla\right)\vec{v}\right] = 
  \vec{\omega_{\mathrm{r}}}\left( \nabla\cdot\vec{v} \right) + \left( \vec{v}\cdot\nabla \right)\vec{\omega_{\mathrm{r}}} - \left(\vec{\omega_{\mathrm{r}}}\cdot\nabla\right)\vec{v} \, ,
  \label{eq:advec_term_2}
\end{equation}

The curl of the pressure term leads to
\begin{equation}
  \nabla\times\left( \frac{\nabla P}{\rho} \right) = -\frac{1}{\rho^{2}}\nabla\rho\times\nabla P \, ,
  \label{eq:baro}
\end{equation}
because $\nabla\times\left( \nabla P \right)=\vec{0}$ (Eq.~\ref{eq:identity_curl_grad}).

When dealing with the gravitational term,
it is useful to realise that the centrifugal
acceleration can be expressed as
$\vec{\Omega}\times\left( \vec{\Omega}\times\vec{r} \right) = \nabla \Phi_{\mathrm{c}}$,
with the centrifugal potential $\Phi_{\mathrm{c}} = - \frac{1}{2}r_{\perp}^{2}\Omega^{2}$.
The curl of a combined force term, $\nabla\times\left[ \nabla\left(\Phi+\Phi_{\mathrm{c}} \right) \right]$,
is zero owing to Eq.~(\ref{eq:identity_curl_grad}).

Finally, we take the curl of the Coriolis acceleration:
\begin{equation}
%  \nabla\times(2\vec{\Omega}\times\vec{v}) = 2\left[\vec{\Omega}(\nabla\cdot\vec{v}) + (\vec{v}\cdot\nabla)\vec{\Omega} - \left( \vec{\Omega}\cdot\nabla \right)\vec{v}\right] \, ,
  \nabla\times(2\vec{\Omega}\times\vec{v}) = 2\left[\vec{\Omega}(\nabla\cdot\vec{v}) - \left( \vec{\Omega}\cdot\nabla \right)\vec{v}\right] \, ,
  \label{eq:coriolis}
\end{equation}
where we removed terms $\sim$$\nabla\cdot\vec{\Omega}$, $\sim$$\nabla\vec{\Omega}$
because $\vec{\Omega}$ in our simulations is constant.

Recollecting all the terms, we can write the relative vorticity equation
\begin{equation}
  \frac{\partial\vec{\omega_{\mathrm{r}}}}{\partial t} + \left( \vec{v}\cdot\nabla \right)\vec{\omega_{\mathrm{r}}} =
  \frac{\mathrm{D}\vec{\omega_{\mathrm{r}}}}{\mathrm{D}t} = \left(\vec{\omega_{\mathrm{a}}}\cdot\nabla  \right)\vec{v} - \vec{\omega_{\mathrm{a}}}\left( \nabla\cdot\vec{v} \right) + \frac{\nabla\rho\times\nabla P}{\rho^{2}} \, ,
  \label{eq:vorticity_appendix}
\end{equation}
where we defined the absolute vorticity
in the inertial frame $\vec{\omega_{\mathrm{a}}}=\vec{\omega_{\mathrm{r}}} + 2\vec{\Omega}$.

\end{appendix}

\end{document}